\documentclass[letterpaper,11pt]{article}
\usepackage[utf8]{inputenc}

\usepackage{geometry}
\usepackage{setspace}
\usepackage{authblk}
\usepackage{amsmath}
\usepackage{amsfonts}
\usepackage{graphicx}
\usepackage{float}
\usepackage{subcaption}

\usepackage[numbers,sort&compress,super]{natbib}

\usepackage{multicol}
\usepackage{multirow}
\usepackage{xcolor}
\usepackage{booktabs}

\usepackage[colorlinks=true, allcolors=blue]{hyperref}

\renewcommand{\figurename}{Fig.}
\renewcommand{\tablename}{Table}

\usepackage{newtxtext,newtxmath}

\usepackage{bbm}

\usepackage{booktabs}
\usepackage{array}
\usepackage[most]{tcolorbox}
\usepackage{caption}

\usepackage{tocloft}
\usepackage{caption}

\begin{document}
\title{\textbf{The Rise of Large Language Models and the Direction and Impact of US Federal Research Funding}}

% \date{}
\date{\vspace{0.5em}\normalsize Forthcoming in \textit{Proceedings of the National Academy of Sciences of the United States of America (PNAS)}}

\author[1,2,3,4]{Yifan Qian}
\author[1,2,3,4]{Zhe Wen}
\author[1,2,3,4]{Alexander C. Furnas}
\author[1,2,3,4]{Yue Bai}
\author[1,2,3,5]{Erzhuo Shao}
\author[1,2,3,4,5,*]{Dashun Wang}
\affil[1]{Center for Science of Science and Innovation, Northwestern University, Evanston, IL, USA}
\affil[2]{Ryan Institute on Complexity, Northwestern University, Evanston, IL, USA}
\affil[3]{Northwestern Innovation Institute, Northwestern University, Evanston, IL, USA}
\affil[4]{Kellogg School of Management, Northwestern University, Evanston, IL, USA}
\affil[5]{McCormick School of Engineering, Northwestern University, Evanston, IL, USA}
\affil[*]{Correspondence to: dashun.wang@kellogg.northwestern.edu}

\maketitle

\section*{Abstract}
Federal research funding shapes the direction, diversity, and impact of the US scientific enterprise. Large language models (LLMs) are rapidly diffusing into scientific practice, holding substantial promise while raising widespread concerns. Despite growing attention to AI use in scientific writing and evaluation, little is known about how the rise of LLMs is reshaping the public funding landscape. Here, we examine LLM involvement at key stages of the federal funding pipeline by combining two complementary data sources: confidential National Science Foundation (NSF) and National Institutes of Health (NIH) proposal submissions from two large US R1 universities, including funded, unfunded, and pending proposals, and the full population of publicly released NSF and NIH awards. We find that LLM use rises sharply beginning in 2023 and exhibits a bimodal distribution, indicating a clear split between minimal and substantive use. Across both private submissions and public awards, higher LLM involvement is consistently associated with lower semantic distinctiveness, positioning projects closer to recently funded work within the same agency. The consequences of this shift are agency-dependent. LLM use is positively associated with proposal success and higher early-stage publication output at NIH, whereas no comparable associations are observed at NSF. Notably, the productivity gains at NIH are concentrated in non-hit papers rather than the most highly cited work. Together, these findings provide large-scale evidence that the rise of LLMs is reshaping how scientific ideas are positioned, selected, and translated into publicly funded research, with implications for portfolio governance, research diversity, and the long-run impact of science.

\newpage
\section*{Introduction}
Federal research funding is the primary mechanism through which the United States converts public resources into scientific knowledge~\cite{bush1945endless,stephan2015economics}. Through competitive grant programs at agencies such as the National Science Foundation (NSF) and the National Institutes of Health (NIH), decisions about the allocation of taxpayer dollars influence which scientific ideas receive support, which fields advance, and which investigators are able to initiate and sustain research programs~\cite{li2017applied,galkina2018contribution,azoulay2019public,fleming2019government,yin2022public,azoulay2025if,furnas2025partisan,wang2025funding}. Thus, the federal funding process plays a central role in shaping the direction, diversity, and long-run impact of the US scientific enterprise and---given America's scientific leadership---the global research landscape as well~\cite{li2017applied,galkina2018contribution,azoulay2019public,fleming2019government,yin2022public,azoulay2025if,furnas2025partisan,wang2025funding}. Understanding forces that influence this process is therefore essential not only for science policy and the rate and direction of scientific progress, but also for the stewardship and accountability of public investment in research~\cite{yin2019quantifying,wang2019early,li2015big,peng2024promotional}.

Despite growing attention to the use of AI in science~\cite{liang2025quantifying,kusumegi2025scientific,liang2024monitoring,liang2025widespread,kobak2025delving,liu2025ai,bao2025there,wang2023scientific,gao2024quantifying,jabarian2025artificial,swanson2025virtual,hao2026artificial,frank2019evolution}, little is known about how the rise of large language models (LLMs) is reshaping the public funding landscape. This gap is consequential, as funding decisions operate upstream of publication, hiring, and recognition, shaping which scientific ideas are pursued long before other forms of evaluation occur. Yet proposal texts, especially unfunded submissions, are typically confidential, making this stage of the funding pipeline difficult to study systematically at scale. As a result, we lack systematic evidence on how LLM adoption enters the funding pipeline, how it relates to proposal evaluation, and how it shapes the research that ultimately receives public support.

Two broader questions follow from recent work on AI in science~\cite{kusumegi2025scientific,hao2026artificial}. One is whether the productivity gains associated with generative AI primarily reflect lower communication costs or broader improvements in scientific execution. Another is whether generative AI matters only for scientific outputs, such as manuscripts and citations, or is already reshaping the upstream processes through which ideas are proposed, evaluated, and selected for public support. Federal research funding provides a useful setting for examining both questions, because proposal review sits at the interface between idea articulation and public investment, yet has remained difficult to observe at scale.

 LLMs represent a potentially transformative development for scientific work~\cite{liang2025quantifying,kusumegi2025scientific,liang2024monitoring,liang2025widespread,kobak2025delving,liu2025ai,bao2025there}. Recent large-scale evidence shows that the diffusion of LLMs has already been accompanied by sharp increases in scientific production across major preprint platforms, underscoring their potential to substantially raise output at scale~\cite{kusumegi2025scientific}. By dramatically reducing the cost of drafting, synthesizing, and revising complex text, LLMs may help scientists articulate ideas more clearly, navigate the growing burden of knowledge, and explore unfamiliar domains more efficiently~\cite{jones2009burden,hill2025pivot}. In principle, such tools could accelerate the pace of discovery, lower barriers to entry, facilitate interdisciplinary recombination, and support more effective pivoting across research directions~\cite{wang2023scientific,hill2025pivot,liu2021understanding,tripodi2025tenure,shao2025sciscigpt,wang2026ai}. From this perspective, LLMs hold the promise of expanding scientific exploration and improving the efficiency with which ideas are generated, refined, and communicated~\cite{bail2024can,musslick2025automating}. 

Yet these same features also raise critical concerns~\cite{bail2024can}. Because LLMs are trained on large corpora of existing scientific text, including published papers and public records of funded projects, they may tend to produce high-probability, discipline-typical language. When used in proposal preparation, such tools may enhance fluency and alignment with prevailing norms while reducing rhetorical and topical variance, making proposed projects more closely resemble those that have recently succeeded~\cite{gao2024quantifying,hao2026artificial,doshi2024generative}. In this way, LLM use could shift the balance between exploration and exploitation in scientific funding, pulling proposals toward the center of existing funding patterns and potentially narrowing the diversity of ideas that are submitted for or receive public support, even as it may lower the barriers to producing competitive applications~\cite{liu2021understanding,march1991exploration}. Recent work highlights this tension more broadly, showing that AI can expand individual scientists' productivity and impact while simultaneously contracting the collective focus of science, underscoring the possibility that gains at the individual level may coexist with reduced diversity at the system level~\cite{hao2026artificial}.

Beyond these concerns, a bottlenecks perspective further suggests caution in extrapolating writing-related productivity gains to downstream scientific outputs~\cite{jones2025artificial}. Discovery is constrained by multiple frictions, from data collection and experimental throughput to coordination, implementation, and more, many of which LLM involvement does not directly resolve. Because the outputs of funded projects depend on both communication and execution, there are reasons to believe that productivity gains observed in manuscript writing may not translate to the impact of funded projects. This distinction makes federal research funding a particularly important context to evaluate how LLM adoption reshapes scientific activity.

Reflecting the policy relevance of this uncertainty, federal agencies have begun to issue guidance on generative AI in proposal preparation. In July 2025, NIH clarified that applications ``substantially developed by AI'' will not be considered applicants' original work (see \href{https://grants.nih.gov/grants/guide/notice-files/NOT-OD-25-132.html}{NOT-OD-25-132}), while NSF has emphasized investigator responsibility for accuracy and originality and encouraged disclosure of generative AI use (see \href{https://www.nsf.gov/news/notice-to-the-research-community-on-ai}{Notice to research community}). These actions underscore both the growing salience of LLMs in public funding and the need for empirical evidence on how they are being used and with what consequences.

Here, we examine LLM use at key stages in the federal funding pipeline by employing two complementary levels of analysis. First, we collect novel datasets containing the complete set of NSF (D1) and NIH (D2) proposal submissions (hereafter private NSF and NIH data) from two large US R1 universities with request start dates from 2021 to 2025, including funded, unfunded, and pending proposals. These data offer us a unique opportunity to observe proposal language at the point of submission. Second, we analyze the full population of publicly released NSF (D3) and NIH (D4) award abstracts (hereafter public NSF and NIH data) over the same period, together with publications supported by these awards. These data allow us to systematically analyze the funded awards (see \textit{Materials and Methods} for details on datasets). Overall, these data enable us to characterize the evolution of LLM use in federal funding and to examine how it relates to the positioning of scientific ideas, their selection for funding, and their translation into publicly supported research output.

\section*{Rapid rise and bimodal distribution of LLM use in US federal research funding}
To estimate LLM use in federal grant proposals and awards, we analyze textual traces in their abstracts. Specifically, we leverage an established detection method proposed by Liang \textit{et al.}~\cite{liang2025quantifying}. Using public grant abstracts with start dates in 2021, which preceded the widespread use of LLMs, we estimate the word distribution of human-written text. We also estimate a corresponding word distribution of LLM-generated text by prompting OpenAI's GPT-3.5-turbo-0125 model to rewrite the same abstracts, simulating LLM involvement in proposal language. Comparing these two distributions allows us to identify systematic differences between human and LLM-modified writing (see \textit{SI Appendix}~\ref{subsection:llmdetection} for details on model training).

We then apply the detection method at both the corpus and individual grant levels. At the corpus level, we pool grant abstracts for each focal month and its two adjacent months, and we estimate the fraction of LLM-modified sentences ($\alpha$) within this combined set in each dataset. At the individual grant level, we estimate $\alpha$ separately for each grant abstract in each dataset. These two levels represent complementary estimates, as the corpus-level estimates leverage large volumes of text to capture aggregate LLM use over time, whereas grant-level estimates enable grant-level analyses that capture heterogeneity across grants, allowing us to examine how LLM use relates to semantic distinctiveness and other grant outputs.

\begin{figure}[H]
\centering
\includegraphics[width=1.0\columnwidth]{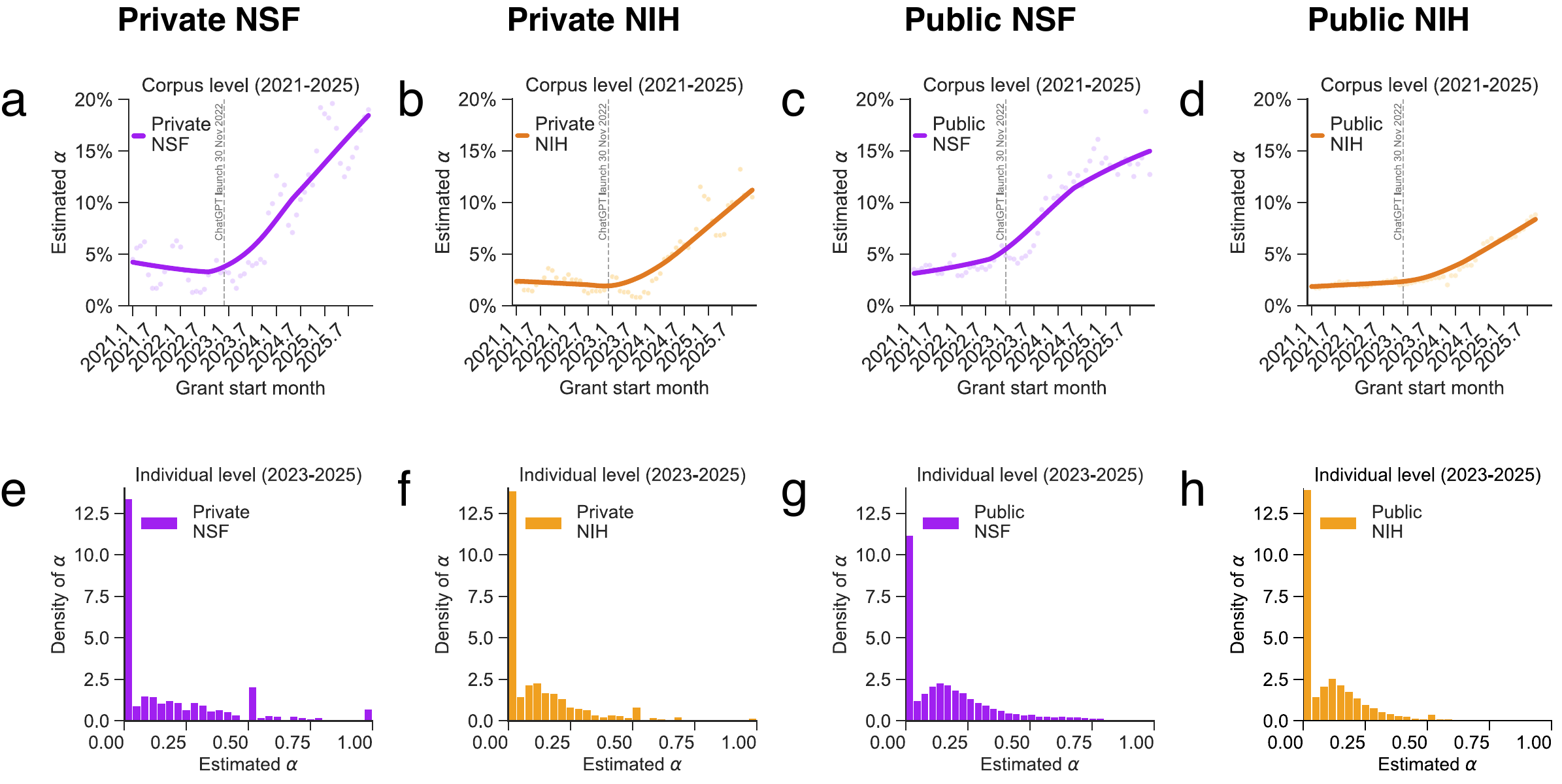}
\caption{\textbf{Rapid rise and bimodal distribution of LLM use in US federal research funding.}  (\textbf{a}-\textbf{d}) Corpus-level estimates of LLM use ($\alpha$) for private and public NSF and NIH grants from 2021 to 2025, computed using rolling three-month windows (points). Solid lines show locally weighted regressions. The vertical dashed line marks November 30, 2022, corresponding to the public release of ChatGPT. (\textbf{e}-\textbf{h}) Distributions of individual-grant $\alpha$ for private and public NSF and NIH proposals and awards with start dates between 2023 and 2025, showing a bimodal pattern consistent with a split between minimal and substantive LLM use across grants.}
\label{fig:Figure1}
\end{figure}

Figures~\ref{fig:Figure1}a-d trace the evolution of LLM involvement ($\alpha$) in NSF and NIH grants across two stages of the funding pipeline, proposal submission and award funding, drawing on confidential proposal submission abstracts (private data) and publicly released award abstracts (public data), respectively. Across all four datasets, $\alpha$ is relatively stable prior to late 2022, and then rises sharply beginning in 2023, coinciding with the widespread availability of ChatGPT. This increase is visible both at the point of submission, when ideas first enter the system, and among funded awards that have passed peer review, indicating that LLM use is not confined to a single stage of the pipeline. Over this period, NIH exhibits the same temporal pattern as NSF but with overall lower average levels of LLM involvement.

At the level of individual grants, Figs.~\ref{fig:Figure1}e-h show that $\alpha$ is not smoothly distributed. Among grants starting in 2023-2025, we observe a bimodal distribution, with one mode near zero and a second centered around roughly 10-15\%. This pattern reveals a clear split: one group of grant abstracts that largely avoids LLM use and another that incorporates LLMs more substantively into proposal preparation. These results indicate that LLM adoption has rapidly become a detectable feature of federal research proposals and awards, but it remains highly uneven across them.

\section*{LLM use and semantic distinctiveness in US federal research funding}
The rapid and uneven adoption of LLMs prompts another critical line of inquiry: Is greater LLM involvement associated with a shift in the kinds of ideas that enter, and ultimately shape, the federal research portfolio? To quantify the positioning of ideas, we measure the semantic distinctiveness of each proposal or award abstract (D1-D4) relative to the agency's recent funding frontier, building on recent work using transformer-based embeddings~\cite{hao2026artificial,scharfmann2025pasteur}. Specifically, for grants starting between 2023 and 2025, we compute each abstract's average cosine distance (using SPECTER2~\cite{singh2023scirepeval} embeddings, an established embedding method for scientific texts) to all abstracts funded in the previous year within the same agency. We then convert these distances into within-year percentiles, where higher values indicate more distinctive ideas relative to other grants in that agency-year. We adopt this distance transformation for interpretability, so that the coefficients can be read as shifts in relative position within the empirical distribution of proposals in the same agency-year.

\begin{figure}[H]
\centering
\includegraphics[width=0.55\columnwidth]{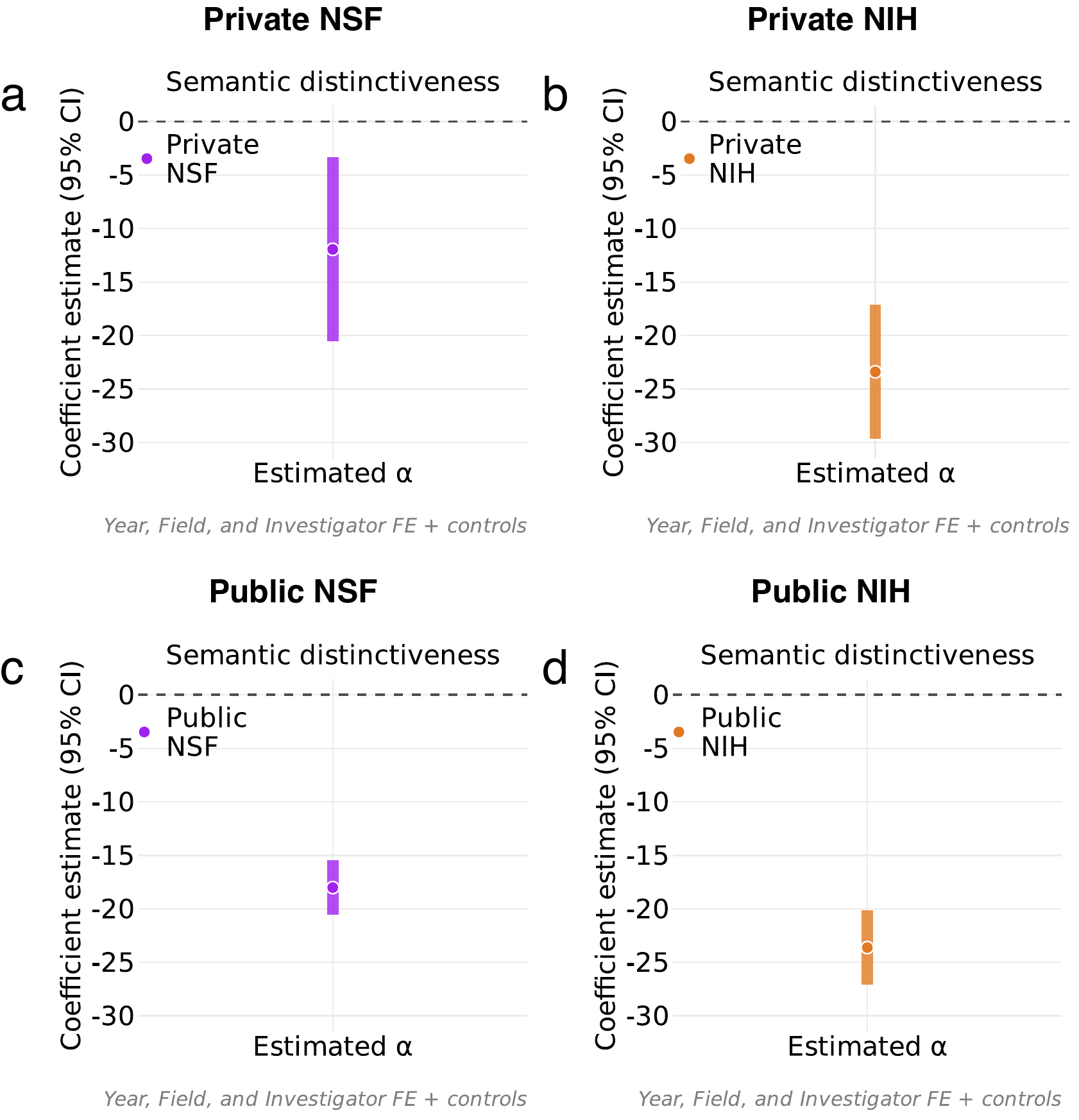}
\caption{\textbf{LLM use and semantic distinctiveness in US federal research funding.}  (\textbf{a}-\textbf{d}) Regression estimates relating grant-level LLM use ($\alpha$) to semantic distance from abstracts funded in the prior year within the same agency, expressed as within-year percentiles. Panels show results separately for private NSF (\textbf{a}), private NIH (\textbf{b}), public NSF (\textbf{c}), and public NIH (\textbf{d}) grants. All regressions include grant start year, field, and investigator fixed effects, as well as controls for funding amount. Points indicate coefficient estimates, and bars denote 95\% confidence intervals. Negative coefficients correspond to proposals and awards that are positioned closer, in semantic space, to recently funded work within the same agency.}
\label{fig:Figure2}
\end{figure}

We relate LLM involvement ($\alpha$) to distinctiveness, estimating regressions that include grant start year, field, and investigator (PI and co-PI) fixed effects, along with controls for requested or awarded funding amount (\textit{SI Appendix}~\ref{subsection:regression}). These fixed effects absorb time-invariant heterogeneity at the field and investigator levels while accounting for common shocks across grant start years. Across all four datasets, including private submissions and public awards for both NSF and NIH, higher $\alpha$ is consistently associated with lower distinctiveness percentiles (Fig.~\ref{fig:Figure2}, \textit{SI Appendix}~\ref{subsection:regression}, \textit{SI Appendix} Tables~\ref{tab:PrivateNSF-Embedding-1yr}-\ref{table:NIH_public_1y_distance}), indicating that proposals and awards with greater LLM involvement are positioned closer to the grants that the agency funded most recently.  For example, for NSF awards, moving from low LLM involvement (25th percentile) to high LLM involvement (75th percentile) corresponds to a $\sim$5-point decrease in the distinctiveness percentile; for NIH awards, the decrease is $\sim$4 points. Because distinctiveness is measured as a within-year percentile rank relative to the full distribution of proposals and awards in the same agency and year, this shift indicates that proposals and awards with higher LLM involvement are repositioned meaningfully among their immediate peers. Importantly, the investigator fixed effects compare each investigator only to themselves across proposals and awards rather than comparing different investigators to one another. Thus, these estimates capture within-investigator changes in positioning: for the same investigator's proposals and awards, when the LLM involvement is higher, their proposals and awards become systematically less distinctive. These findings suggest that as LLM use rises, the proposal and award portfolio shifts toward the center of recently funded work, providing empirical evidence of a consistent reduction in semantic distinctiveness in federal research funding that is already occurring across both agencies. Note that this analysis exploits within-investigator variation in LLM involvement, reporting conditional correlations within investigators, not causal estimates. We therefore interpret LLM involvement as associated with the positioning, selection, and downstream output of federally funded research, not as a cause of these outcomes.

Does this association merely reflect LLM-induced stylistic smoothing, whereby LLMs homogenize surface-level language across proposals, or a substantive convergence in the ideas themselves? To distinguish between these channels, we apply our distinctiveness pipeline to LLM-rewritten versions of baseline abstracts from 2021, holding the underlying scientific content constant while substantially altering the surface language. The average semantic distinctiveness of original and LLM-rewritten abstracts is nearly indistinguishable for both NSF and NIH (\textit{SI Appendix}~\ref{subsection:robustness_distinctiveness_stylistic}, \textit{SI Appendix} Fig.~\ref{fig:Figure_comment2.8}). This finding suggests that our distinctiveness measure is not simply driven by LLM-mediated linguistic changes, and that the evidence is consistent with convergence in substantive positioning rather than surface language alone.

\section*{LLM use and US federal research proposal success}
\begin{figure}[H]
\centering
\includegraphics[width=0.6\columnwidth]{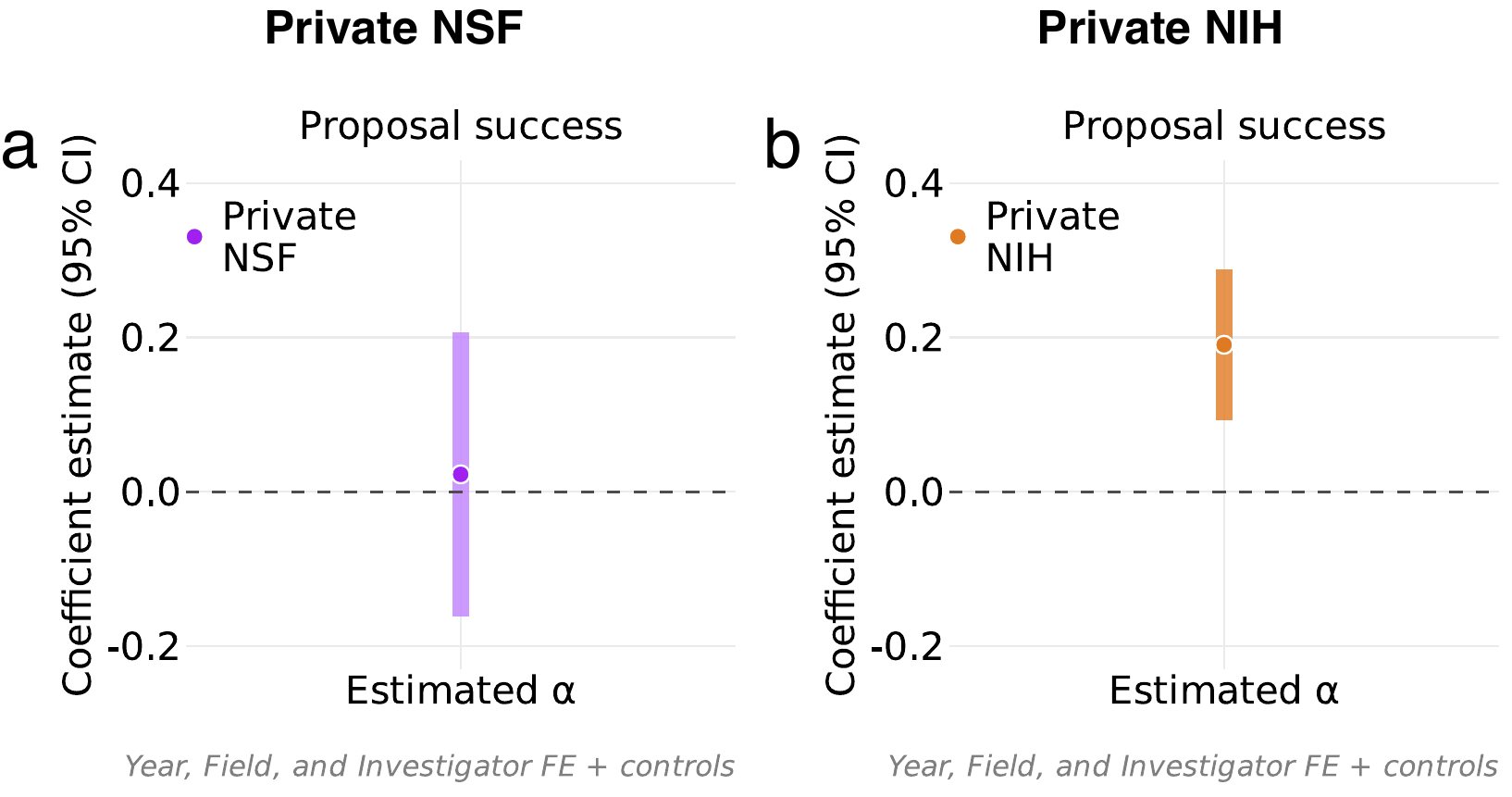}
\caption{\textbf{LLM use and US federal research proposal success.} Based on private NSF and NIH proposal submissions from two large US R1 universities, this figure examines the relationship between LLM use at submission ($\alpha$) and proposal success. (\textbf{a}) Regression estimates for NSF submissions. (\textbf{b}) Corresponding estimates for NIH submissions. All regressions include proposal request start year, field, and investigator fixed effects, as well as controls for requested funding amount. Points indicate coefficient estimates, and bars denote 95\% confidence intervals.}
\label{fig:Figure3}
\end{figure}

We next ask whether LLM adoption is associated with a first-order consequence in the funding pipeline: which proposals actually receive funding. Using confidential submissions that include both funded and unfunded proposals (D1-D2), we regress proposal success on proposal-level LLM involvement ($\alpha$) for request start years 2023-2025, estimating models with request start year, field, and investigator fixed effects and controlling for requested funding (Fig.~\ref{fig:Figure3}, \textit{SI Appendix}~\ref{subsection:regression}, \textit{SI Appendix} Tables~\ref{table:NSF_private_success}-\ref{table:NIH_private_success}). These fixed effects absorb time-invariant heterogeneity across research areas and investigators, while accounting for common shocks across years. For example, the investigator fixed effects hold constant time-invariant differences in investigators' ability, style, and topic choice, asking whether an investigator's proposals are more likely to be funded when their LLM involvement is higher.

We find no statistically significant association between $\alpha$ and proposal success for NSF proposals. In contrast, for NIH submissions, $\alpha$ is positively and significantly associated with proposal success. Specifically, moving from low LLM involvement (25th percentile) to high LLM involvement (75th percentile) corresponds to a $\sim$4-percentage-point jump in NIH funding probability. Overall, these results indicate that the consequences of LLM adoption for proposal selection are agency-dependent: while LLM use does not systematically advantage proposals at NSF, it is associated with a higher likelihood of being funded at NIH, even for the same investigators. This agency-dependent selection pattern raises the question of translation: conditional on being funded, is LLM adoption associated with changes in downstream outputs from these funded projects?

\section*{LLM use and US federal research funding outputs}
\begin{figure}[H]
\centering
\includegraphics[width=0.65\columnwidth]{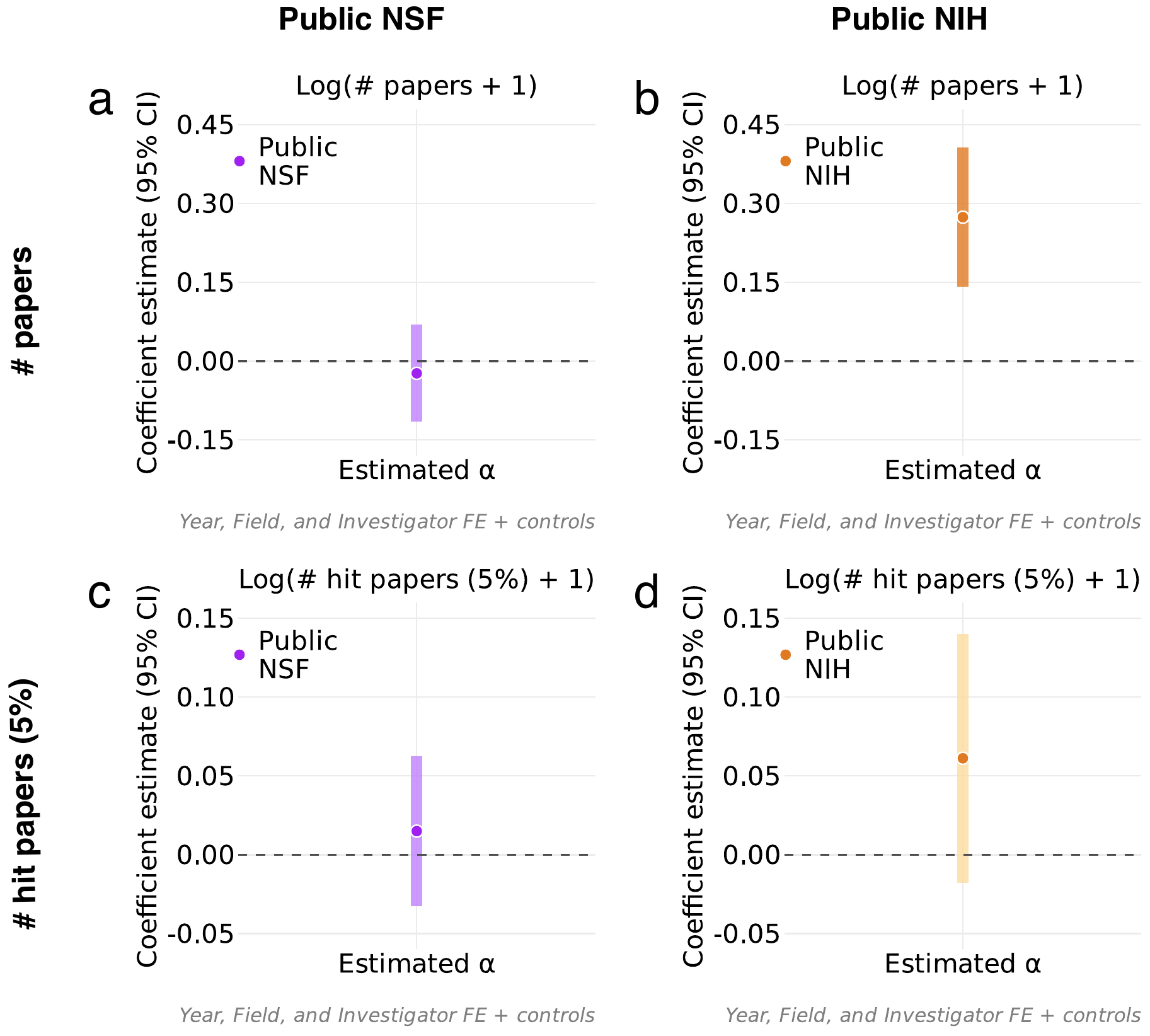}
\caption{\textbf{LLM use and US federal research funding outputs.} (\textbf{a}-\textbf{b}) Regression estimates relating grant-level LLM use ($\alpha$) to the total number of resulting publications for NSF (\textbf{a}) and NIH (\textbf{b}) grants. (\textbf{c}-\textbf{d}) Corresponding estimates for high-impact outputs, where a ``hit'' paper is defined as one whose citations fall within the top 5\% of all papers published worldwide in the same year and field. All regressions include grant start year, field, and investigator fixed effects, as well as controls for funding amount. Points indicate coefficient estimates, and bars denote 95\% confidence intervals.}
\label{fig:Figure4}
\end{figure}

Using the public NSF and NIH award data (D3-D4), we next relate grant-level LLM involvement ($\alpha$) to subsequent publication output for grants starting in 2023-2024, estimating regression models that control for funding amount and include grant start year, field, and investigator fixed effects (Fig.~\ref{fig:Figure4}, \textit{SI Appendix}~\ref{subsection:regression}, \textit{SI Appendix} Tables~\ref{table:NSF_public_num_of_papers}-\ref{table:NIH_public_num_of_hit_1pct_papers}).

We again find no statistically significant relationship between $\alpha$ and publication output for NSF awards (Fig.~\ref{fig:Figure4}a). However, for NIH awards, higher $\alpha$ is strongly associated with more resulting publications (Fig.~\ref{fig:Figure4}b). Moving from low LLM involvement (25th percentile) to high LLM involvement (75th percentile) predicts $\sim$5\% more publications for NIH grants. Moreover, this NIH productivity premium appears mostly concentrated in publication volume rather than high-impact outputs: when we restrict our analyses to highly cited papers (e.g., top 5\% papers in Figs.~\ref{fig:Figure4}c-d; top 1\% in \textit{SI Appendix} Fig.~\ref{fig:Figure4-SI-hitpaper1percent}), the relationship attenuates and is no longer statistically significant at conventional levels. Taken together, these results suggest that the downstream consequences of LLM adoption are also agency-dependent. At NIH, higher $\alpha$ is associated with more follow-on publications, but the additional output is concentrated in non-hit papers rather than the most highly cited work. We emphasize that these estimates are based on grants starting in 2023-2024 and therefore capture only the first one to two years of post-award publication output, reflecting early-stage rather than long-run productivity. Descriptive analyses of older cohorts suggest that such early windows appear to capture a meaningful share of a grant’s eventual output, and that early output is overall predictive of longer-run output (\textit{SI Appendix}~\ref{subsection:publication_trajectories}, \textit{SI Appendix}~Figs.~\ref{fig:Figure_comment1.6_over_time}-\ref{fig:Figure_comment1.6_2011_regression}). Nonetheless, because high-impact work may unfold over longer horizons, the relationship between LLM involvement and long-run outcomes remains an important open question for future work.

These productivity patterns both align with and qualify recent evidence of large output gains following LLM diffusion in manuscript production~\cite{kusumegi2025scientific}. While we observe a positive association between LLM involvement and publication output for NIH-funded grants, the magnitude of this association is smaller than the large increases documented in preprint production and is strongly agency-dependent, with no detectable relationship at NSF. In the federal funding context, where projects are selected through peer review and outputs reflect execution constraints and portfolio objectives rather than writing speed alone, LLM adoption appears to translate into additional publications primarily at NIH, and predominantly among non-hit papers, while showing no detectable relationship with publication output at NSF.

\section*{Robustness and sensitivity checks}
We conduct a series of robustness and sensitivity checks across numerous dimensions. First, to better characterize proposals and awards with higher LLM involvement, we follow prior work~\cite{kusumegi2025scientific} and compute the Flesch Reading Ease score for each abstract as a measure of writing complexity (\textit{SI Appendix}~\ref{subsection:writing_complexity}). Consistent with prior work~\cite{kusumegi2025scientific}, we find that higher LLM involvement is positively associated with greater writing complexity in both proposals and awards (\textit{SI Appendix} Fig.~\ref{fig:Figure1-SI-WordComplexity}).

Second, recent work has shown that the use of promotional language in grant proposals is positively associated with funding outputs and related proposal characteristics~\cite{peng2024promotional}. To assess the extent to which our results may be driven by promotional words, we conduct a robustness analysis that explicitly removes promotional words at all stages of our text processing pipeline. The resulting LLM-use measures remain highly similar to the original estimates across all datasets, indicating that our results are robust to the exclusion of promotional words (\textit{SI Appendix}~\ref{subsection:promotional}, \textit{SI Appendix} Fig.~\ref{fig:Figure1-SI-PromotionalWords}). 

Third, in the main analysis of semantic distinctiveness, we define the comparison set using grants funded in the preceding year. We redefine the comparison set using funded grants starting in the prior two years, finding the results are robust (\textit{SI Appendix}~\ref{subsection:diffwindow}, \textit{SI Appendix} Fig.~\ref{fig:Figure2-SI-2years-embedding}).

Fourth, while our primary measure of semantic distinctiveness relies on cosine distances between embedding vectors, we also compute distances using L2 metrics and obtain consistent results (\textit{SI Appendix}~\ref{subsection:diffdistance}, \textit{SI Appendix} Fig.~\ref{fig:Figure2-SI-L2distance-embedding}). In addition, to ensure that our findings are not driven by the percentile normalization, we replicate the main analysis using raw cosine distances without the within-year percentile transformation and find highly consistent results (\textit{SI Appendix}~\ref{subsection:diffdistance}, \textit{SI Appendix} Fig.~\ref{fig:Figure_comment1.5}).

Fifth, in our primary specifications, we use ordinary least squares regressions. We re-estimate proposal success using logistic regressions and publication outputs using negative binomial regressions and continue to observe the same associations between $\alpha$ and outcomes (\textit{SI Appendix}~\ref{subsection:alternativeregression}, \textit{SI Appendix} Figs.~\ref{fig:Figure3-SI-logit}-\ref{fig:Figure4-SI-negbino}).

Sixth, we further examine robustness by field using public award data where sample sizes are sufficiently large to support field-level analyses and find broadly consistent results (\textit{SI Appendix}~\ref{subsection:robustnessbyfield}, \textit{SI Appendix} Figs.~\ref{fig:Figure1-SI-byField-corpus-NSF}-\ref{fig:Figure4-SI-byField}). We also conduct analogous robustness checks by subagency using the public award data and again find broadly consistent patterns (\textit{SI Appendix}~\ref{subsection:robustnessbysubagency}, \textit{SI Appendix} Figs.~\ref{fig:Figure1-SI-byAgency-corpus-NSF}-\ref{fig:Figure4-SI-byAgency}).

Seventh, to ensure that our findings are not driven by the idiosyncrasies of a single language model, we replicate the entire detection and analysis pipeline using Claude, Gemini, and Llama as alternative reference LLMs. The main analyses yield broadly consistent results across model families (\textit{SI Appendix}~\ref{subsection:alternative_llm}, \textit{SI Appendix} Figs.~\ref{fig:Figure_comment2.2_timeline}-\ref{fig:Figure_comment2.2_Fig4}).

Eighth, because the detection method relies on word-frequency statistics, we systematically evaluate its sensitivity to preprocessing choices by varying the tokenization scheme, stopword handling, and punctuation normalization across nine alternative pipelines. The resulting $\alpha$ estimates remain highly correlated with the baseline, and the main results are broadly consistent under alternative pipelines (\textit{SI Appendix}~\ref{subsection:sensitivity_preprocessing}, \textit{SI Appendix} Figs.~\ref{fig:Figure_comment2.3_correlation}-\ref{fig:Figure_comment2.3_Fig4}).

% Ninth, to ensure that our findings are not driven by the brevity of grant abstracts, we restrict the sample to abstracts with at least 100, 150, and 250 words and repeat the main analyses, and obtain broadly consistent results (\textit{SI Appendix}~\ref{subsection:abstract_length}, \textit{SI Appendix} Figs.~\ref{fig:Figure_comment2.4_Fig2}-\ref{fig:Figure_comment2.4_Fig4}).

% Tenth, we compare estimated LLM-involvement rates between single-investigator and multi-investigator proposals and awards, and include a single-investigator versus multi-investigator indicator as a control variable in all regression specifications. The results remain consistent (\textit{SI Appendix}~\ref{subsection:single_multi_investigator}, \textit{SI Appendix} Figs.~\ref{fig:Figure_comment2.5_raw_plot}-\ref{fig:Figure_comment2.5_Fig4}).

Ninth, because AI-related concepts have become increasingly prevalent in scientific research~\cite{gao2024quantifying}, grant proposals may naturally contain more AI-method terminology over time, independent of LLM involvement in writing. To rule out the possibility that our estimator captures topic-driven AI vocabulary growth rather than LLM-driven shifts, we remove all AI-related terms identified in prior work~\cite{gao2024quantifying} at every stage of the estimation pipeline and re-estimate $\alpha$. The resulting measures remain highly correlated with the original estimates, and all main regression results are robust to this exclusion (\textit{SI Appendix}~\ref{subsection:ai_scientific_vocabulary}, \textit{SI Appendix} Figs.~\ref{fig:Figure_comment2.6_ngram_correlation}-\ref{fig:Figure_comment2.6_ngram_Fig4}).

% Twelfth, to assess sensitivity to predictor scaling, we standardize all predictors by converting them to z-scores and re-estimate the main regressions, finding consistent results (\textit{SI Appendix}~\ref{subsection:standardized_predictors}, \textit{SI Appendix} Figs.~\ref{fig:Figure_comment2.7_Fig2}-\ref{fig:Figure_comment2.7_Fig4}). Full regression tables reporting coefficients and standard errors for all variables, including control variables, are provided in \textit{SI Appendix}~Tables~\ref{table:semantic_distinctiveness_standardized}-\ref{table:papers_standardized}.

Tenth, to increase the interpretability of the detection method, we apply the generalized word-shift graph framework~\cite{gallagher2021generalized} to decompose the aggregate divergence into per-word contribution scores. This decomposition reveals that the LLM signal is driven by a concentrated set of evaluative and elaborative terms, while human-written text is characterized by simpler, more functional vocabulary (\textit{SI Appendix}~\ref{subsection:word_shift}, \textit{SI Appendix} Figs.~\ref{fig:Figure_comment2.9_Proportion}-\ref{fig:Figure_comment2.9_JSD}). Our results are further robust to varying text-length thresholds, single- vs. multi-investigator, and standardizing predictors (\textit{SI Appendix}~\ref{subsection:abstract_length}-\ref{subsection:standardized_predictors}, \textit{SI Appendix}~Figs.~\ref{fig:Figure_comment2.4_Fig2}-\ref{fig:Figure_comment2.7_Fig4}, and \textit{SI Appendix}~Tables~\ref{table:Table_comment2.7_Fig2}-\ref{table:Table_comment2.7_Fig4}).

\section*{Discussion}
Taken together, our findings reveal that the rapid diffusion of LLMs is already reshaping key stages of the US federal research funding pipeline, from proposal submission to downstream scientific output. By combining confidential proposal submissions with the full population of NSF and NIH awards, we provide systematic evidence that LLM adoption is associated with (i) how scientific ideas are positioned relative to recently funded work, (ii) whether proposals clear the funding threshold, and (iii) the publication output that funded projects go on to produce. These findings speak directly to the two open questions raised at the outset. First, the early effects of generative AI in science may arise less from accelerating the execution of research than from reshaping how ideas are articulated, packaged, and aligned with prevailing norms, while the harder bottlenecks of project design, coordination, and implementation remain: the publication output gains we observe are modest and context-dependent, absent at NSF and concentrated in non-hit papers at NIH, in contrast to the large productivity increases documented at the manuscript level. Second, whereas prior work has largely studied the finished outputs of science, our findings show that generative AI already operates upstream, at the idea-generation stage, where scientific ideas are first articulated and compete for public investment. Together, these results reposition generative AI not merely as a tool that changes how scientific text is produced, but as a force that is beginning to reshape the articulation and allocation of ideas in the federal research funding system.

A central concern in science policy is maintaining a diverse and exploratory research portfolio, one that supports both cumulative progress and the pursuit of unconventional ideas. Across both agencies and both stages (proposal submission and award funding), higher LLM involvement is consistently associated with lower semantic distinctiveness: proposals with greater LLM traces are positioned closer to recently funded work within the same agency. Importantly, our analysis suggests that the convergence we observe is not simply a byproduct of LLM-induced surface-level rewriting, but reflects shifts in the substantive positioning of proposals and awards. We further find that investigators with greater LLM involvement have, so far, largely stayed within their existing lines of inquiry: proposals with greater LLM involvement are not systematically more distant from an investigator’s own prior proposals within the same agency (\textit{SI Appendix}~\ref{subsection:llm_involvement}, \textit{SI Appendix} Fig.~\ref{fig:Figure_commnet1.3_pivot_within_investigator}). In other words, rather than moving investigators away from their established research trajectories, greater LLM involvement appears to be associated with proposals that more closely resemble what the agencies have funded in the past. A portfolio that is closer to recent funding patterns may reflect improved clarity, tighter alignment with reviewer expectations, or lower transaction costs in articulating a fundable project. But it also implies reduced exploration in the idea landscape, which matters for public funders explicitly tasked with sustaining high-variance discovery, with implications for long-run impact and sustainability of science. 

Importantly, the downstream consequences of LLM adoption are agency-dependent. In the confidential submission data, LLM involvement appears unrelated to proposal success at NSF but positively associated with funding at NIH. In the public award data, we similarly detect no relationship between LLM involvement and publication output at NSF, whereas at NIH, higher LLM involvement predicts greater publication output, though concentrated in non-hit papers rather than the most highly cited work. Notably, these agency-dependent patterns do not appear to be explained solely by differences in field composition. For example, we restrict the comparison to a common field—Biomedical \& Clinical Sciences—across agencies and extend it to include the Department of Defense (DOD). Within this shared field, we continue to observe distinct relationships between LLM involvement and downstream outputs across agencies (\textit{SI Appendix}~\ref{subsection:within_field}, \textit{SI Appendix} Fig.~\ref{fig:Figure_comment1.2_Fig4}). Overall, these results suggest that the observed differences likely reflect an interaction between disciplinary context and agency-specific institutional structure, rather than a simple NSF-NIH contrast or a pure field effect. One potential interpretation is that NIH funding and review norms may more strongly reward incremental, executable projects that yield multiple publications, and LLM-assisted drafting may help proposals conform to those established templates. The contrast also raises questions regarding specific features of agency review, such as NIH’s study section system, the role of preliminary data and demonstrated feasibility in R01 review, and differences in how agencies weight innovation versus feasibility, and whether these agency-specific features may have played a role. We emphasize, however, that these institutional accounts are speculative: our analysis documents agency-dependent associations but does not test the mechanisms behind them, and we caution against reading the agency contrast as evidence of specific channels. In addition, publication counts capture a larger share of NIH outputs than NSF outputs, which often include software, datasets, instrumentation, and training. It is also possible that high-impact work unfolds over longer horizons than our follow-up window (\textit{SI Appendix}~\ref{subsection:publication_trajectories}, \textit{SI Appendix} Figs.~\ref{fig:Figure_comment1.6_over_time}-\ref{fig:Figure_comment1.6_2011_regression}). Disentangling these possibilities and, more generally, explaining why LLM adoption translates into selection and output effects in some contexts but not others, is a key direction for future research.

This context dependence is especially striking when contrasted with recent large-scale evidence that LLM diffusion has been accompanied by sharp increases in manuscript production~\cite{kusumegi2025scientific}. Taken together, the manuscript setting and the funding-to-output setting suggest a simple but consequential hypothesis: a meaningful share of the productivity gains associated with LLM adoption may operate through accelerating communication---the speed and ease with which researchers articulate, package, and iterate on ideas---rather than accelerating execution, which remains constrained by project design, coordination, data collection, and other bottlenecks that LLM-enabled writing does not directly resolve. This interpretation is necessarily tentative, given limitations of publication-based output measures and the time required for high-impact work to accrue. Nonetheless, the contrast highlights an important implication for science policy: the benefits of generative AI may accrue first and most strongly where writing is the binding constraint, whereas their effects on the composition and ultimate impact of publicly funded research also hinge on institutional incentives and execution constraints.

Our within-investigator, across-grant design adds a distinct perspective to a growing literature on how AI adoption is reshaping scientific production. 
Existing work has documented that LLM adoption is associated with changes in productivity and referencing behavior in manuscripts and published papers, and more broadly that AI-augmented research can contract the collective focus of science~\cite{kusumegi2025scientific,hao2026artificial}. But these analyses largely rely on comparisons across researchers, across papers, or across AI versus non-AI work. By contrast, our design leverages variation in LLM involvement across a given investigator’s own proposals and awards, allowing us to examine whether more intensive LLM use within the same investigator is associated with differences in semantic positioning, proposal success, and downstream output. At the same time, this design has limitations. While within-investigator comparisons hold constant time-invariant differences across investigators, they cannot eliminate time-varying strategic choices in when and how investigators use LLMs. For example, investigators may rely more heavily on LLMs for proposals that are already more incremental, executable, or aligned with agency priorities; in that case, higher LLM involvement may partly reflect the type of project being pursued rather than an LLM-induced shift toward the center of the funding portfolio. The relationships we report should therefore be interpreted as conditional correlations within investigators, not causal effects. In addition, the confidential proposal data come from two large US R1 universities. Although these data provide rare access to funded, unfunded, and pending proposals at the point of submission, two institutions cannot capture the full diversity of US research settings. Broadening institutional coverage remains an important direction for future work.

Beyond the analytical design, our study helps address a persistent blind spot in the literature: how generative AI enters the funding pipeline at the moment when ideas are first articulated for review. Because unfunded proposals are rarely observable at scale, most prior work has had to infer behavior from public records of funded projects or published papers. By directly observing confidential submissions alongside the population of funded awards, we can separate proposal preparation from post-award editing and examine selection outcomes, not just the characteristics of funded projects. It is important to note that our measurement strategy—textual traces of LLM involvement in proposal abstracts—captures only one dimension of how LLMs may be used in proposal preparation. Accordingly, our estimates should be interpreted as measuring how LLMs shape the articulation and positioning of ideas in proposal narratives, not the full scope of LLM use in research development.

Finally, because federal grants deploy public funds, the rise of LLMs in proposals has implications not only for scientific efficiency but also for accountability and trust. If LLM adoption systematically pulls proposals toward safer, more easily justifiable ideas, it could quietly tilt public portfolios toward exploitation even as agencies emphasize the value of high-risk, high-reward exploratory research. At the same time, heavy reliance on LLMs in documents intended to reflect investigators’ original ideas raises questions about authorship, responsibility, and the integrity of the funding process. By quantifying LLM involvement at scale and documenting its associations with idea positioning, selection, and downstream output, our study provides an empirical foundation for evidence-informed policy, ranging from disclosure norms and reviewer training to mechanisms that protect high-variance exploration and preserve portfolio diversity as generative AI becomes increasingly embedded in scientific practice.

\section*{Materials and Methods}
\subsection*{Datasets}
This study integrates four complementary datasets capturing distinct stages of the US federal research funding pipeline at the National Science Foundation (NSF) and the National Institutes of Health (NIH). Two datasets consist of confidential proposal submissions obtained directly from universities (D1-D2), while two other datasets are constructed from publicly released award records (D3-D4). The data collection protocol on private NSF and NIH proposals from two US universities was reviewed and approved by the Institutional Review Board of Northwestern University (IRB no. STU00222200). Together, these data enable us to characterize the evolution of LLM use in federal funding and to examine how it relates to the positioning of scientific ideas, their selection for funding, and their translation into publicly supported research output.

% % \subsection*{Private NSF and NIH proposal data (D1-D2)}
D1 and D2 comprise confidential NSF and NIH proposal submissions collected from two large US R1 research universities, under approved Institutional Review Board (IRB) protocols. Data from the first university were obtained in May 2025, and data from the second university were obtained in November 2025. These data include the full population of proposals submitted to NSF (D1) and NIH (D2) with request start dates between 2021 and 2025, covering funded, unfunded, and pending proposals. The inclusion of unfunded proposals enables analyses that are not possible using public data alone. In total, D1 contains 1.6K NSF proposal submissions, and D2 contains 4.1K NIH proposal submissions over the study period (2021-2025).

For each proposal, the datasets include the proposal abstract text, which serves as the primary textual input for estimating LLM involvement. In addition, the data include proposal-level metadata used in the empirical analyses, including the funding agency (NSF or NIH), request start year, requested funding amount, funding outcome (funded, unfunded, or pending), and investigator names. We link investigators to Dimensions using names and affiliations and use Dimensions researcher IDs as unique investigator identifiers. One university provides principal investigator (PI) names only, whereas the other provides both PI and co-PI names. We assign scientific fields to each proposal by identifying, among grants funded by the same agency, the grant whose abstract has the smallest cosine distance to the proposal abstract based on SPECTER2 embeddings, and assigning the corresponding grant fields from Dimensions~\citep{herzog2020dimensions} to the proposal. Field assignments use Dimensions' main (level-0) field classification (e.g., Information and Computing Sciences, Biological Sciences), which defines 22 fields. These variables allow us to estimate proposal-level LLM involvement, relate it to semantic distinctiveness, and examine proposal success while controlling for funding amount and investigator, field, and year fixed effects.

% \subsection*{Public NSF and NIH award data (D3-D4)}
D3 and D4 consist of the full population of publicly released NSF and NIH funded awards constructed from Dimensions. Dimensions (\url{https://www.dimensions.ai/}) is a large-scale, integrated research information platform that uniquely enables systematic linking of research grants to their resulting publications through funding acknowledgment data, making it particularly well suited for studying downstream scientific outputs of funded grants. D3 includes NSF awards with start dates between 2021 and 2025, and D4 includes NIH awards with start dates over the same period, aligning the observation window with the private proposal data. The award data were downloaded from Dimensions in November 2025. In total, D3 contains 57K NSF awards, and D4 contains 74K NIH awards.

For each funded award, the datasets include the award abstract text, which serves as the primary textual input for estimating LLM involvement at the award stage. In addition, the data include award-level metadata used in the empirical analyses, including the funding agency (NSF or NIH), award start year, awarded funding amount, scientific fields, and investigator names and identifiers. Both PI and co-PI identifiers are provided directly by Dimensions. Scientific fields are also directly obtained from Dimensions using its main (level-0) field classification system (e.g., Information and Computing Sciences, Biological Sciences), in which Dimensions defines 22 fields.

To examine downstream scientific output, we link funded awards in D3 and D4 to their resulting publications in Dimensions, including both preprints and published articles. When a paper has both a preprint and a published version, we retain only the published version to avoid double counting. Using these linked records, we construct measures of total publication output and high-impact output based on field- and year-normalized citation percentiles, where citations are measured as cumulative citations through November 2025. These measures allow us to relate award-level LLM involvement to semantic distinctiveness relative to recently funded work within each agency and to assess downstream research outputs while controlling for investigator, field, and year fixed effects.

The combination of private proposal submissions (D1-D2) and public award records (D3-D4) enables a unified analysis across multiple stages of the funding process. The private data capture idea articulation and selection at submission, including proposals that were never funded, while the public data capture post-selection outcomes, including funded project descriptions and subsequent publications.

% \newpage
\section*{Acknowledgments}
We thank Yian Yin for thoughtful comments. We thank all members of the Center for Science of Science and Innovation (CSSI) at Northwestern University for helpful discussions, and Alyse Freilich for her careful editing and valuable feedback. This work is supported by the National Science Foundation under Award Number 2404035. The authors used large language models (GPT-5 and Claude Opus 4) to improve the language, style, and readability of the text in this manuscript. The authors are fully responsible for the content.

\section*{Author contributions}
Y.Q. and D.W. jointly conceived the project, Y.Q., A.F., and D.W. designed the experiments, Y.Q. collected data, Y.Q. and Z.W. performed empirical analyses with the help from A.F., Y.B., and E.S., all authors collaboratively interpreted results, Y.Q., A.F., and D.W. wrote the manuscript, and all authors edited the manuscript.

\section*{Competing interests}
The authors declare no competing interests.

\section*{Data and code availability}
D1 and D2, internal confidential grant proposal datasets from two US R1 universities, require a data use agreement with the universities; interested researchers should contact the authors for further information. D3 and D4 are bibliometric datasets from Dimensions~\cite{herzog2020dimensions}. Those who are interested in raw data from Dimensions should contact Digital Science directly. The de-identified data necessary to reproduce main plots and statistical analyses are freely available at \href{https://northwestern-cssi.github.io/llmfunding/}{https://northwestern-cssi.github.io/llmfunding/}.
% D1 and D2, internal confidential grant proposal datasets from two US R1 universities, require a data use agreement with the universities; interested researchers should contact the authors for further information. D3 and D4 are bibliometric datasets from Dimensions~\cite{herzog2020dimensions} (\href{https://www.dimensions.ai}{https://www.dimensions.ai}). Those who are interested in raw data from Dimensions should contact Digital Science directly. The de-identified data necessary to reproduce main plots and statistical analyses are freely available at \href{https://northwestern-cssi.github.io/llmfunding/}{https://northwestern-cssi.github.io/llmfunding/}.

% \section*{Code availability}
% Code necessary to reproduce all plots will be made freely available.

\newpage
\begin{center}
{\Large\bfseries Supporting Information for \\ The Rise of Large Language Models and the Direction and Impact of US Federal Research Funding}
\end{center}

\renewcommand{\figurename}{Fig.}
\renewcommand{\tablename}{Table}
\renewcommand{\thefigure}{S\arabic{figure}}
\renewcommand{\thetable}{S\arabic{table}}

\renewcommand{\thesection}{S\arabic{section}}
\renewcommand{\thesubsection}{S\arabic{section}.\arabic{subsection}}
\renewcommand{\thesubsubsection}{S\arabic{section}.\arabic{subsection}.\arabic{subsubsection}}

\section{Methods}
\subsection{LLM detection\label{subsection:llmdetection}}
Following recent work \cite{liang2025quantifying,kusumegi2025scientific,liu2025ai}, we quantify the extent of large language model (LLM) use in grant proposal and award abstracts using the distributional LLM-quantification framework introduced by Liang \textit{et al.}~\cite{liang2025quantifying}. We estimate separate detection models for NSF and NIH grants to account for agency-specific writing styles and topical distributions. Below we describe the NSF pipeline; the NIH pipeline is constructed analogously.

We begin by assembling a reference corpus of 6{,}000 NSF grant abstracts, a random subset of publicly available NSF grant abstracts with start dates in 2021, prior to the widespread adoption of LLMs. These abstracts are used to estimate the word distribution of human-written text and to construct a corresponding LLM-generated distribution by prompting OpenAI's GPT-3.5-turbo-0125 model to rewrite the same abstracts, simulating LLM involvement in proposal language. The prompts used, which were adapted from those in Liang \textit{et al.}~\cite{liang2025quantifying}, are shown below. Comparing these two distributions allows us to identify systematic differences between human and LLM-modified writing.

\newtcolorbox{promptbox}{
  colback=gray!12,      
  colframe=gray!12,     
  boxrule=0pt,          
  arc=0pt,              
  left=10pt,right=10pt,top=8pt,bottom=8pt,
  width=\linewidth,
  enhanced,
}
\begin{figure}[htbp]
\centering

\begin{promptbox}
\ttfamily\small
The aim here is to reverse-engineer the author's writing process by taking a grant abstract and compressing it into a more concise form. This process simulates how an author might distill their thoughts and key points into a structured, yet not overly condensed form. Now as a first step, given a grant abstract, reverse-engineer it into a list of bullet points.
\end{promptbox}

\vspace{3pt}

\begin{promptbox}
\ttfamily\small
Following the initial step of reverse-engineering the author's writing process by compressing a grant abstract, you now enter the second phase. Here, your objective is to expand upon the concise version previously crafted. This stage simulates how an author elaborates on the distilled thoughts and key points, enriching them into a detailed, structured narrative similar in length to the original abstract. Given the concise output from the previous step, your task is to develop it into a fully fleshed-out text.
\end{promptbox}

\end{figure}

More specifically, the framework consists of the following steps applied to the NSF datasets (D1 and D3):

\begin{enumerate}
\item \textbf{Problem formulation:} denote $X$ as each single sentence, $\mathcal{P}$ and $\mathcal{Q}$ as the population-level distributions of human-written and LLM-modified sentences, respectively. The mixture distribution is given by
\begin{equation}
\mathcal{D}_{\alpha}(X) = (1 - \alpha)\mathcal{P}(x) + \alpha \mathcal{Q}(x),
\end{equation}
where $\alpha$ is the mixture weight of the LLM-modified sentence distribution in the observed data. The goal is to estimate $\alpha$ based on observed sentences $\{X_i\}_{i=1}^{N} \sim \mathcal{D}_{\alpha}(X)$, where $i$ is an integer index of observed sentences.

\item \textbf{Parameterization:} to make $\alpha$ identifiable, the framework models the distributions of token occurrences in human-written and LLM-modified sentences, denoted as $\mathcal{P}_T$ and $\mathcal{Q}_T$, respectively, for a chosen list of tokens $T = \{t_i\}_{i=1}^{M}$. The occurrence probabilities of each token in human-written and LLM-modified sentences, $p_t$ and $q_t$, are used to parameterize $\mathcal{P}_T$ and $\mathcal{Q}_T$:
\begin{equation}
\mathcal{P}_T(X) = \prod_{t \in T} p_t^{\mathbbm{1}\{t \in X\}} (1 - p_t)^{\mathbbm{1}\{t \notin X\}},
\qquad
\mathcal{Q}_T(X) = \prod_{t \in T} q_t^{\mathbbm{1}\{t \in X\}} (1 - q_t)^{\mathbbm{1}\{t \notin X\}}.
\end{equation}

\item 
\textbf{Estimation:} the occurrence probabilities $p_t$ and $q_t$ are estimated using collections of known human-written and LLM-modified sentences, $\{X_j^{P}\}_{j=1}^{n_{P}}$ and $\{X_j^{Q}\}_{j=1}^{n_{Q}}$, respectively, here $j$ is an integer index of documents with known sources

\begin{equation}
\hat{p}_t = \frac{1}{n_{P}} \sum_{j=1}^{n_{P}} \mathbbm{1}\{t \in X_j^{P}\},
\qquad
\hat{q}_t = \frac{1}{n_{Q}} \sum_{j=1}^{n_{Q}} \mathbbm{1}\{t \in X_j^{Q}\}.
\end{equation}

\item \textbf{Inference:} the fraction $\alpha$ is estimated by the maximum likelihood estimator (MLE) on the observed sentences under the mixture distribution
$
\hat{\mathcal{D}}_{\alpha,T}(X) = (1 - \alpha)\hat{\mathcal{P}}_T(X) + \alpha \hat{\mathcal{Q}}_T(X):
$
\begin{equation}
\hat{\alpha}^{\mathrm{MLE}}_T
=
\arg\max_{\alpha \in [0,1]}
\sum_{i=1}^{N}
\log \left( (1 - \alpha)\hat{\mathcal{P}}_T(X_i) + \alpha \hat{\mathcal{Q}}_T(X_i) \right).
\end{equation}
\end{enumerate}

\subsection{Regression models\label{subsection:regression}}
To examine the relationship between LLM involvement in grants and semantic distinctiveness, proposal success, and downstream scientific outputs, we estimate a series of ordinary least squares (OLS) linear regression models separately for NSF and NIH grants. Across all model specifications, we use a consistent set of explanatory and control variables, varying only the dependent variable to reflect the outcome of interest.

\medskip
\textbf{Dependent variables}: We consider three classes of outcome variables, denoted $y_{ijft}$, where $i$ indexes individual proposals or awards, $j$ indexes investigators, $f$ denotes the scientific field, and $t$ indicates the proposal request start year or award start year. The outcomes are defined as follows:
\begin{itemize}
\item \textbf{Semantic distinctiveness}: Defined as the percentile rank (ranging from 0 to 100, with higher values indicating greater distinctiveness) of the average cosine distance between the SPECTER2~\cite{singh2023scirepeval} embedding of a grant abstract and the embeddings of all grants funded by the same agency in the preceding year. This measure captures how semantically distinct a grant is relative to recently funded work and is available for both public and private datasets (D1-D4).

\item \textbf{Proposal success}: Defined as a binary indicator equal to one if a proposal is funded and zero if unfunded. This outcome is observed only in the private proposal dataset (D1-D2).

\item \textbf{Federal research funding outputs}: Defined as the logarithm of the number of publications resulting from a funded grant plus one, $\log(\#\text{ papers} + 1)$. In additional specifications, we examine high-impact output, where a ``hit'' paper is defined as a publication whose citation count falls within the top 5\% of all papers published worldwide in the same field and year; the corresponding outcome is $\log(\#\text{ hit papers} + 1)$. Measures of downstream output are constructed using public award data only (D3-D4).
\end{itemize}

\medskip
\textbf{Explanatory variables}: The key explanatory variable, $\alpha_i$, measures the extent of LLM use in the grant abstract associated with proposal or award $i$ and ranges from 0 to 1.

\textbf{Control variables}: To account for systematic differences across investigators, research fields, and time, all specifications include investigator fixed effects $\eta_j$, field fixed effects $\phi_f$, and grant start year fixed effects $\mu_t$. We additionally control for $\log(\text{Funding}_i)$, defined as the logarithm of the total requested funding amount for proposals or the awarded funding amount for funded grants.

Depending on the choice of outcome variable, we estimate the following specification:
\begin{equation}
y_{ijft}
=
\beta\, \alpha_i
+ \delta\, \log(\textit{Funding}_i)
+ \mu_t
+ \phi_f
+ \eta_j
+ \varepsilon_{ijft}
\end{equation}

The dependent variable $y_{ijft}$ alternatively corresponds to semantic distinctiveness, proposal success, or downstream research output. Standard errors are clustered at the investigator level.

\section{Robustness and sensitivity checks}
\subsection{Robustness of semantic distinctiveness to LLM-induced stylistic changes\label{subsection:robustness_distinctiveness_stylistic}}
We applied our distinctiveness pipeline to the LLM-rewritten versions of the training abstracts used to calibrate our detection model. These are 6,000 randomly sampled public NSF and NIH grant abstracts with start dates in 2021. The rewriting prompt instructs the LLM to first distill the original abstract into key points and then re-expand them into a full narrative, preserving the core scientific content while altering the surface-level language and style. This design ensures that comparing the original and rewritten versions isolates the effect of LLM-induced stylistic changes on our distinctiveness measure, holding the underlying ideas constant. For each abstract, we computed SPECTER2 embeddings for both the original version and its LLM-rewritten counterpart, and then measured each grant's average cosine distance to all grants funded in the prior year (2020) within the same agency, following the same procedure used in Fig.~2.
\\~\\
The results (Fig.~\ref{fig:Figure_comment2.8}) reveal that the average semantic distinctiveness of original abstracts is nearly indistinguishable from that of LLM-rewritten abstracts for both NSF and NIH. LLM rewriting---which substantially alters surface-level language, as confirmed by our detection method---does not meaningfully shift where an abstract is positioned in the semantic space defined by SPECTER2 embeddings.

\subsection{Publication trajectories and early observation windows\label{subsection:publication_trajectories}}
In the main analysis (Fig.~4), we study grants starting in 2023 and 2024, for which we can currently observe approximately two years and one year of post-start publication outcomes, respectively. To help contextualize what portion of downstream research output may be observable within such windows, we provide descriptive analyses of publication trajectories following the start of a grant.
\\~\\
Specifically, we analyze older grant cohorts (e.g., grants starting in 2011, 2015, and 2020) and track the average number of publications produced per grant as a function of years since the grant start year. These older cohorts allow us to observe longer post-start publication windows and therefore characterize the typical lifecycle of grant-funded research. As shown in Fig.~\ref{fig:Figure_comment1.6_over_time}, publication output rises rapidly after a grant begins and typically peaks around 2-4 years, before gradually declining. Importantly, the publication trajectory is highly similar across cohorts.
\\~\\
To quantify how much output is captured within early windows, we focus on grants starting in 2011 (for which we observe the full 14-year publication window) and compute, for each grant, the fraction of papers produced within the first two years relative to the total number produced over 14 years. As shown in Fig.~\ref{fig:Figure_comment1.6_2011_raw}(a-b), for NSF grants the distribution is roughly evenly split between grants that produce less than half versus at least half of their eventual publications within two years. For NIH grants, approximately one third of grants already generate at least 50\% of their lifetime publications within the first two years. We further repeat the analysis using a one-year window. As shown in Fig.~\ref{fig:Figure_comment1.6_2011_raw}(c-d), approximately 35\% of NSF grants and 25\% of NIH grants generate at least 50\% of their lifetime publications within the first year. Taken together, these results indicate that even short observation windows capture a meaningful portion of downstream research output, helping contextualize the early publication outcomes observed for the more recent grant cohorts analyzed in Fig.~4.
\\~\\
Finally, we examine whether early publication output is predictive of longer-run research output. Using grants starting in 2011, we regress the total number of papers produced over the full 14-year window on the number of papers produced within the first two years after the grant start, controlling for investigator fixed effects, field fixed effects, and funding amount. As shown in Fig.~\ref{fig:Figure_comment1.6_2011_regression}(a-b), early publication output is strongly and positively associated with longer-run output for both NSF and NIH grants. We obtain similar results when using the number of papers produced within the first year as the predictor, as shown in Fig.~\ref{fig:Figure_comment1.6_2011_regression}(c-d). These results suggest that early publication outcomes provide a meaningful signal of the eventual research output of a grant, further supporting the use of short observation windows to study early downstream outcomes.
\\~\\
Overall, these analyses show that early publication windows capture a meaningful share of downstream research output and provide informative signals of longer-run grant productivity. Together, they help contextualize the interpretation of early publication outcomes observed for the recent grant cohorts analyzed in Fig.~4.

\subsection{Writing complexity\label{subsection:writing_complexity}}
To illustrate the differences between the abstracts with higher and lower levels of LLM involvement (i.e., $\alpha$)~\cite{kusumegi2025scientific}, we compute the Flesch Reading Ease score for each abstract as a measure of writing complexity~\cite{flesch1948new}. The Flesch score is a readability metric that captures syntactic complexity through sentence length, and lexical complexity through the average number of syllables per word. Since higher values of the original Flesch Reading Ease score correspond to lower complexity, we multiply the score by $-1$ so that higher values consistently indicate higher writing complexity (Equation~\ref{eq:flesch}). This transformation facilitates a more intuitive interpretation and aligns the direction of the measure with our analysis. Consistent with recent findings on preprints~\cite{kusumegi2025scientific}, we find that higher LLM involvement is positively associated with greater writing complexity in both proposals and awards (Fig.~\ref{fig:Figure1-SI-WordComplexity}).

\begin{equation}
\label{eq:flesch}
\text{Writing complexity}
=
1.015 \cdot \frac{\#\,\text{total words}}{\#\,\text{total sentences}}
+
84.6 \cdot \frac{\#\,\text{total syllables}}{\#\,\text{total words}}
-
206.835
\end{equation}

\subsection{Promotional words\label{subsection:promotional}}

Recent work has shown that the use of promotional language in grant proposals is positively associated with funding outcomes and related proposal characteristics~\cite{peng2024promotional,millar2022trends}. To assess the extent to which our results may be driven by promotional wording, we conduct a robustness analysis that explicitly removes promotional terms ($139$ words identified in prior studies~\cite{peng2024promotional,millar2022trends}) at all stages of our estimation pipeline.

Specifically, we re-estimate individual-level LLM use ($\alpha$) after excluding promotional words throughout the entire pipeline. Promotional terms are removed from the original abstracts prior to LLM rewriting and from the LLM-rewritten abstracts themselves.

Figure~\ref{fig:Figure1-SI-PromotionalWords} compares the original $\alpha$ estimates with those obtained after promotional-word removal for private NSF, private NIH, public NSF, and public NIH samples. Across all settings, the recalculated $\alpha$ remains highly correlated with the original measure, with Pearson correlation coefficients ranging from $r=0.95$ to $r=0.97$.

The consistently high correlations indicate that removing promotional language has little effect on the resulting LLM-use estimates. Overall, this analysis shows that our measure of LLM use remains stable when promotional wording is excluded, supporting the robustness of our results to this alternative text preprocessing choice.

\subsection{Different prior windows for measuring semantic distinctiveness\label{subsection:diffwindow}}
In the main analysis of semantic distinctiveness, we define the comparison set using grants funded in the preceding one year. Our results are robust to alternative choices of the comparison window. Specifically, when we construct the reference set using grants that started in the prior two years, the estimated relationships remain qualitatively and quantitatively similar (Fig.~\ref{fig:Figure2-SI-2years-embedding}).

\subsection{Different distance metrics for measuring semantic distinctiveness\label{subsection:diffdistance}}
In the main analysis of semantic distinctiveness, we use cosine distance to measure distance between embedding vectors. We also compute distances using L2 metrics and obtain consistent results (Fig.~\ref{fig:Figure2-SI-L2distance-embedding}). Furthermore, in the main analysis of semantic distinctiveness, we convert cosine distances into within-year percentile ranks before estimating the models. We adopt this transformation primarily for interpretability. Converting distances into percentiles allows the regression coefficients to be interpreted as shifts in the relative position of a proposal within the empirical distribution of semantic distances among proposals and awards in the same agency-year. To ensure that our results are not driven by this transformation, we replicate Fig.~2 using the raw cosine distance without percentile normalization. The results are highly consistent with the main findings. As shown in Fig.~\ref{fig:Figure_comment1.5}, the estimated coefficients remain negative and statistically significant across all four datasets, indicating that proposals involving LLM use are positioned closer to the grants most recently funded by the agency.

\subsection{Alternative regression models\label{subsection:alternativeregression}}
While our primary specifications use ordinary least squares regressions, we re-estimate proposal success using logistic regressions (Fig.~\ref{fig:Figure3-SI-logit}) and publication outputs using negative binomial regressions, and we continue to observe the same associations between $\alpha$ and outcomes  (Fig.~\ref{fig:Figure4-SI-negbino}).

\subsection{Robustness by field for public data\label{subsection:robustnessbyfield}}
To examine whether our main findings are consistent across award fields in the public datasets, we focus on fields where sample sizes are sufficiently large to support field-level analysis and re-estimate all main results separately by field. Specifically, for awards with start dates between 2023 and 2025, we calculate the number of grants in each field and retain those with more than 2,000 awards. All remaining fields are merged into a single category labeled ``Others''.

For NSF, in addition to ``Others'', the fields (ordered by the number of grants) are Information and Computing Sciences, Engineering, Education, Biological Sciences, Earth Sciences, Physical Sciences, and Chemical Sciences. For NIH, in addition to ``Others'', the fields (ordered by the number of grants) are Biomedical and Clinical Sciences, Biological Sciences, Health Sciences, and Psychology.

Using this grouping, we re-run the related analyses reported in the main text separately by field. We find broadly consistent results and the corresponding results are shown in Figs.~\ref{fig:Figure1-SI-byField-corpus-NSF}-\ref{fig:Figure4-SI-byField}.

\subsection{Robustness by subagency for public data\label{subsection:robustnessbysubagency}}
To assess whether our main findings are consistent across award subagencies in the public datasets, we focus on subagencies where sample sizes are sufficiently large to permit subagency-level analysis and re-estimate all main results separately by subagency. Specifically, for awards with start dates between 2023 and 2025, we compute the number of grants in each subagency. For NSF, all subagencies are retained. For NIH, we retain subagencies with more than 2{,}000 awards and merge all remaining subagencies into a single category labeled ``Others''.

For NSF, the subagencies (ordered by the number of grants) are Directorate for Mathematical \& Physical Sciences (MPS), Directorate for Computer \& Information Science \& Engineering (CISE), Directorate for Engineering (ENG), Directorate for Geosciences (GEO), Directorate for STEM Education (EDU), Directorate for Biological Sciences (BIO), Directorate for Technology, Innovation and Partnerships (TIP), Directorate for Social, Behavioral \& Economic Sciences (SBE), and Office of the Director (OD). For NIH, in addition to ``Others'', the subagencies (ordered by the number of grants) are National Cancer Institute (NCI), National Institute of Allergy and Infectious Diseases (NIAID), National Heart Lung and Blood Institute (NHLBI), National Institute of General Medical Sciences (NIGMS), National Institute of Neurological Disorders and Stroke (NINDS), National Institute on Aging (NIA), National Institute of Diabetes and Digestive and Kidney Diseases (NIDDK), and Eunice Kennedy Shriver National Institute of Child Health and Human Development (NICHD).

Using this grouping, we re-run the related analyses reported in the main text separately by subagency. We find broadly consistent results and the corresponding results are shown in Figs.~\ref{fig:Figure1-SI-byAgency-corpus-NSF}-\ref{fig:Figure4-SI-byAgency}.

\subsection{Alternative reference LLMs\label{subsection:alternative_llm}}
Because different LLMs vary in their preferred vocabulary and stylistic tendencies, it is essential to demonstrate that our results are not driven by the idiosyncrasies of a single model. Our primary detection model uses GPT-3.5 as the reference LLM, which is motivated by the adoption landscape during our study period: ChatGPT, powered by GPT-3.5, was the first widely accessible LLM and rapidly became the dominant tool used by researchers, particularly in 2023 when LLM adoption in grant writing first surged. It therefore represents the most natural reference model for detecting LLM-modified text in our data. Nonetheless, to ensure that our results generalize beyond this single model, we replicate our entire detection and analysis pipeline using alternative LLMs from distinct model families: Claude 3 Haiku (Anthropic), Gemini 2.5 Flash Lite (Google), and Llama 2 7B Chat (Meta). Claude and Gemini are the earliest API-accessible models in their respective families available at the time of this analysis (Fig.~\ref{fig:Figure_comment2.2_timeline}).
\\~\\
For each alternative model, we re-ran the complete estimation procedure: rewriting the reference corpus of grant abstracts using the alternative LLM, re-estimating the human and LLM word-frequency distributions, and recomputing grant-level $\alpha$ for all four datasets. This ensures that the alternative $\alpha$ estimates reflect the linguistic signatures of each specific model rather than those of GPT-3.5. We first assess how strongly the alternative $\alpha$ estimates correlate with the original GPT-3.5-based estimates. As shown in Fig.~\ref{fig:Figure_comment2.2_correlation}, across all four datasets, $\alpha$ values derived from Claude 3 Haiku, Gemini 2.5 Flash Lite, and Llama 2 7B Chat are positively and strongly correlated with the GPT-3.5-based estimates. These correlations indicate that the detection framework captures a shared signal of LLM involvement that is not specific to any single model's linguistic fingerprint. Grants identified as having high (or low) LLM involvement under one model tend to be similarly classified under the others, confirming that the underlying signal is robust to the choice of reference LLM. We then replicate all main regression analyses (Figs.~2-4) using the alternative $\alpha$ measures. The results are broadly consistent across model families (Figs.~\ref{fig:Figure_comment2.2_Fig2}-\ref{fig:Figure_comment2.2_Fig4}).
\\~\\
That said, we note some differences across models. For example, in the semantic distinctiveness analysis, the Gemini-based $\alpha$ estimates yield no statistically significant results. One likely contributor to these differences is temporal mismatch between the reference LLM and the models actually used by grant writers. As shown in Fig.~\ref{fig:Figure_comment2.2_timeline}, the earliest API-accessible models for Claude and Gemini became available in 2024, whereas our data cover grants with start dates beginning in 2023---a period during which ChatGPT was by far the most widely used LLM among researchers. The GPT-3.5-based detection model is therefore the most naturally aligned with the linguistic signatures present in our study period, and the stronger results under this baseline likely reflect a closer match between the reference model and the model most widely used during proposal preparation. Despite this temporal gap, the broad consistency of results across model families reinforces confidence that the core findings are not likely to be artifacts of a single model's idiosyncrasies.

\subsection{Sensitivity to preprocessing choices\label{subsection:sensitivity_preprocessing}}
Word-frequency approaches can be sensitive to preprocessing choices, so demonstrating robustness across alternative pipelines is important for ensuring confidence in the estimated LLM involvement measures. To address this, we systematically evaluate the sensitivity of our $\alpha$ estimates to three key dimensions of text preprocessing: tokenization schemes, stopword handling, and normalization of punctuation.
\\~\\
\paragraph{Current preprocessing pipeline.}
Our current preprocessing pipeline closely follows the approach introduced by Liang \textit{et al}.~\cite{liang2025quantifying}, ensuring consistency with the existing literature and facilitating comparability across studies. Specifically, we apply sentence segmentation using spaCy's \texttt{en\_core\_web\_lg} model; tokenization using a lowercase regex pattern that captures only alphanumeric tokens; removal of pure digits; retention of all non-numeric words (i.e., no stopword removal); and a frequency filter requiring each word to appear in both the human and AI reference corpora (minimum 5 occurrences in human text and 3 in AI text). Because the regex tokenizer captures only word characters, punctuation is effectively removed during tokenization.
\\~\\
\paragraph{Sensitivity analysis design.}
We examine three dimensions of preprocessing: tokenization, stopword handling, and punctuation treatment. For tokenization, we compare (i)~the current regex tokenizer, (ii)~a simple whitespace-based tokenizer, and (iii)~spaCy's NLP tokenizer. For stopword handling, we compare retaining all words versus removing standard English stopwords. For punctuation, we compare removing punctuation (current approach), retaining punctuation as tokens, and normalizing punctuation (e.g., standardizing Unicode variants) before retaining it. These three dimensions define a broader space of possible preprocessing combinations. Not all combinations are meaningful (for example, punctuation-related choices are irrelevant when punctuation is already removed by the tokenizer). After excluding such cases, we construct nine alternative preprocessing pipelines that are logically distinct and well-defined, in addition to our current approach. These include both one-at-a-time changes and combinations that simultaneously vary multiple dimensions. Importantly, for each pipeline, we re-run the entire estimation procedure: re-tokenizing the reference corpora, re-estimating word-level probability distributions, re-processing all grant abstracts, and recomputing $\alpha$ for each grant. This ensures that any differences arise solely from preprocessing choices.
\\~\\
\paragraph{Results.}
Across all preprocessing variants, the resulting $\alpha$ estimates remain highly consistent with the current approach. Correlations between the current and alternative $\alpha$ values are high across both NSF and NIH datasets, generally exceeding 0.90 and often above 0.98. Three main patterns emerge. First, tokenization choice has minimal impact: switching to an NLP tokenizer or even a simple whitespace tokenizer produces nearly identical results. Second, punctuation handling is highly robust: whether punctuation is removed, retained, or normalized has little effect on $\alpha$. Third, stopword removal introduces the largest variation, but even in this case the results remain highly correlated with the current approach. Even when combining multiple changes simultaneously, the estimates remain stable. To further assess agreement, we visualize the relationship between current and alternative $\alpha$ estimates using binned comparisons in Fig.~\ref{fig:Figure_comment2.3_correlation}. Across all variants and both datasets, the results closely follow the 45-degree line, indicating strong agreement throughout the full distribution of $\alpha$---rather than only at extreme values. We also re-estimate our main regression analyses (Figs.~2-4) under these alternative preprocessing pipelines. The estimated coefficients and overall patterns remain highly consistent with the main results (Figs.~\ref{fig:Figure_comment2.3_Fig2}-\ref{fig:Figure_comment2.3_Fig4}). Together, these analyses demonstrate that our $\alpha$ measure is robust to a wide range of preprocessing choices across all three dimensions. The high degree of agreement across pipelines confirms that our findings are not driven by any specific tokenization scheme, stopword list, or normalization of punctuation.

\subsection{AI-related scientific vocabulary\label{subsection:ai_scientific_vocabulary}}
Because AI-related concepts have become increasingly prevalent in scientific research~\cite{gao2024quantifying}, grant proposals may naturally contain more AI-method terminology over time, independent of any LLM involvement in writing. To assess whether our LLM-use estimator captures the growing scientific integration of AI rather than genuine LLM-driven stylistic shifts, we conduct a robustness analysis that mirrors the promotional-language check. Specifically, we compile the list of AI-related concepts identified in prior work~\cite{gao2024quantifying} and remove these terms at all stages of our estimation pipeline. AI-related terms are excluded from the original abstracts prior to LLM rewriting, from the LLM-rewritten abstracts, and from all grant abstracts before estimating $\alpha$. This ensures that any AI-method vocabulary that might reflect substantive research content rather than LLM-driven stylistic shifts cannot contribute to the estimated LLM signal.
\\~\\
Figure~\ref{fig:Figure_comment2.6_ngram_correlation} compares the original $\alpha$ estimates with those obtained after removing AI-related concepts. Across all four datasets, the recalculated $\alpha$ remains highly correlated with the original measure, consistent with what we observe for promotional language. We further replicate the main regression analyses (Figs.~2-4) after removing AI-related concepts. As shown in Figs.~\ref{fig:Figure_comment2.6_ngram_Fig2}-\ref{fig:Figure_comment2.6_ngram_Fig4}, the results remain highly consistent with the main findings. Together, these analyses demonstrate that our measure of LLM involvement is robust to the exclusion of AI-related scientific vocabulary.

\subsection{Word-shift graph decomposition of the LLM detection signal\label{subsection:word_shift}}
To increase the interpretability of the LLM detection method, we apply the generalized word-shift graph framework developed by Gallagher \textit{et al}.~\cite{gallagher2021generalized} to decompose the aggregate divergence between human-written and LLM-modified grant abstracts into per-word contribution scores. Because our estimator relies on corpus-level word-frequency shifts to quantify the prevalence of LLM-modified text, this framework provides a natural way to identify which specific words contribute most strongly to the observed signal.
\\~\\
We report two complementary decompositions. Fig.~\ref{fig:Figure_comment2.9_Proportion} presents word-shift graphs based on proportion differences, which rank words by their directional contribution to the overall frequency shift between human-written and LLM-modified abstracts. Fig.~\ref{fig:Figure_comment2.9_JSD} presents an analogous decomposition using Jensen-Shannon divergence (JSD), which captures symmetric distributional differences and highlights words that are most informative for distinguishing between the two text sources.
\\~\\
Several consistent patterns emerge across both agencies and both decomposition methods. Words that contribute most strongly to the detected LLM signal on the AI side tend to be evaluative or elaborative terms (e.g., endeavor, crucial, comprehensive, innovative for NSF; delve, crucial, intricate, pivotal for NIH), whereas the most distinguishing words on the human side tend to be simpler and more functional (e.g., develop, work, result, help for NSF; develop, increase, work, identify for NIH). Together, these decompositions increase the transparency of the detection method by revealing the specific lexical features underlying the estimated LLM signal.

\subsection{Abstract length sensitivity\label{subsection:abstract_length}}
Because the LLM detection method relies on word-frequency statistics estimated from individual abstracts, shorter texts contain fewer observations per estimate, which could in principle introduce noise or systematic bias in the estimated $\alpha$. To assess whether abstract brevity drives our findings, we conduct a length-based subsample analysis, restricting the sample to abstracts with at least $100$, $150$, and $250$ words and repeating the main analyses underlying Figs.~2-4 within each subsample. If shorter abstracts were driving our results through inflated or noisier $\alpha$ estimates, we would expect the coefficients to attenuate as we progressively exclude shorter texts. As shown in Figs.~\ref{fig:Figure_comment2.4_Fig2}-\ref{fig:Figure_comment2.4_Fig4}, across all length thresholds the results remain consistent with the main findings, with no systematic attenuation of effect sizes. These results demonstrate that our findings are not driven by the brevity of grant abstracts, and that the LLM detection method produces stable estimates across the range of abstract lengths observed in our data.

\subsection{Single-investigator versus multi-investigator proposals and awards}
We compare estimated LLM-modification rates between single-investigator and multi-investigator proposals and awards. As shown in Fig.~\ref{fig:Figure_comment2.5_raw_plot}, in most cases there are no clear differences in average $\alpha$ between single-investigator and multi-investigator proposals or awards, with the exception of the public NSF dataset where multi-investigator proposals exhibit higher $\alpha$. We therefore include an indicator for single-investigator versus multi-investigator as a control variable in the regression analysis to account for this potential source of variation. After including this control, we re-estimate the regression models underlying Figs.~2-4. The results (Figs.~\ref{fig:Figure_comment2.5_Fig2}-\ref{fig:Figure_comment2.5_Fig4}) remain highly consistent with the main findings, indicating that controlling for differences between single-investigator and multi-investigator proposals or awards does not affect the overall conclusions of the analysis.

\subsection{Standardized predictors\label{subsection:standardized_predictors}}
In the main analysis, we do not normalize or standardize the predictors. To assess sensitivity to predictor scaling, we conduct an additional robustness check in which all predictors are standardized by converting them to $z$-scores before estimating the regressions. Repeating the main analyses underlying Figs.~2-4 with these standardized predictors yields consistent results (Figs.~\ref{fig:Figure_comment2.7_Fig2}-\ref{fig:Figure_comment2.7_Fig4}), further confirming the robustness of the findings. In addition, full regression tables---including coefficients and standard errors for all variables, including control variables---are reported in Tables~\ref{table:Table_comment2.7_Fig2}-\ref{table:Table_comment2.7_Fig4}.

\subsection{LLM involvement and within-investigator topical distance\label{subsection:llm_involvement}}
We conduct an additional analysis measuring how far each proposal departs from the investigator’s own recent research trajectory. Specifically, for each investigator who submitted at least one proposal in 2023-2025, we compute the cosine distance between that proposal and the investigator’s own proposals submitted within the previous two years within the same agency (NSF or NIH). This provides a measure of how much the topical content of a proposal deviates from the investigator’s recent proposal history.
\\~\\
We then estimate regressions in which the dependent variable is this within-investigator cosine distance. The key explanatory variable is the estimated degree of LLM involvement in the proposal ($\alpha$). The specification includes investigator fixed effects, proposal year fixed effects, and field fixed effects, and controls for proposal funding amount. Standard errors are clustered at the investigator level. The results are shown in Fig.~\ref{fig:Figure_commnet1.3_pivot_within_investigator}. Across both NSF and NIH, the estimated relationship between LLM involvement and within-investigator cosine distance is statistically indistinguishable from zero. In other words, proposals with higher LLM involvement are not systematically farther from an investigator’s own prior proposals. These findings suggest that LLM use does not lead investigators to submit proposals that deviate more strongly from their own recent research trajectory.

\subsection{Within-field, cross-agency analysis\label{subsection:within_field}}
We conduct a more stringent analysis that directly compares agencies within a common field. Specifically, we focus on Biomedical \& Clinical Sciences, which is the largest field in NIH's portfolio and has non-trivial representation in NSF and the Department of Defense (DOD) as well, and extend the analysis to include DOD as an additional agency beyond NSF and NIH. Within this restricted sample, we estimate regression models analogous to our main specifications in Fig. 4, where the coefficients capture the association between LLM involvement ($\alpha$) and research outcomes separately for each agency.
\\~\\
The results (Fig.~\ref{fig:Figure_comment1.2_Fig4}) reveal that, even within the same subfield, the relationship between $\alpha$ and research outcomes differs across agencies. For total publication output (panel a), $\alpha$ is negatively associated with publications for NSF-funded projects and positively and significantly associated for NIH-funded projects, while the coefficient for DOD is not statistically distinguishable from zero. For high-impact publications (panel b), a similar pattern emerges: the coefficient for NSF is negative, the coefficient for NIH is positive, and the DOD estimate is again not statistically significant. These results show that all three agencies exhibit distinct relationships between LLM involvement and downstream research outcomes, even when restricting to the same scientific domain. These within-field agency differences indicate that the patterns we document in the main analysis cannot be attributed solely to field composition.

\subsection{Within-investigator variation and robustness to single-field unit of analysis}
Our main regression specifications include investigator fixed effects, which rely on within-investigator variation for identification. To clarify the extent of within-investigator variation underlying these fixed-effects specifications, we report descriptive statistics on the distribution of proposals and awards per investigator across the four datasets. A substantial share of investigators appears multiple times: approximately 50\% in the private NSF data and 60\% in the private NIH data, and these repeat investigators account for roughly 80\% and 86\% of all proposals, respectively. In the public data, the share of repeat investigators is smaller---approximately 24\% (NSF) and 21\% (NIH)---but they still account for a large fraction of observations ($\sim$54\% and $\sim$42\% of awards, respectively). Moreover, because our unit of analysis is defined at the proposal-PI-field-year level, a given proposal or award can be associated with multiple fields. As a result, even investigators with a single proposal or award contribute multiple observations through field-level variation and are not dropped by the inclusion of investigator and field fixed effects. The share of observations retained in the fixed-effects regression sample remains high: approximately 91\% (private NSF), 96\% (private NIH), 77\% (public NSF), and 76\% (public NIH). Results are also robust to restricting to one field per proposal or award (Figs.~\ref{fig:Figure_comment1.7_Fig2}-\ref{fig:Figure_comment1.7_Fig4}).

\newpage
\section*{Supplementary Figures}
\addcontentsline{toc}{section}{Supplementary Figures}
\setcounter{figure}{0}
\setcounter{table}{0}

\begin{figure}[htbp!]
\centering
\includegraphics[width=0.8\columnwidth]{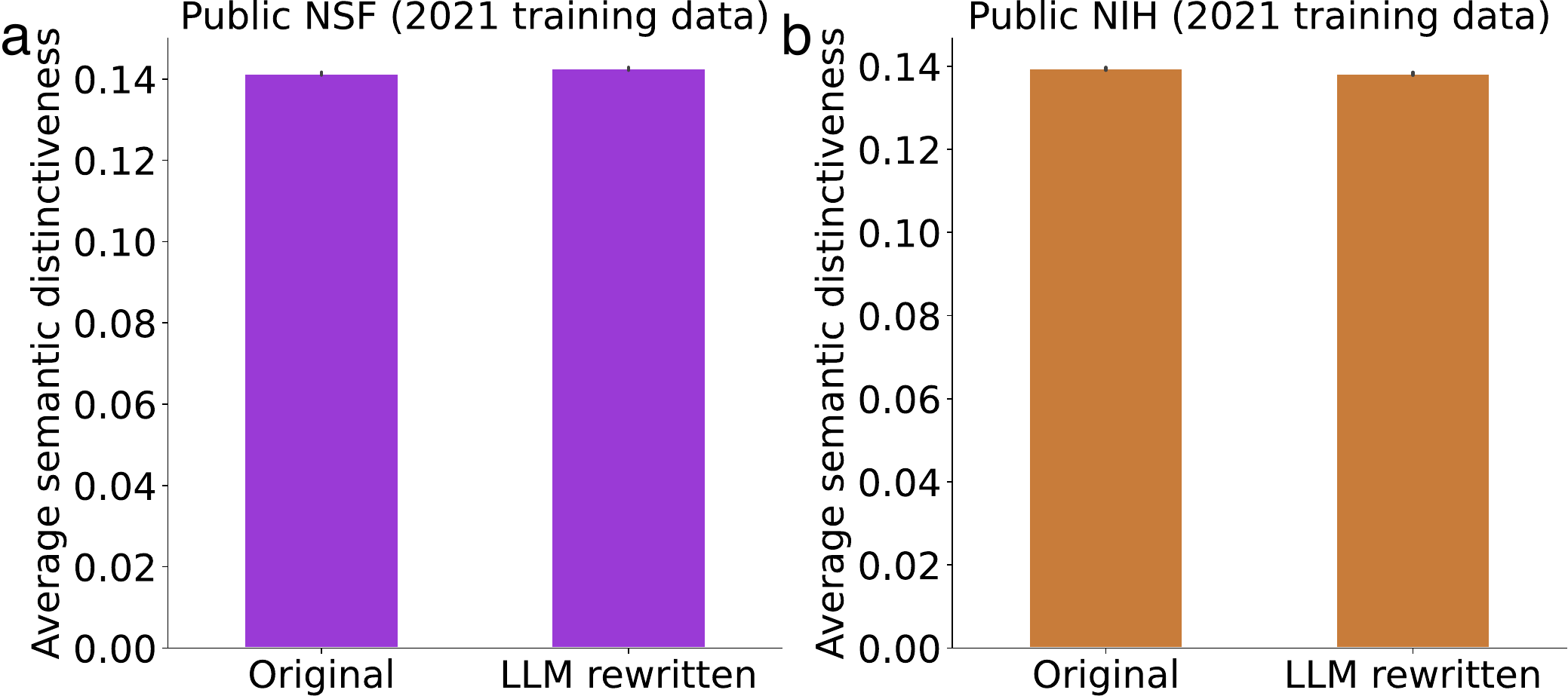}
\caption[Semantic distinctiveness is robust to LLM-induced stylistic changes.]{\textbf{Semantic distinctiveness is robust to LLM-induced stylistic changes.}  (\textbf{a}-\textbf{b}) Average cosine distance to grants funded in the prior year for original and LLM-rewritten versions of 6,000 grant abstracts with start dates in 2021, shown separately for NSF (\textbf{a}) and NIH (\textbf{b}).}
\label{fig:Figure_comment2.8}
\end{figure}

\newpage
\begin{figure}[htbp!]
\centering
\includegraphics[width=0.8\columnwidth]{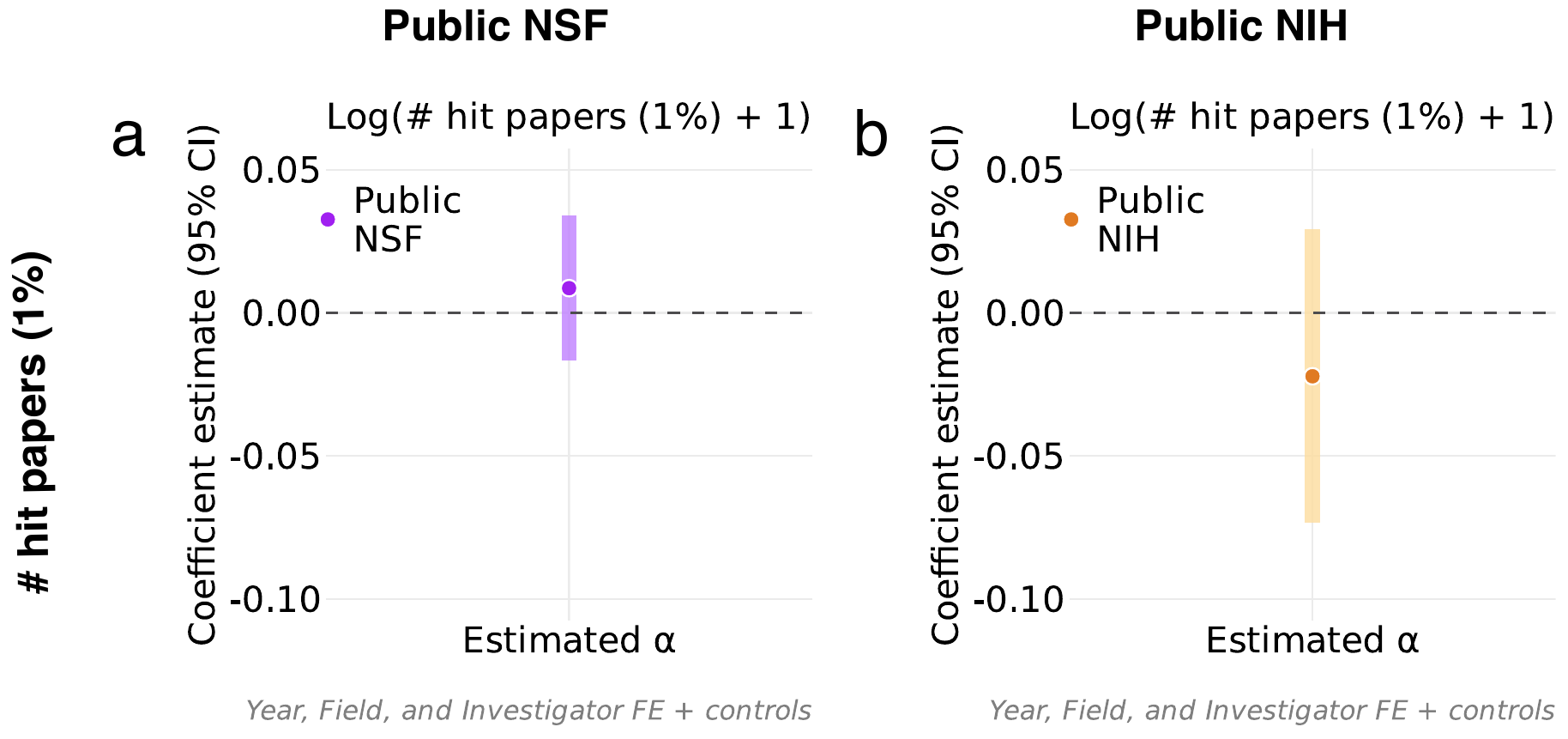}
\caption[LLM use and downstream output in US federal research funding.]{\textbf{LLM use and downstream output in US federal research funding.}  (\textbf{a}-\textbf{b}) Regression estimates relating grant-level LLM use ($\alpha$) to high-impact outputs, where a ``hit'' paper is defined as one whose citations fall within the top 1\% of all papers published worldwide in the same year and field. All regressions include grant start year, field, and investigator fixed effects, as well as controls for funding amount. Points indicate coefficient estimates, and bars denote 95\% confidence intervals.}
\label{fig:Figure4-SI-hitpaper1percent}
\end{figure}

\newpage
\begin{figure}[htbp!]
\centering
\includegraphics[width=0.8\columnwidth]{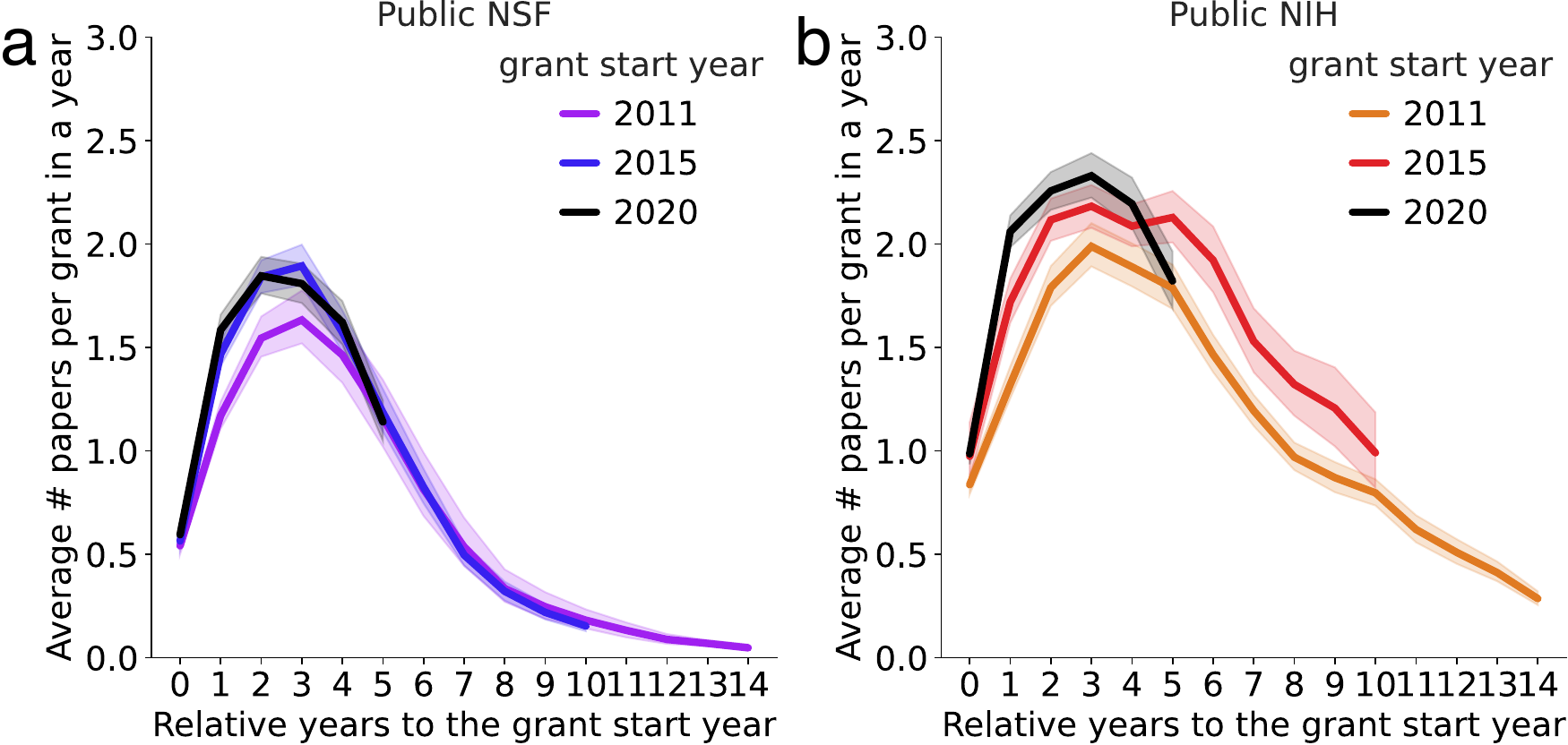}
\caption[Publication trajectories following grant start.]{\textbf{Publication trajectories following grant start.}  Average number of publications produced per grant as a function of years since the grant start year, shown separately for (\textbf{a}) NSF and (\textbf{b}) NIH grants. Each line represents a different grant start cohort (2011, 2015, and 2020). Publication output rises rapidly after a grant begins, typically peaks around 2-4 years, and then gradually declines. The trajectory is highly similar across cohorts.}
\label{fig:Figure_comment1.6_over_time}
\end{figure}

\newpage
\begin{figure}[htbp!]
\centering
\includegraphics[width=0.8\columnwidth]{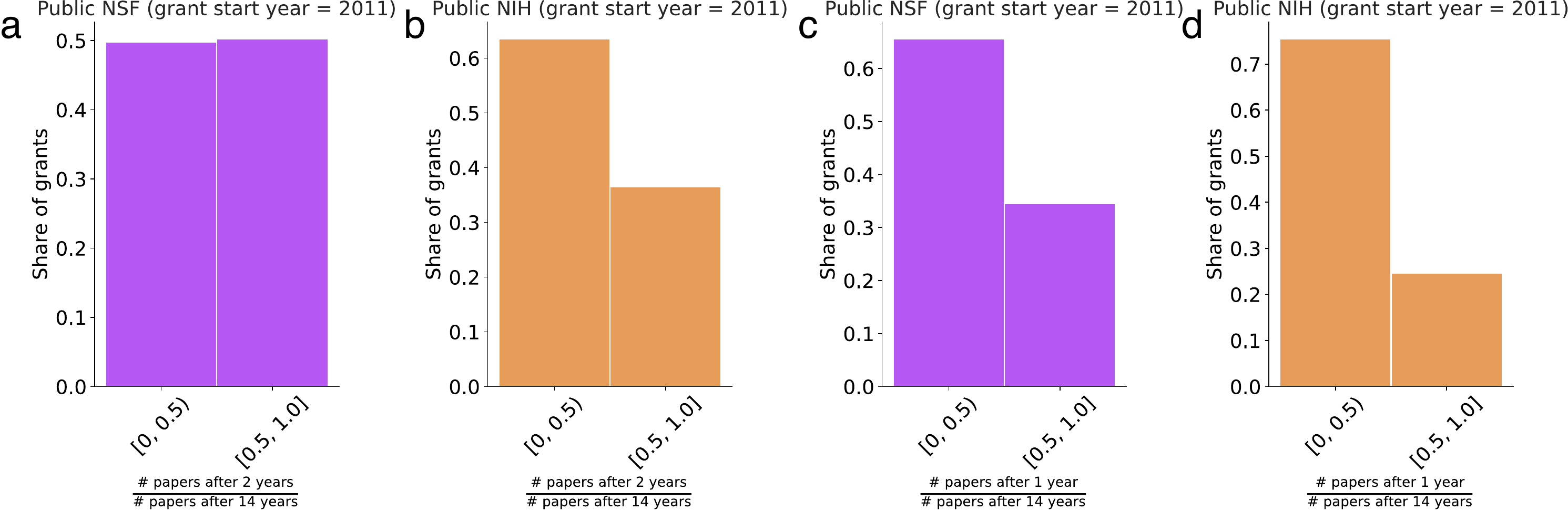}
\caption[Share of lifetime publications captured within early observation windows.]{\textbf{Share of lifetime publications captured within early observation windows.}  Distribution of the fraction of total publications (over 14 years) produced within the first two years after the grant start for (\textbf{a}) NSF and (\textbf{b}) NIH grants starting in 2011. (\textbf{c}-\textbf{d}) Corresponding distributions using a one-year window for (\textbf{c}) NSF and (\textbf{d}) NIH grants.}
\label{fig:Figure_comment1.6_2011_raw}
\end{figure}

\newpage
\begin{figure}[htbp!]
\centering
\includegraphics[width=0.8\columnwidth]{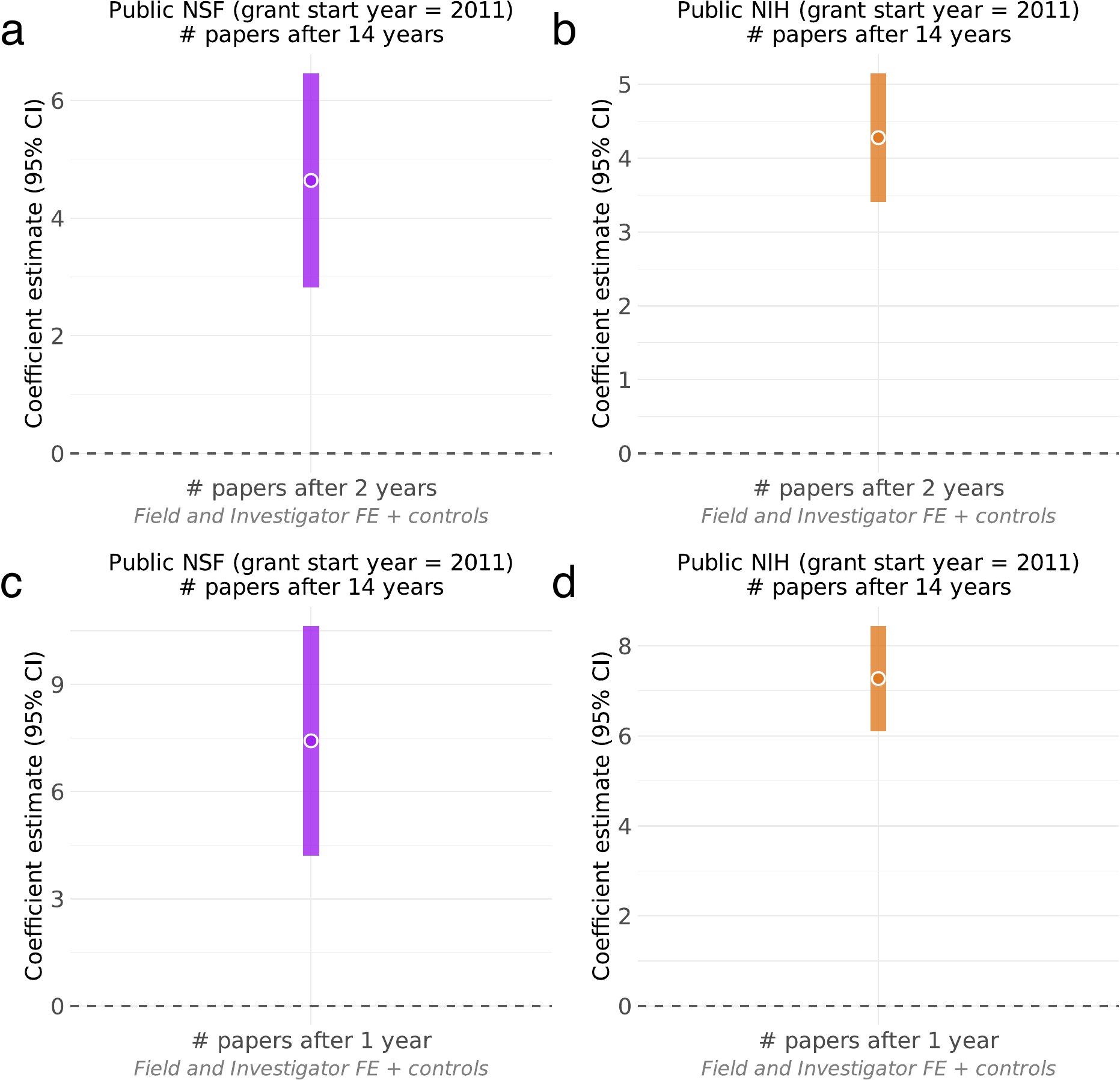}
\caption[Early publication output predicts long-run publication output.]{\textbf{Early publication output predicts long-run publication output.}  Regression estimates of the association between early and long-run publication output for grants starting in 2011. (\textbf{a}-\textbf{b}) The number of papers produced within the first two years after the grant start predicts the total number of papers produced over the full 14-year window for (\textbf{a}) NSF and (\textbf{b}) NIH grants, controlling for investigator fixed effects, field fixed effects, and funding amount. (\textbf{c}-\textbf{d}) Corresponding estimates using the number of papers produced within the first year as the predictor for (\textbf{c}) NSF and (\textbf{d}) NIH grants. All regressions include investigator and field fixed effects, as well as controls for funding amount. Points indicate coefficient estimates, and bars denote 95\% confidence intervals.}
\label{fig:Figure_comment1.6_2011_regression}
\end{figure}

\newpage
\begin{figure}[htbp!]
\centering
\includegraphics[width=0.8\columnwidth]{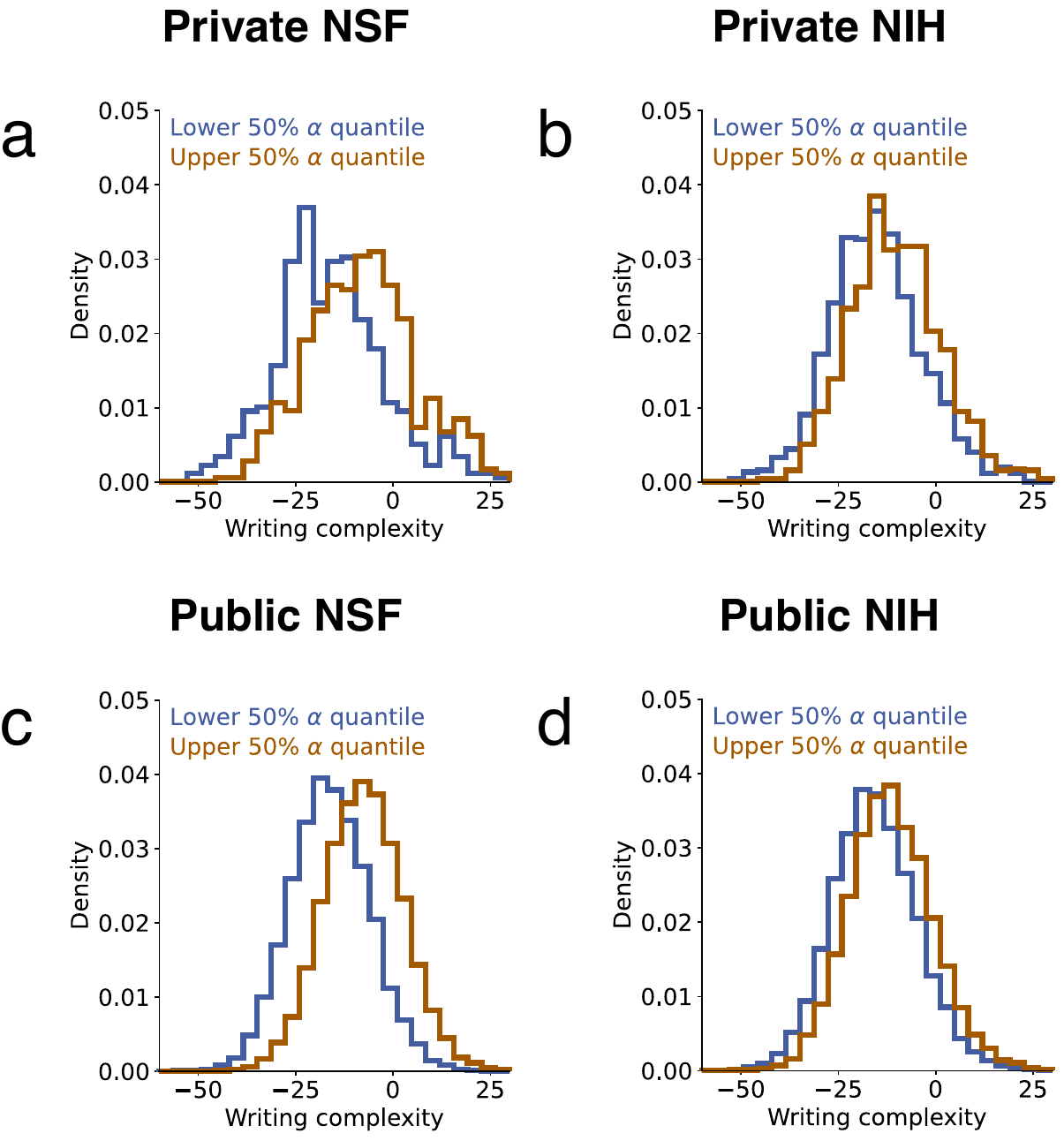}
\caption[Distributions of writing complexity by LLM use intensity.]{
\textbf{Distributions of writing complexity by LLM use intensity.}
This figure uses grants with start years between 2023 and 2025 across four datasets. Writing complexity is measured based on the Flesch Reading Ease score (see Section~\ref{subsection:writing_complexity}). Grants are split by the median $\alpha$ value into lower 50\% (blue) and upper 50\% (brown) quantiles. Panels show results separately for private NSF (\textbf{a}), private NIH (\textbf{b}), public NSF (\textbf{c}), and public NIH (\textbf{d}) grants. Across all four datasets, grants with higher $\alpha$ values show systematically higher writing complexity. In addition, within each panel, the differences between the two groups are statistically significant (two-sided Welch's t-test, $p < 0.001$ in all cases).}
\label{fig:Figure1-SI-WordComplexity}
\end{figure}

\newpage
\begin{figure}[htbp!]
\centering
\includegraphics[width=0.8\columnwidth]{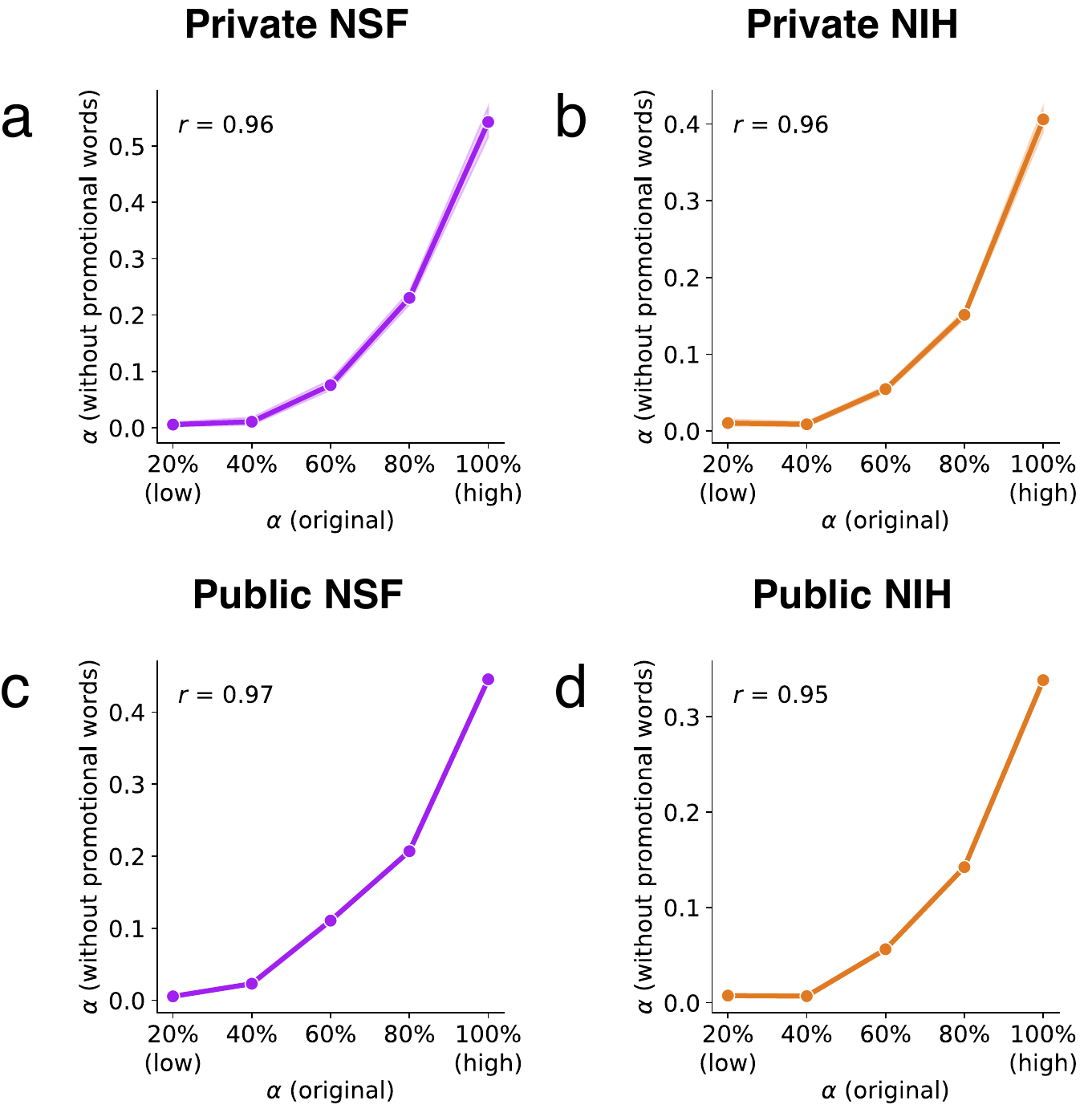}
\caption[LLM-use estimates after removing promotional words.]{\textbf{LLM-use estimates after removing promotional words.} This figure assesses whether our measure of individual-level LLM use ($\alpha$) is driven by promotional language. Panels report the relationship between the original $\alpha$ and a recalculated $\alpha$ after removing promotional words for private NSF (\textbf{a}), private NIH (\textbf{b}), public NSF (\textbf{c}), and public NIH (\textbf{d}) samples. Points denote estimates at different $\alpha$ quantiles, with lines connecting adjacent quantiles. Across all samples, the recalculated measure remains highly correlated with the original estimate ($r$ = 0.95-0.97, $p < 0.001$ in all cases), where $r$ denotes the Pearson correlation coefficient between the original and recalculated $\alpha$ across quantiles.}
\label{fig:Figure1-SI-PromotionalWords}
\end{figure}

\newpage
\begin{figure}[htbp!]
\centering
\includegraphics[width=0.8\columnwidth]{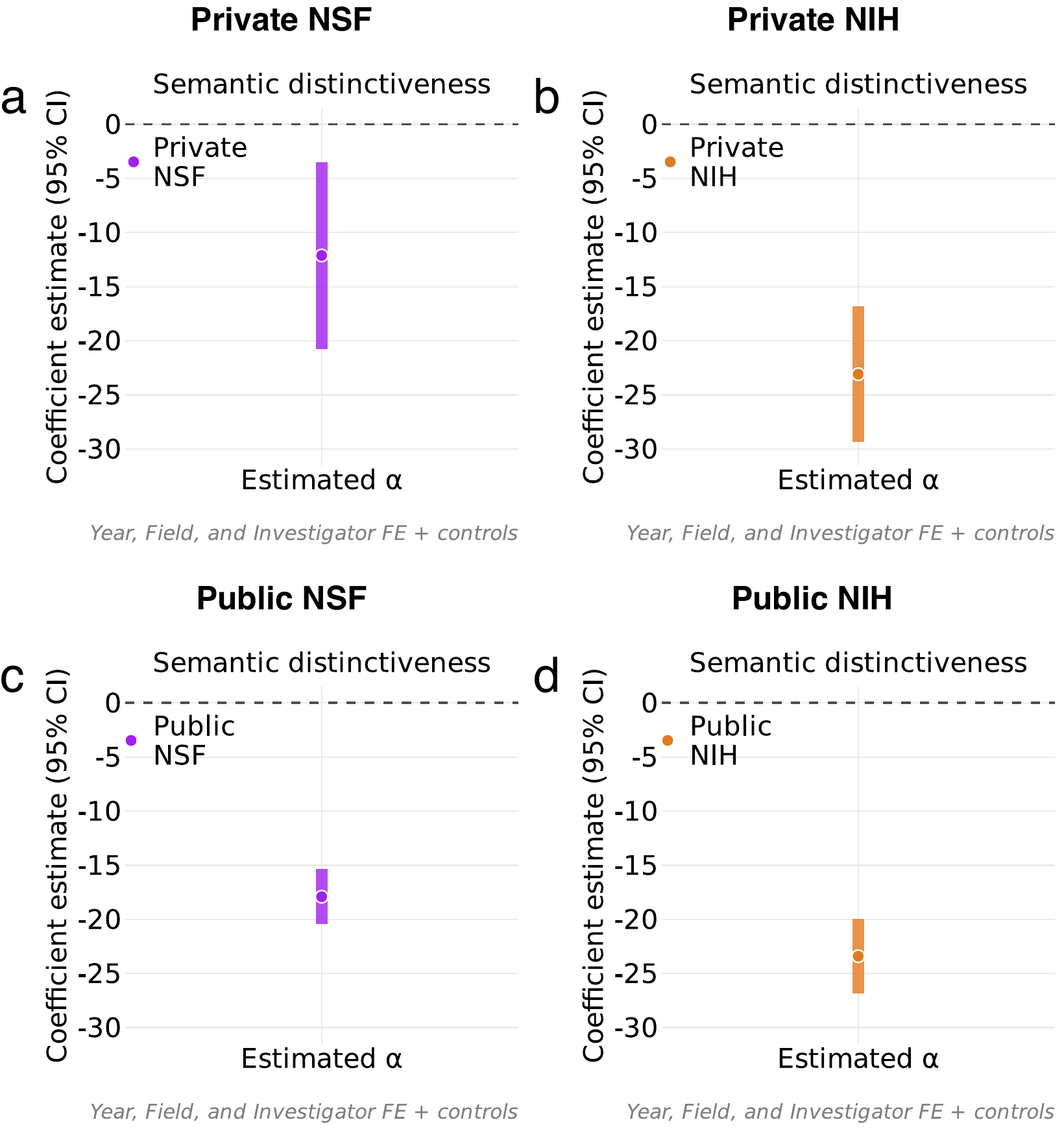}
\caption[LLM use and semantic distinctiveness in US federal research funding.]{\textbf{LLM use and semantic distinctiveness in US federal research funding.} Regression estimates relating grant-level LLM use ($\alpha$) to semantic distance from abstracts funded in the prior two years within the same agency, expressed as within-year percentiles. Panels show results separately for private NSF (\textbf{a}), private NIH (\textbf{b}), public NSF (\textbf{c}), and public NIH (\textbf{d}) grants. All regressions include grant start year, field, and investigator fixed effects, as well as controls for funding amount. Points indicate coefficient estimates, and bars denote 95\% confidence intervals. Negative coefficients correspond to proposals and awards that are positioned closer, in semantic space, to recently funded work within the same agency.}
\label{fig:Figure2-SI-2years-embedding}
\end{figure}

\begin{figure}[htbp!]
\centering
\includegraphics[width=0.8\columnwidth]{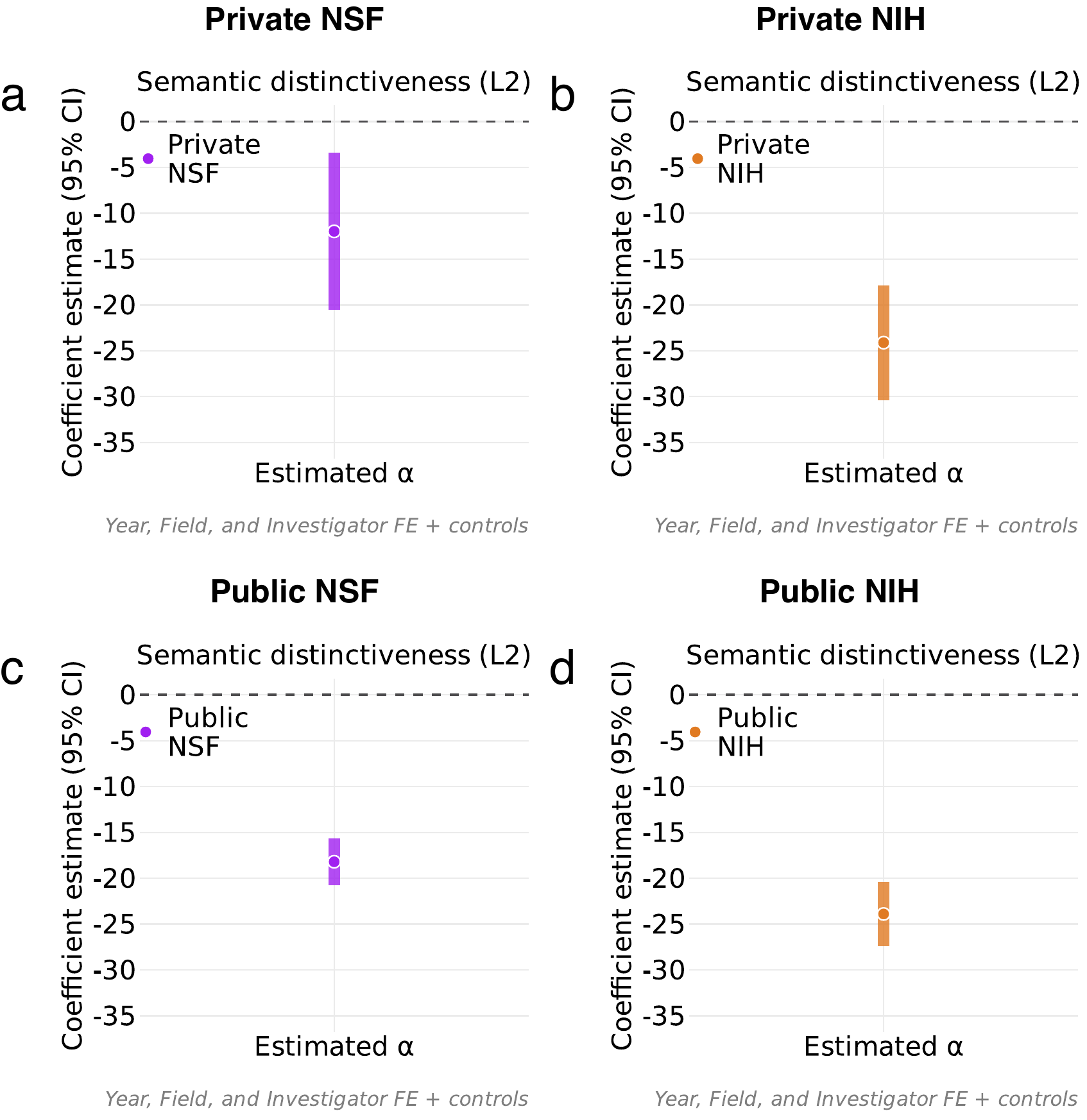}
\caption[LLM use and semantic distinctiveness in US federal research funding using L2 distance.]{\textbf{LLM use and semantic distinctiveness in US federal research funding using L2 distance.}  (\textbf{a}-\textbf{d}) Regression estimates relating grant-level LLM use ($\alpha$) to semantic distance from abstracts funded in the prior year within the same agency, expressed as within-year percentiles. Panels show results separately for private NSF (\textbf{a}), private NIH (\textbf{b}), public NSF (\textbf{c}), and public NIH (\textbf{d}) grants. All regressions include grant start year, field, and investigator fixed effects, as well as controls for funding amount. Points indicate coefficient estimates, and bars denote 95\% confidence intervals. Negative coefficients correspond to proposals and awards that are positioned closer, in semantic space, to recently funded work within the same agency.}
\label{fig:Figure2-SI-L2distance-embedding}
\end{figure}

\begin{figure}[htbp!]
\centering
\includegraphics[width=0.8\columnwidth]{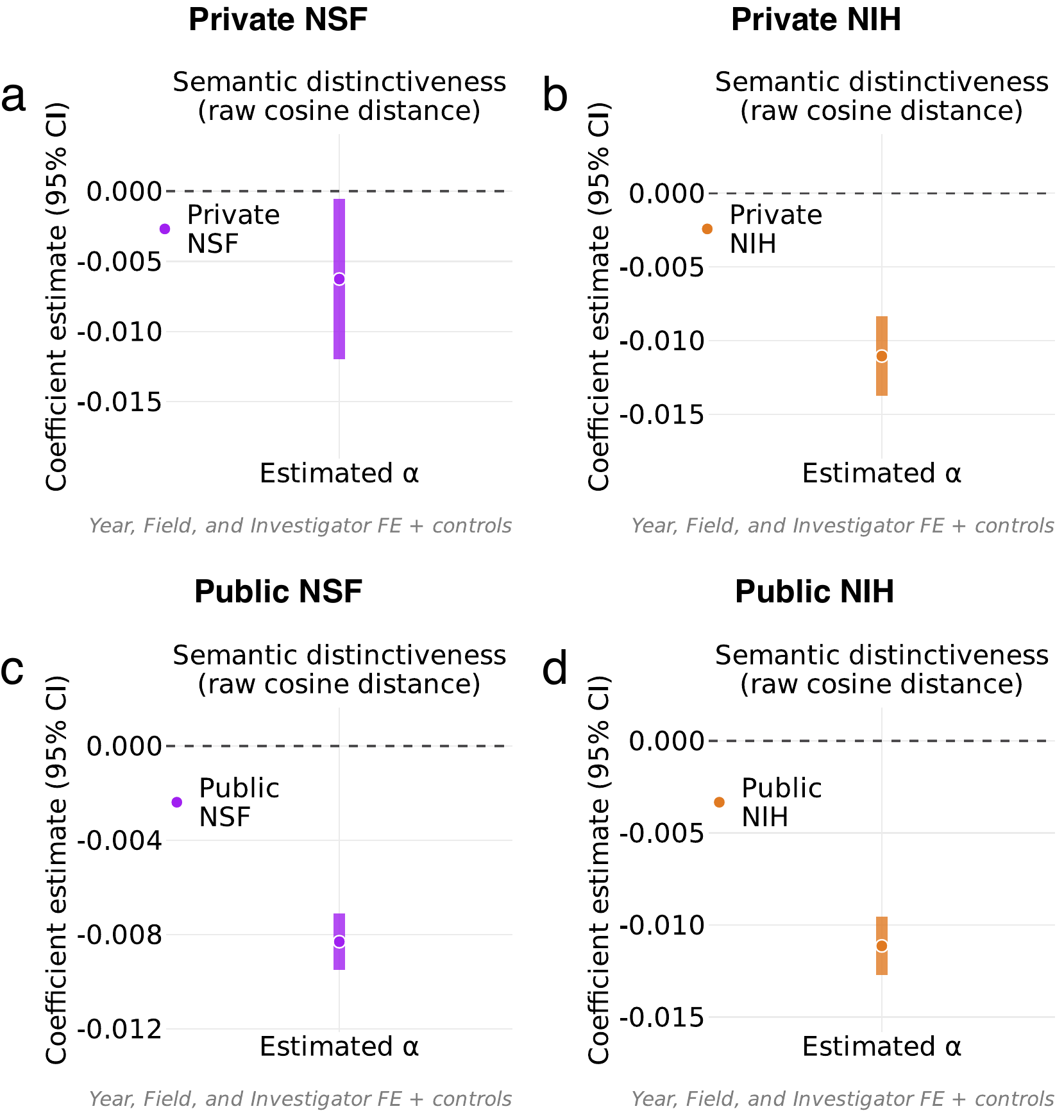}
\caption[LLM use and semantic distinctiveness in US federal research funding using raw cosine distance.]{\textbf{LLM use and semantic distinctiveness in US federal research funding using raw cosine distance.}  (\textbf{a}-\textbf{d}) Regression estimates relating grant-level LLM use ($\alpha$) to semantic distance from abstracts funded in the prior year within the same agency. Panels show results separately for private NSF (\textbf{a}), private NIH (\textbf{b}), public NSF (\textbf{c}), and public NIH (\textbf{d}) grants. All regressions include grant start year, field, and investigator fixed effects, as well as controls for funding amount. Points indicate coefficient estimates, and bars denote 95\% confidence intervals. Negative coefficients correspond to proposals and awards that are positioned closer, in semantic space, to recently funded work within the same agency.}
\label{fig:Figure_comment1.5}
\end{figure}

\newpage
\begin{figure}[htbp!]
\centering
\includegraphics[width=0.8\columnwidth]{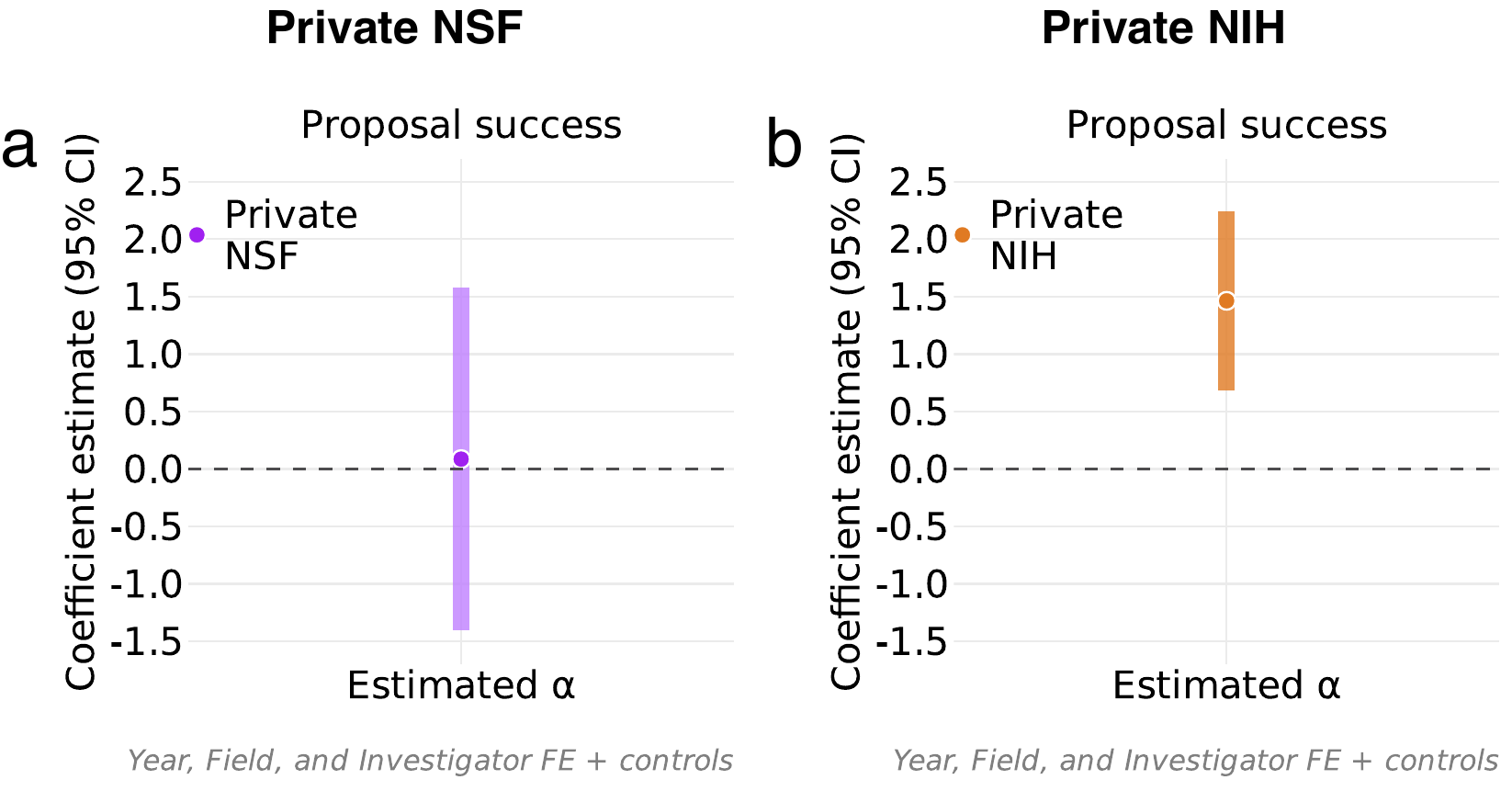}
\caption[LLM use and proposal success in US federal research funding.]{\textbf{LLM use and proposal success in US federal research funding.} Using private NSF and NIH proposal submissions from two large US R1 universities, this figure examines the relationship between LLM use at submission ($\alpha$) and funding success. (\textbf{a}) Logistic regression estimates for NSF submissions. (\textbf{b}) Corresponding estimates for NIH submissions. All regressions include proposal request start year, field, and investigator fixed effects, as well as controls for requested funding amount. Points indicate coefficient estimates, and bars denote 95\% confidence intervals.}
\label{fig:Figure3-SI-logit}
\end{figure}

\newpage
\begin{figure}[htbp!]
\centering
\includegraphics[width=0.8\columnwidth]{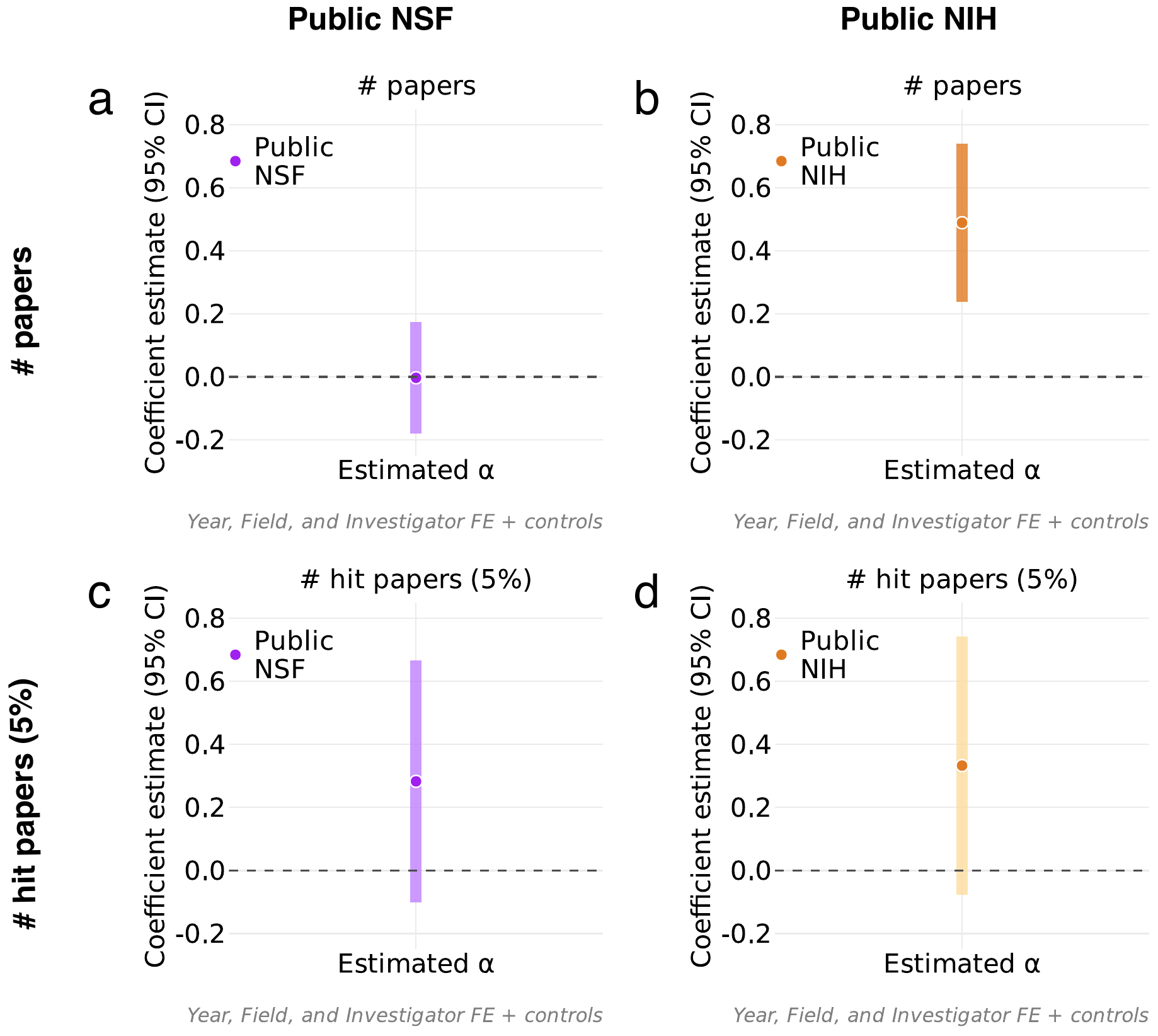}
\caption[LLM use and federal research funding outputs.]{\textbf{LLM use and downstream output in US federal research funding.}  (\textbf{a}-\textbf{b}) Negative binomial regression estimates relating grant-level LLM use ($\alpha$) to the total number of resulting publications for NSF (\textbf{a}) and NIH (\textbf{b}) grants. (\textbf{c}-\textbf{d}) Corresponding estimates for high-impact outputs, where a ``hit'' paper is defined as one whose citations fall within the top 5\% of all papers published worldwide in the same year and field. All regressions include grant start year, field, and investigator fixed effects, as well as controls for funding amount. Points indicate coefficient estimates, and bars denote 95\% confidence intervals.}
\label{fig:Figure4-SI-negbino}
\end{figure}

\newpage
\begin{figure}[htbp!]
\centering
\includegraphics[width=0.8\columnwidth]{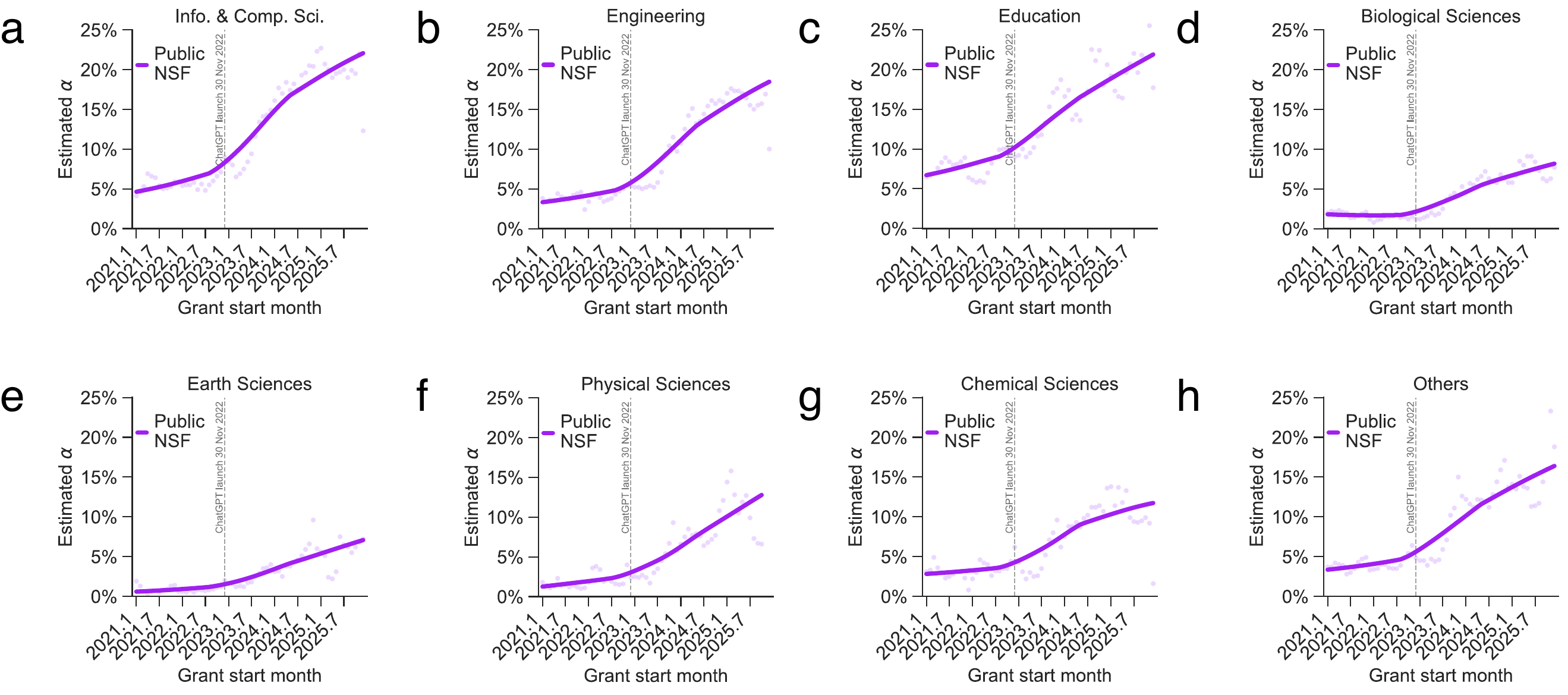}
\caption[Rapid rise of LLM use in US federal research funding by field in public NSF.]{\textbf{Rapid rise of LLM use in US federal research funding by field in public NSF data.} (\textbf{a}-\textbf{h}) Corpus-level estimates of LLM use ($\alpha$) for NSF grants from 2021 to 2025, computed separately by field. The estimates are computed using rolling three-month windows. Solid lines show locally weighted regressions within each field. The vertical dashed line marks November 30, 2022, corresponding to the public release of ChatGPT.}
\label{fig:Figure1-SI-byField-corpus-NSF}
\end{figure}

\newpage
\begin{figure}[htbp!]
\centering
\includegraphics[width=0.8\columnwidth]{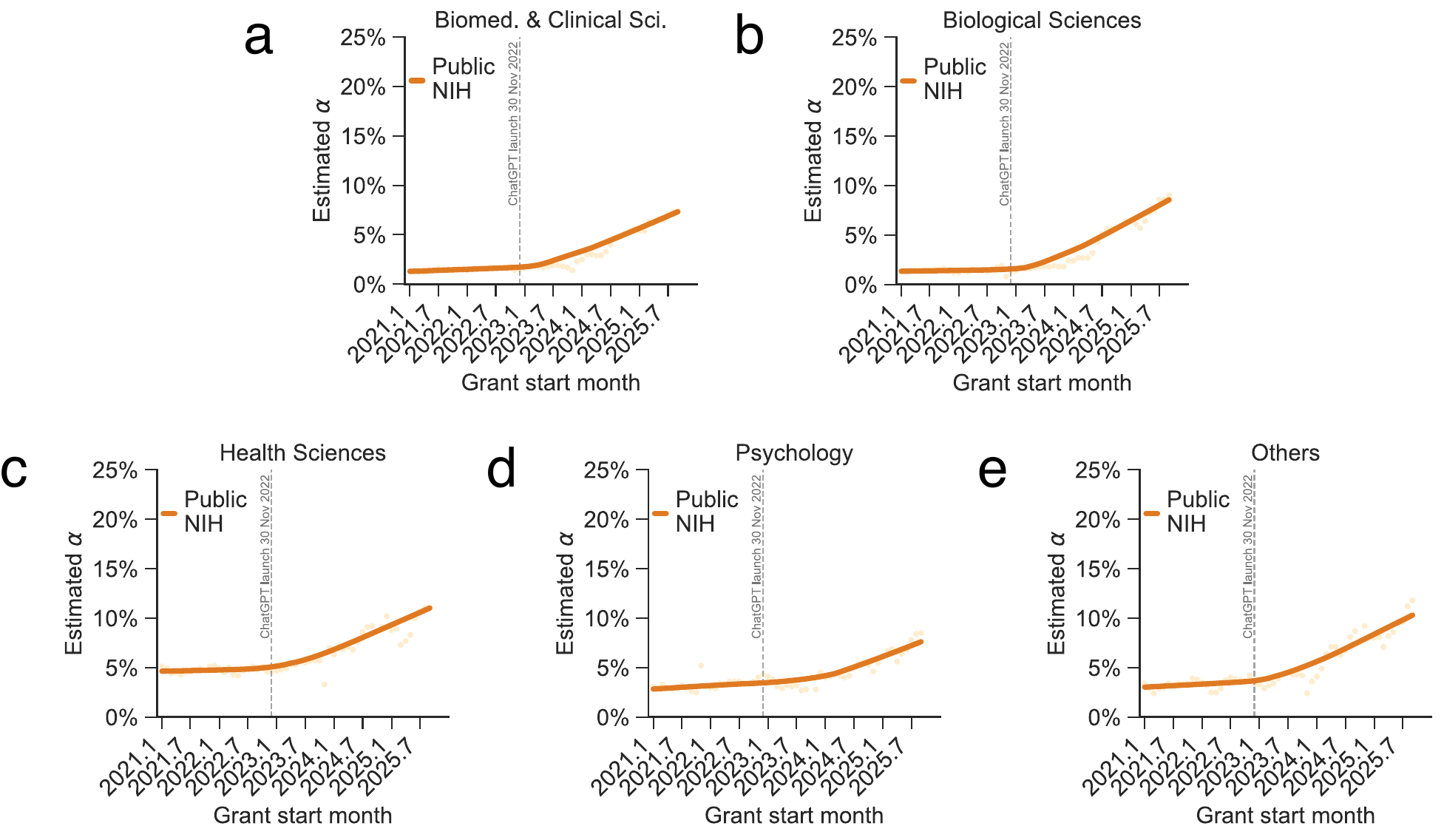}
\caption[Rapid rise of LLM use in US federal research funding by field in public NIH data.]{\textbf{Rapid rise of LLM use in US federal research funding by field in public NIH data.}  (\textbf{a}-\textbf{e}) Corpus-level estimates of LLM use ($\alpha$) for public NIH grants from 2021 to 2025, computed separately by field. The estimates are computed using rolling three-month windows. Solid lines show locally weighted regressions within each field. The vertical dashed line marks November 30, 2022, corresponding to the public release of ChatGPT.}
\label{fig:Figure1-SI-byField-corpus-NIH}
\end{figure}

\newpage
\begin{figure}[htbp!]
\centering
\includegraphics[width=0.8\columnwidth]{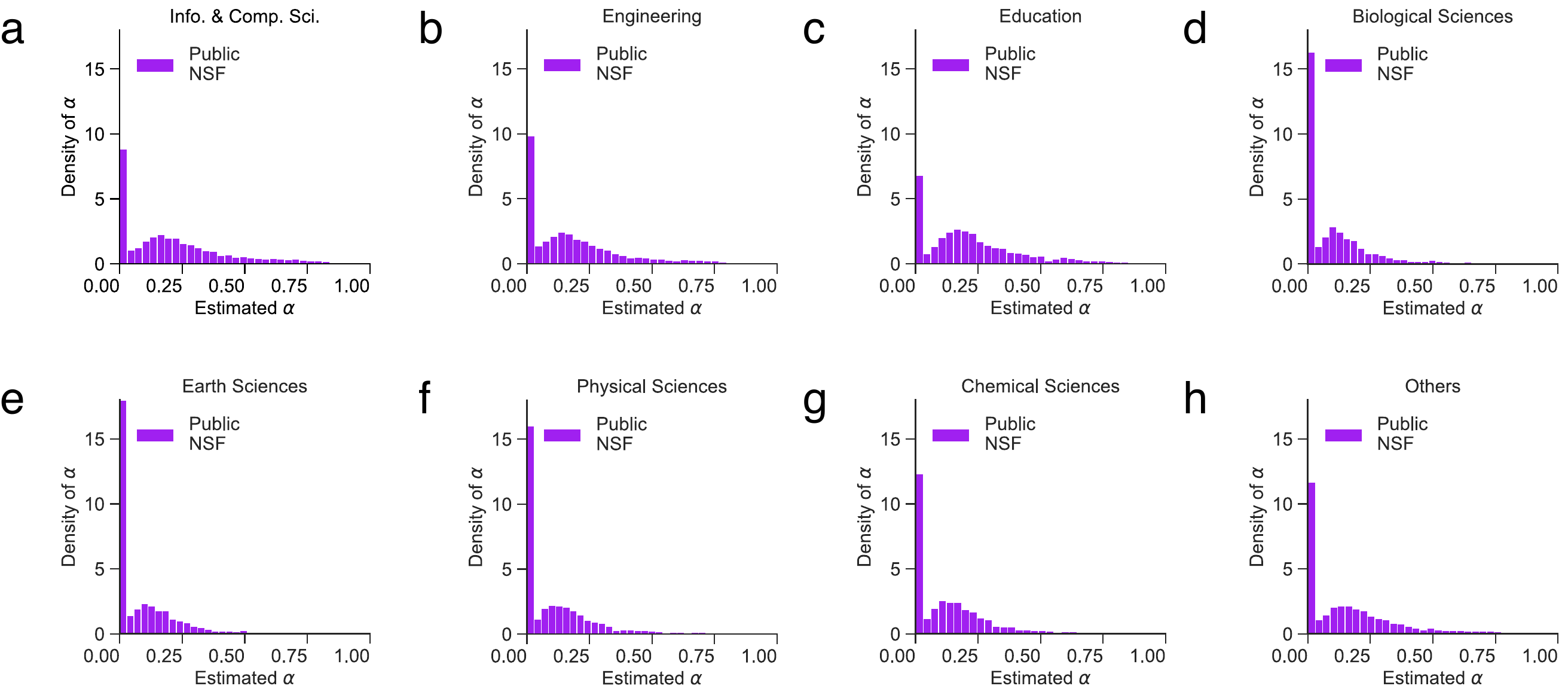}
\caption[Bimodal distribution of LLM use in US federal research funding by field in public NSF.]{\textbf{Bimodal distribution of LLM use in US federal research funding by field in public NSF data.}  (\textbf{a}-\textbf{h}) Distributions of individual-grant $\alpha$ for NSF awards with start dates between 2023 and 2025, shown separately by field. Each field reveals a bimodal pattern consistent with a split between minimal and substantive LLM use across grants.}
\label{fig:Figure1-SI-byField-individual-NSF}
\end{figure}

\newpage
\begin{figure}[htbp!]
\centering
\includegraphics[width=0.8\columnwidth]{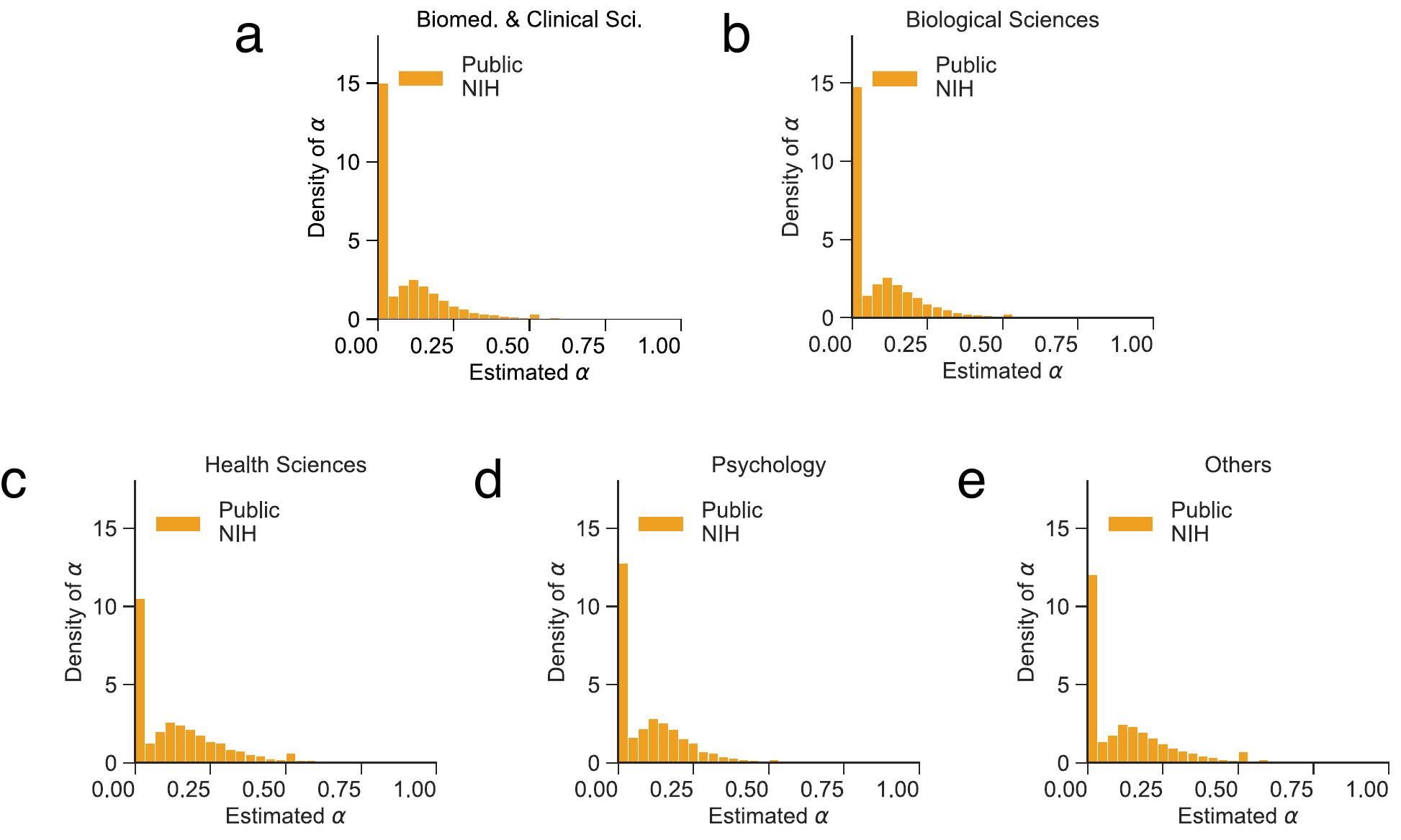}
\caption[Bimodal distribution of LLM use in US federal research funding by field in public NIH.]{\textbf{Bimodal distribution of LLM use in US federal research funding by field in public NIH data.}  (\textbf{a}-\textbf{e}) Distributions of individual-grant $\alpha$ for NIH awards with start dates between 2023 and 2025, shown separately by field. Each field reveals a bimodal pattern consistent with a split between minimal and substantive LLM use across grants.}
\label{fig:Figure1-SI-byField-individual-NIH}
\end{figure}

\newpage
\begin{figure}[htbp!]
\centering
\includegraphics[width=0.8\columnwidth]{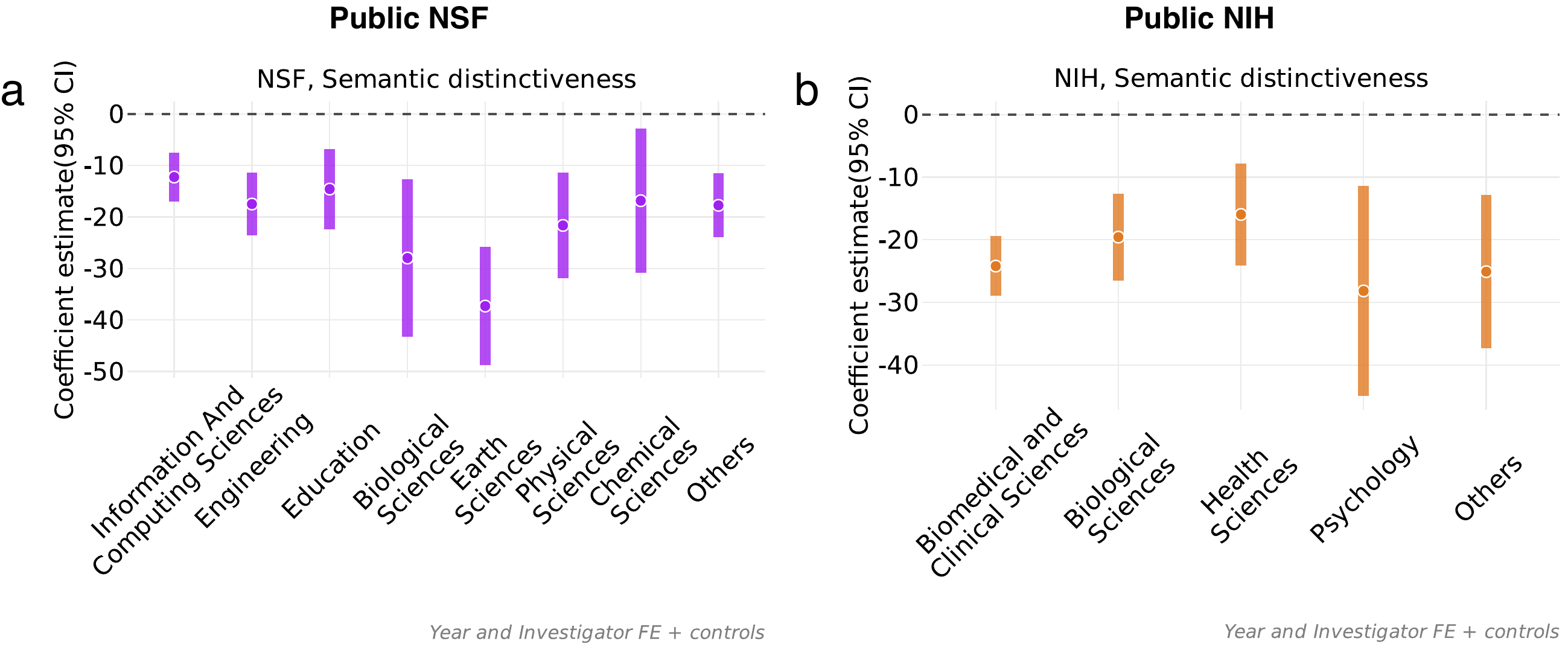}
\caption[LLM use and semantic distinctiveness in US federal research funding by field.]{\textbf{LLM use and semantic distinctiveness in US federal research funding by field.}  (\textbf{a}-\textbf{b}) Regression estimates relating grant-level LLM use ($\alpha$) to semantic distance from abstracts funded in the prior year within the same agency. Panels show results separately for public NSF and public NIH grants. All regressions include grant start year and investigator fixed effects, as well as controls for funding amount. Points indicate coefficient estimates, and bars denote 95\% confidence intervals.}
\label{fig:Figure2-SI-byField-Public}
\end{figure}

\newpage
\begin{figure}[htbp!]
\centering
\includegraphics[width=0.8\columnwidth]{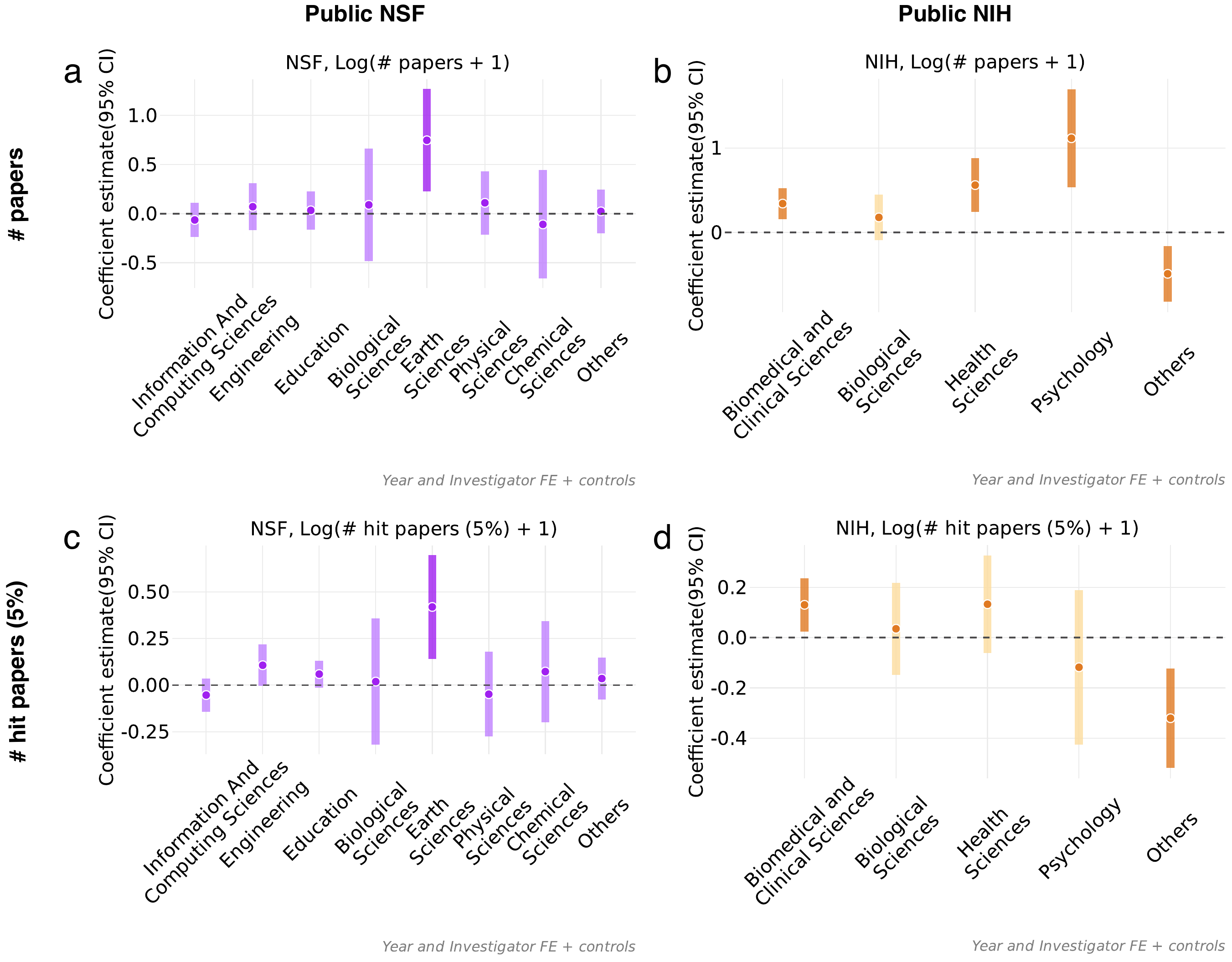}
\caption[LLM use and federal research funding outputs by field.]{\textbf{LLM use and federal research funding outputs by field.}  (\textbf{a-b}) Regression estimates relating grant-level LLM use ($\alpha$) to the total number of resulting publications for NSF and NIH grants. (\textbf{c-d}) Corresponding estimates for high-impact outputs, where a ``hit'' paper is defined as one whose citations fall within the top 5\% of all papers published worldwide in the same year and field. All regressions include grant start year and investigator fixed effects, as well as controls for funding amount. Points indicate coefficient estimates, and bars denote 95\% confidence intervals.}
\label{fig:Figure4-SI-byField}
\end{figure}

\newpage
\begin{figure}[htbp!]
\centering
\includegraphics[width=0.8\columnwidth]{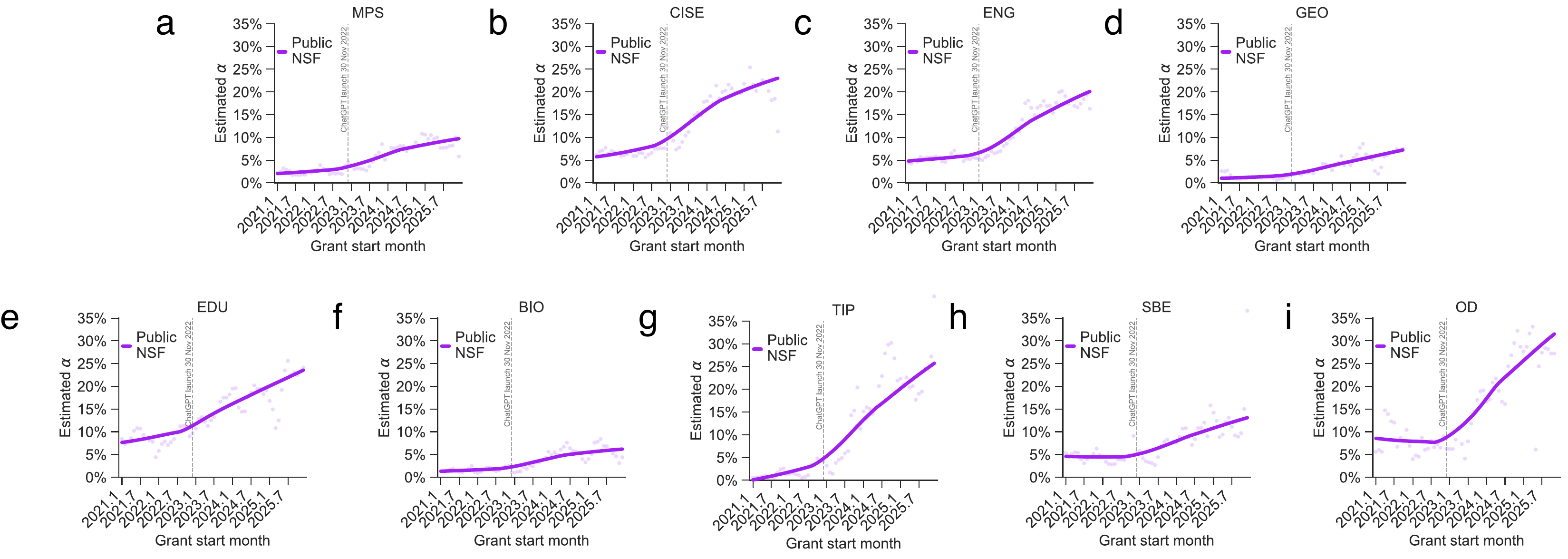}
\caption[Rapid rise of LLM use in US federal research funding by subagency in public NSF data.]{\textbf{Rapid rise of LLM use in US federal research funding by subagency in public NSF data.}  (\textbf{a}-\textbf{i}) Corpus-level estimates of LLM use ($\alpha$) for NSF grants from 2021 to 2025, computed separately by subagency. The estimates are computed using rolling three-month windows. Solid lines show locally weighted regressions within each subagency. The vertical dashed line marks November 30, 2022, corresponding to the public release of ChatGPT.}
\label{fig:Figure1-SI-byAgency-corpus-NSF}
\end{figure}

\newpage
\begin{figure}[htbp!]
\centering
\includegraphics[width=0.8\columnwidth]{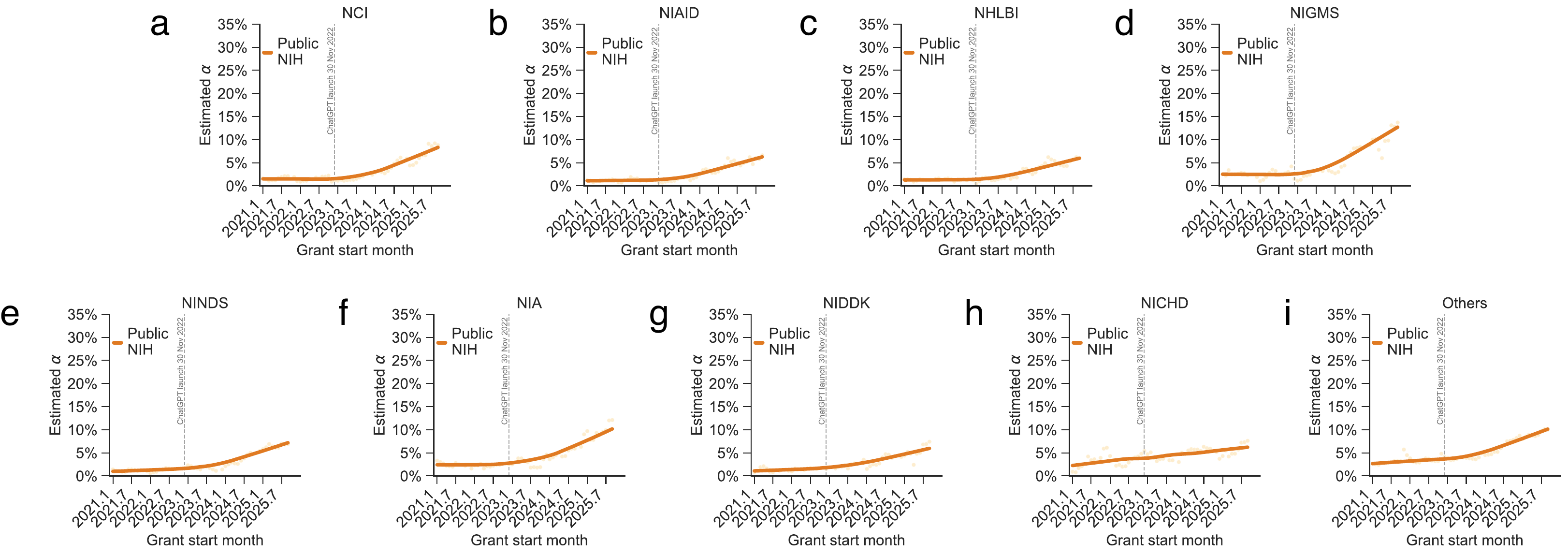}
\caption[Rapid rise of LLM use in US federal research funding by subagency in public NIH data.]{\textbf{Rapid rise of LLM use in US federal research funding by subagency in public NIH data.}  (\textbf{a}-\textbf{i}) Corpus-level estimates of LLM use ($\alpha$) for NIH grants from 2021 to 2025, computed separately by subagency. The estimates are computed using rolling three-month windows. Solid lines show locally weighted regressions within each subagency. The vertical dashed line marks November 30, 2022, corresponding to the public release of ChatGPT.}
\label{fig:Figure1-SI-byAgency-corpus-NIH}
\end{figure}

\newpage
\begin{figure}[htbp!]
\centering
\includegraphics[width=0.8\columnwidth]{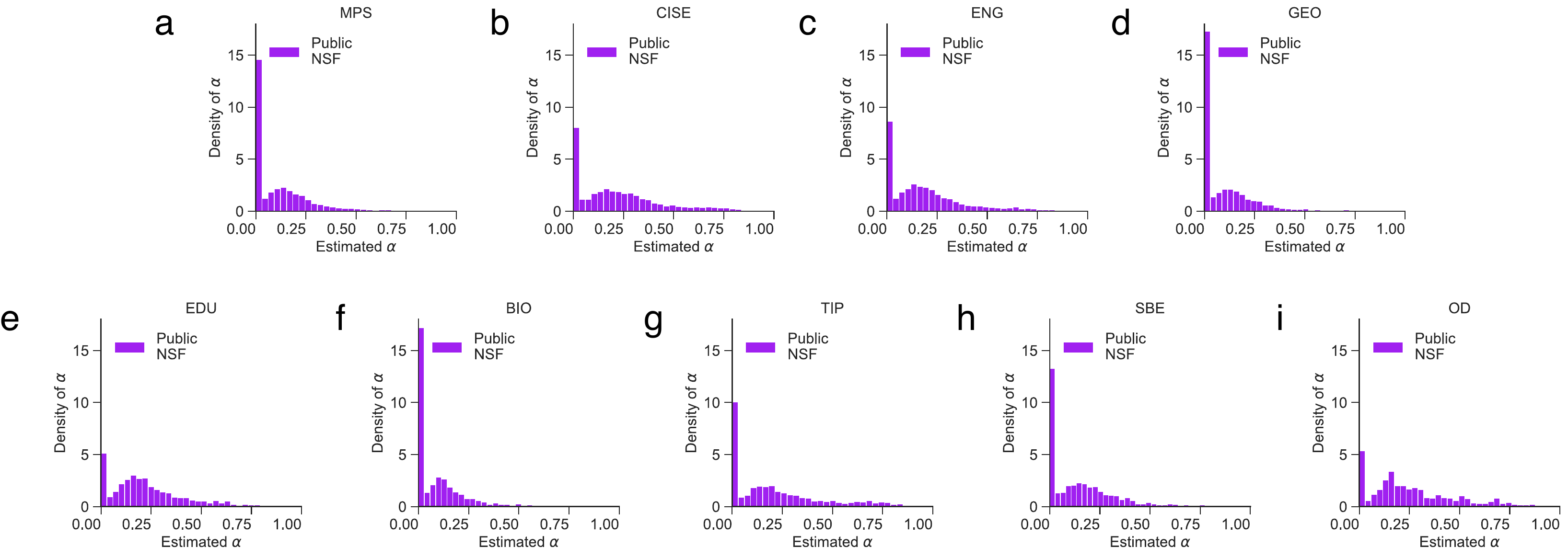}
\caption[Bimodal distribution of LLM use in US federal research funding by subagency in public NSF data.]{\textbf{Bimodal distribution of LLM use in US federal research funding by subagency in public NSF data.}  (\textbf{a}-\textbf{i}) Distributions of individual grant $\alpha$ for NSF awards with start dates between 2023 and 2025, shown separately by subagency. Each subagency reveals a bimodal pattern consistent with a split between minimal and substantive LLM use across grants.}
\label{fig:Figure1-SI-byAgency-individual-NSF}
\end{figure}

\newpage
\begin{figure}[htbp!]
\centering
\includegraphics[width=0.8\columnwidth]{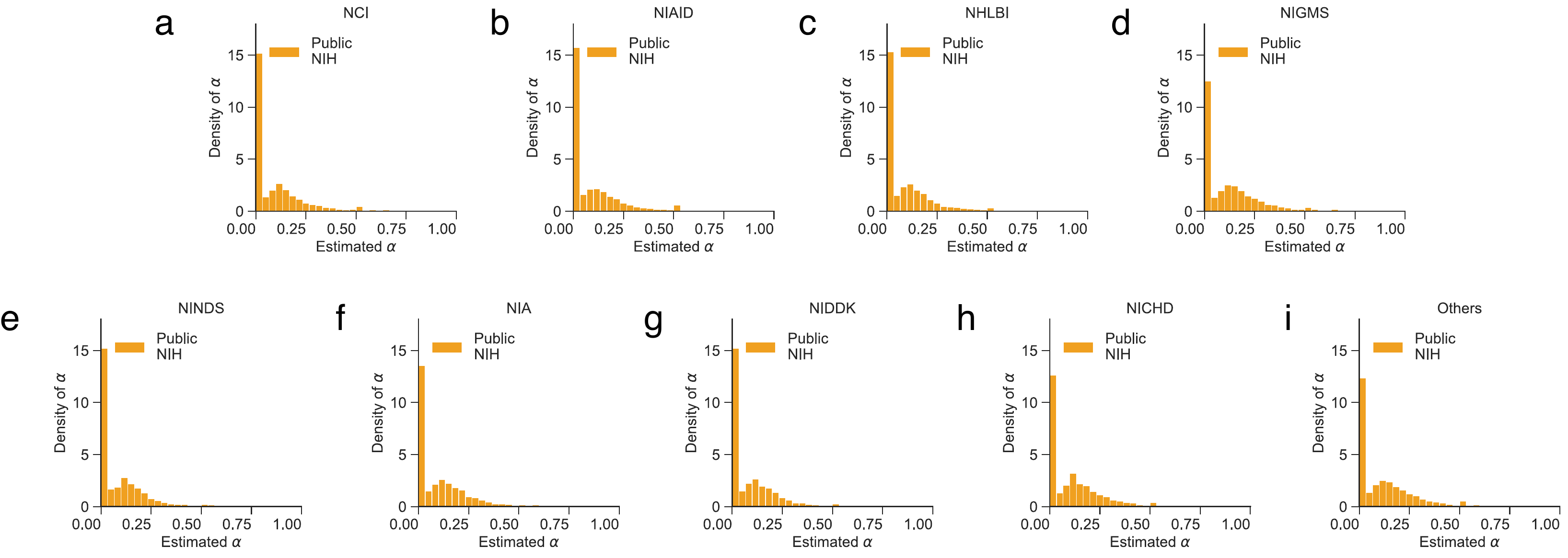}
\caption[Bimodal distribution of LLM use in US federal research funding by subagency in public NIH data.]{\textbf{Bimodal distribution of LLM use in US federal research funding by subagency in public NIH data.}  (\textbf{a}-\textbf{i}) Distributions of individual-grant $\alpha$ for NIH awards with start dates between 2023 and 2025, shown separately by subagency. Each subagency reveals a bimodal pattern consistent with a split between minimal and substantive LLM use across grants.}
\label{fig:Figure1-SI-byAgency-individual-NIH}
\end{figure}

\newpage
\begin{figure}[htbp!]
\centering
\includegraphics[width=0.8\columnwidth]{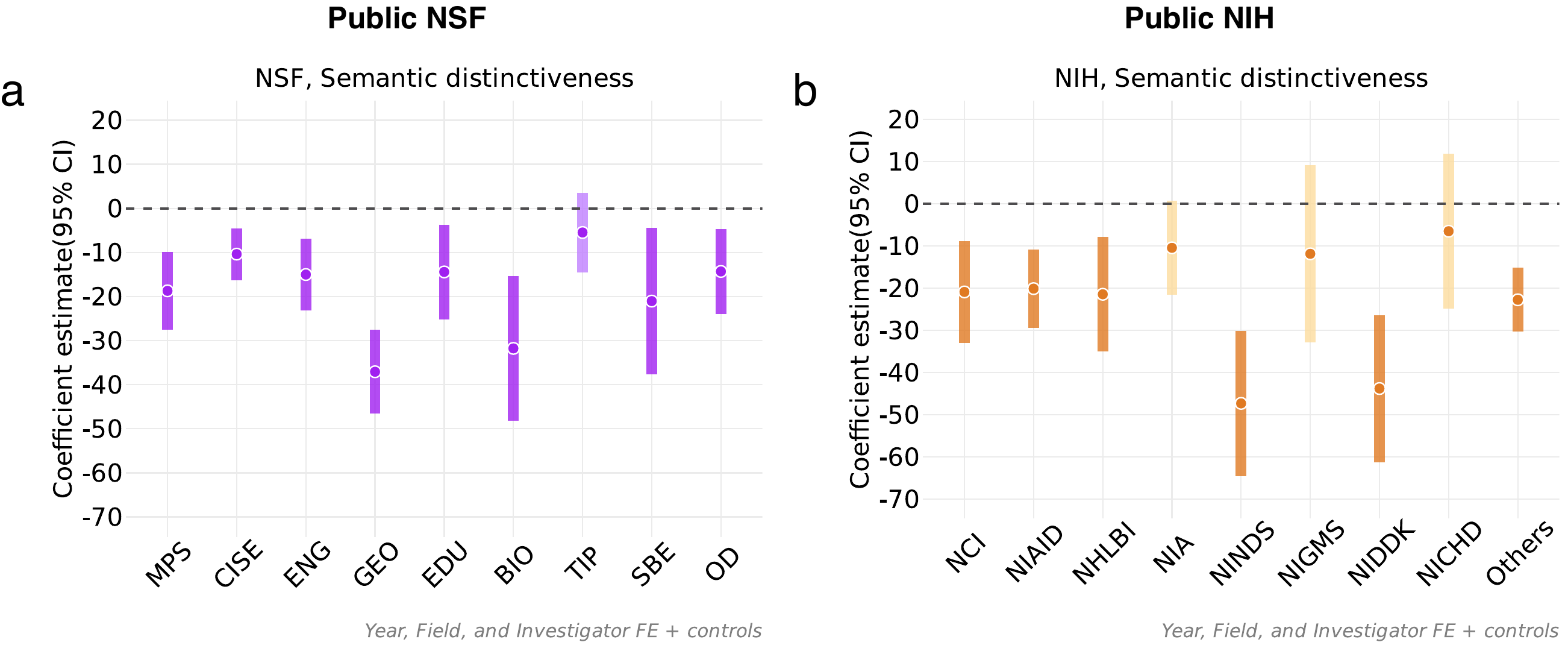}
\caption[LLM use and semantic distinctiveness in US federal research funding by subagency.]{\textbf{LLM use and semantic distinctiveness in US federal research funding by subagency.}  (\textbf{a}-\textbf{b}) Regression estimates relating grant-level LLM use ($\alpha$) to semantic distance from abstracts funded in the prior year within the same agency. Panels show results separately for public NSF and public NIH grants. All regressions include grant start year, field, and investigator fixed effects, as well as controls for funding amount. Points indicate coefficient estimates, and bars denote 95\% confidence intervals.}
\label{fig:Figure2-SI-byAgency-Public}
\end{figure}

\newpage
\begin{figure}[htbp!]
\centering
\includegraphics[width=0.8\columnwidth]{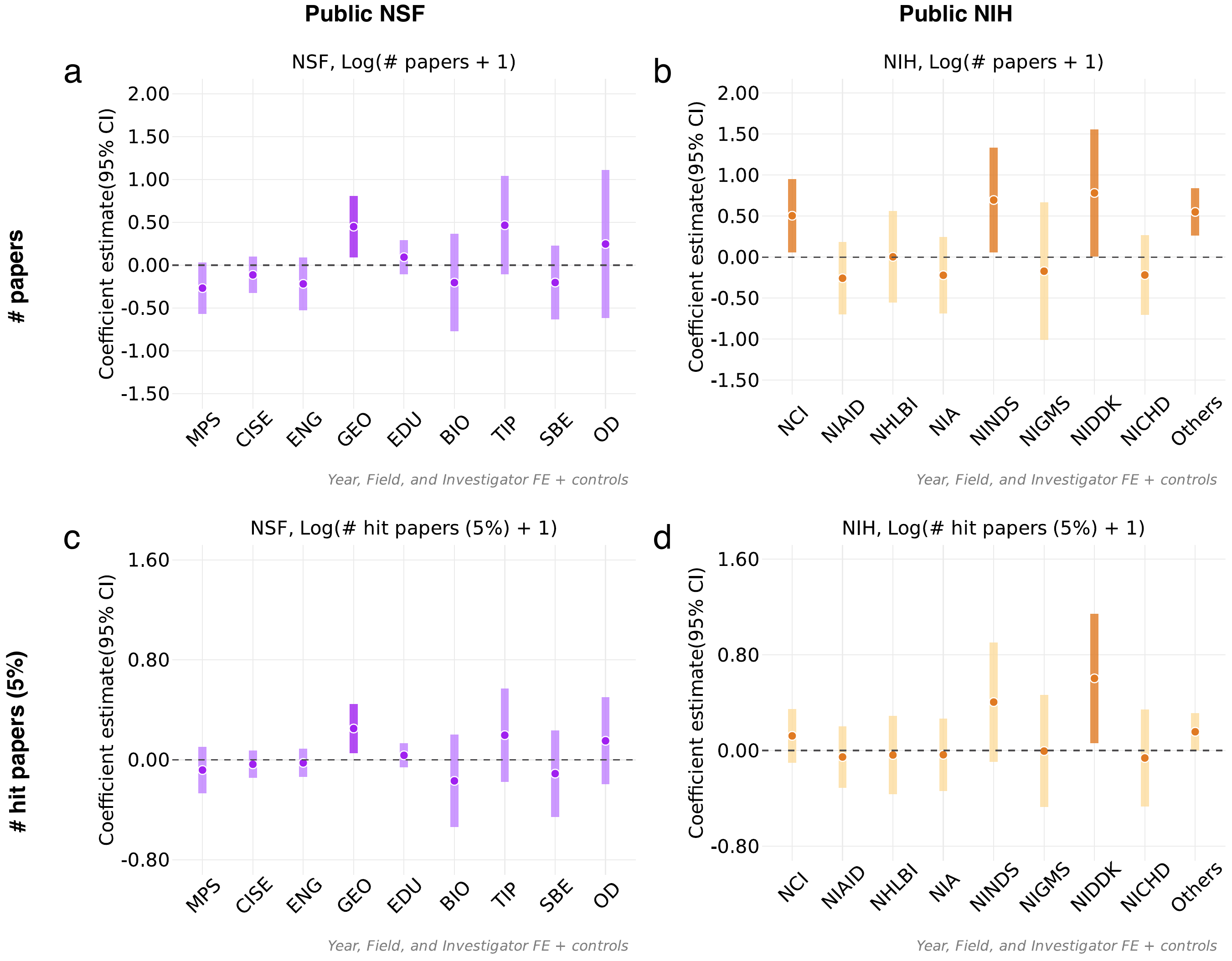}
\caption[LLM use and federal research funding outputs by subagency.]{\textbf{LLM use and federal research funding outputs by subagency.}  (\textbf{a-b}) Regression estimates relating grant-level LLM use ($\alpha$) to the total number of resulting publications for NSF and NIH grants. (\textbf{c-d}) Corresponding estimates for high-impact outputs, where a ``hit'' paper is defined as one whose citations fall within the top 5\% of all papers published worldwide in the same year and field. All regressions include grant start year, field, and investigator fixed effects, as well as controls for funding amount. Points indicate coefficient estimates, and bars denote 95\% confidence intervals.}
\label{fig:Figure4-SI-byAgency}
\end{figure}

\newpage
\begin{figure}[htbp!]
\centering
\includegraphics[width=0.8\columnwidth]{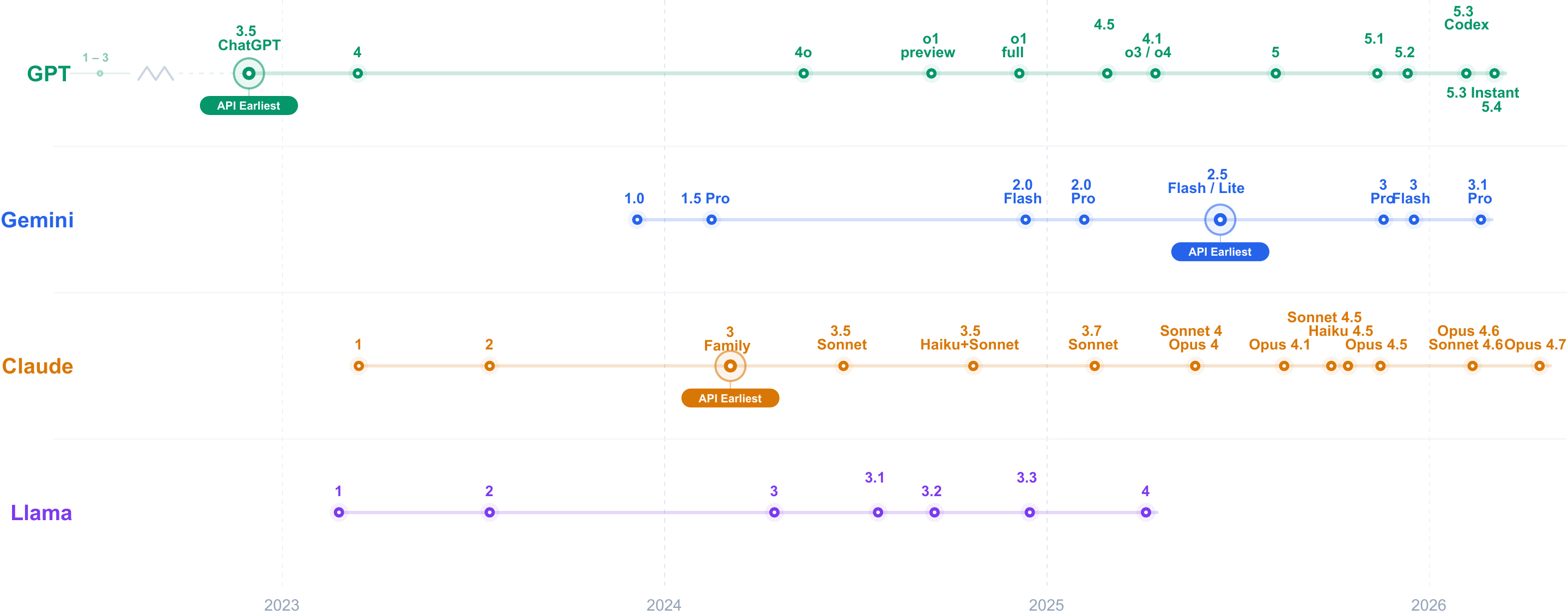}
\caption[Timeline of major LLM releases across model families.]{\textbf{Timeline of major LLM releases across model families.} This figure shows the release dates of GPT, Gemini, Claude, and Llama families and API availability of models from the GPT, Gemini, and Claude families.}
\label{fig:Figure_comment2.2_timeline}
\end{figure}

\newpage
\begin{figure}[htbp!]
\centering
\includegraphics[width=0.8\columnwidth]{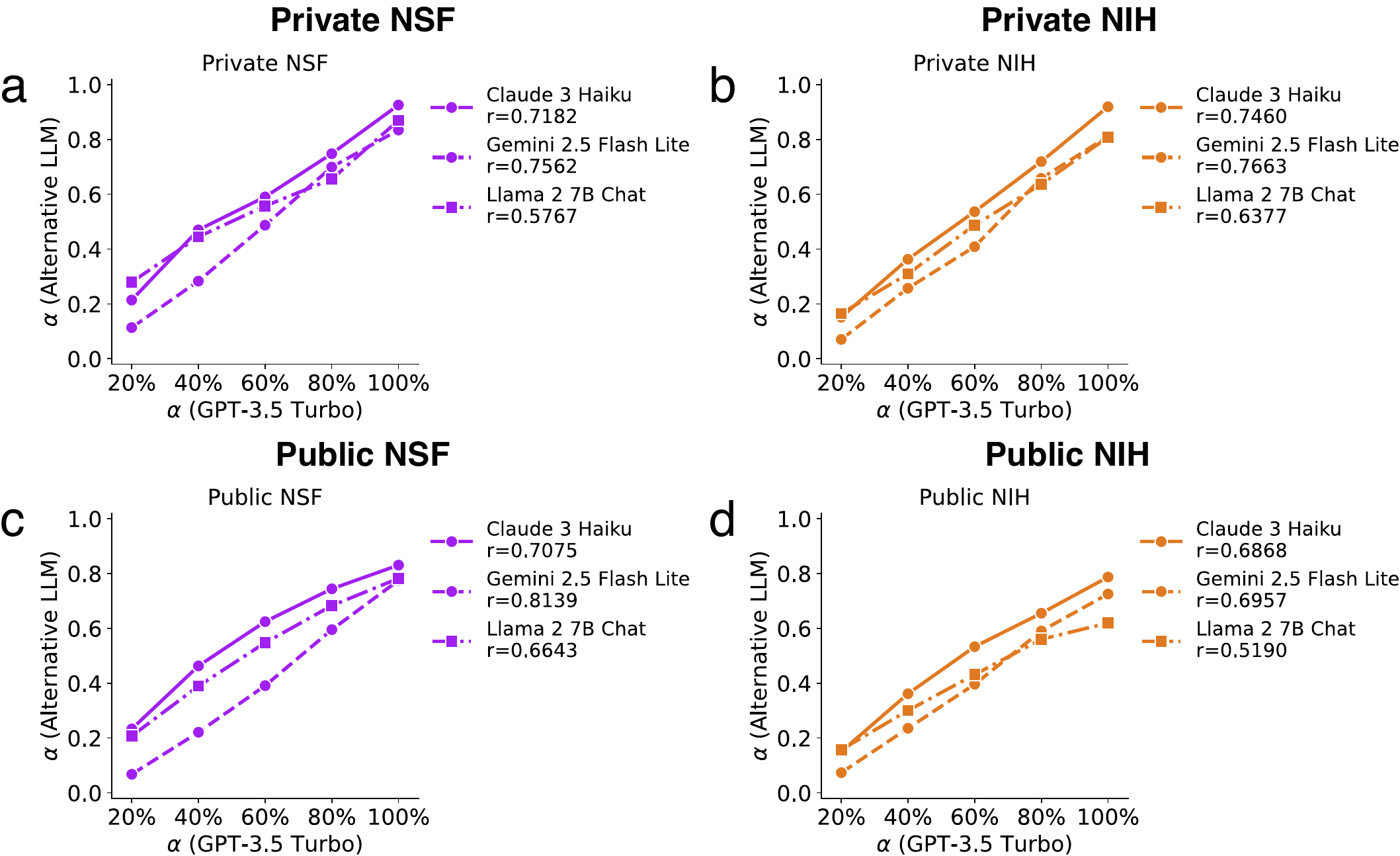}
\caption[Grant-level $\alpha$ estimates are strongly correlated across alternative reference LLMs.]{\textbf{Grant-level $\alpha$ estimates are strongly correlated across alternative reference LLMs.} Each panel plots the GPT-3.5-based $\alpha$ estimate against the $\alpha$ estimate derived from an alternative reference LLM (Claude 3 Haiku, Gemini 2.5 Flash Lite, and Llama 2 7B Chat) for (\textbf{a})~private NSF, (\textbf{b})~private NIH, (\textbf{c})~public NSF, and (\textbf{d})~public NIH grants.}
\label{fig:Figure_comment2.2_correlation}
\end{figure}

\newpage
\begin{figure}[htbp!]
\centering
\includegraphics[width=0.8\columnwidth]{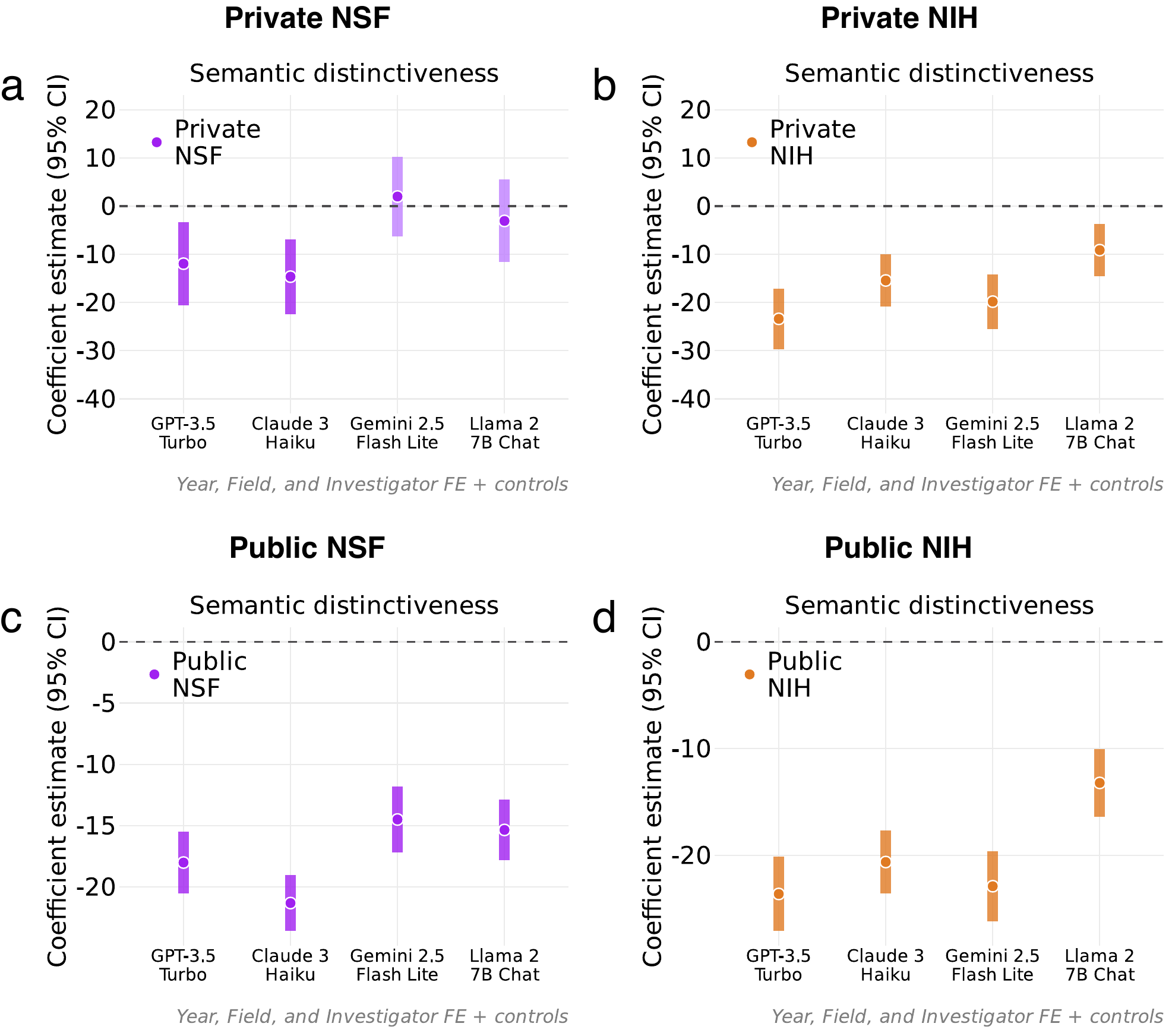}
\caption[LLM use and semantic distinctiveness using alternative reference LLMs.]{\textbf{LLM use and semantic distinctiveness using alternative reference LLMs.} Regression estimates relating grant-level LLM use ($\alpha$) to semantic distinctiveness for (\textbf{a})~private NSF, (\textbf{b})~private NIH, (\textbf{c})~public NSF, and (\textbf{d})~public NIH grants. For each dataset, estimates are shown separately for each alternative reference LLM (Claude 3 Haiku, Gemini 2.5 Flash Lite, and Llama 2 7B Chat). All regressions include grant start year, field, and investigator fixed effects, as well as controls for funding amount. Points indicate coefficient estimates, and bars denote 95\% confidence intervals.}
\label{fig:Figure_comment2.2_Fig2}
\end{figure}

\newpage
\begin{figure}[htbp!]
\centering
\includegraphics[width=0.8\columnwidth]{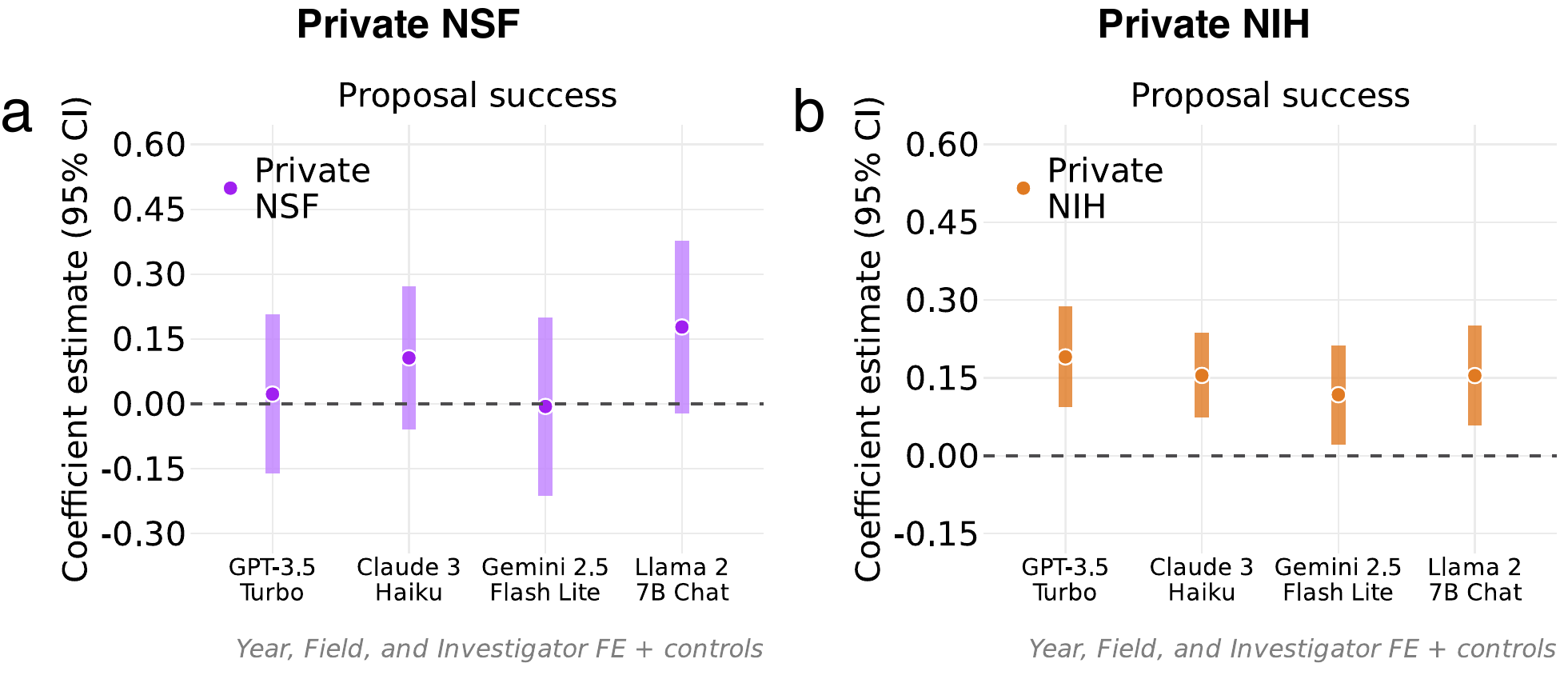}
\caption[LLM use and proposal success using alternative reference LLMs.]{\textbf{LLM use and proposal success using alternative reference LLMs.} Regression estimates relating grant-level LLM use ($\alpha$) to proposal success for (\textbf{a})~private NSF and (\textbf{b})~private NIH submissions. For each dataset, estimates are shown separately for each alternative reference LLM (Claude 3 Haiku, Gemini 2.5 Flash Lite, and Llama 2 7B Chat). All regressions include proposal request start year, field, and investigator fixed effects, as well as controls for requested funding amount. Points indicate coefficient estimates, and bars denote 95\% confidence intervals.}
\label{fig:Figure_comment2.2_Fig3}
\end{figure}

\newpage
\begin{figure}[htbp!]
\centering
\includegraphics[width=0.8\columnwidth]{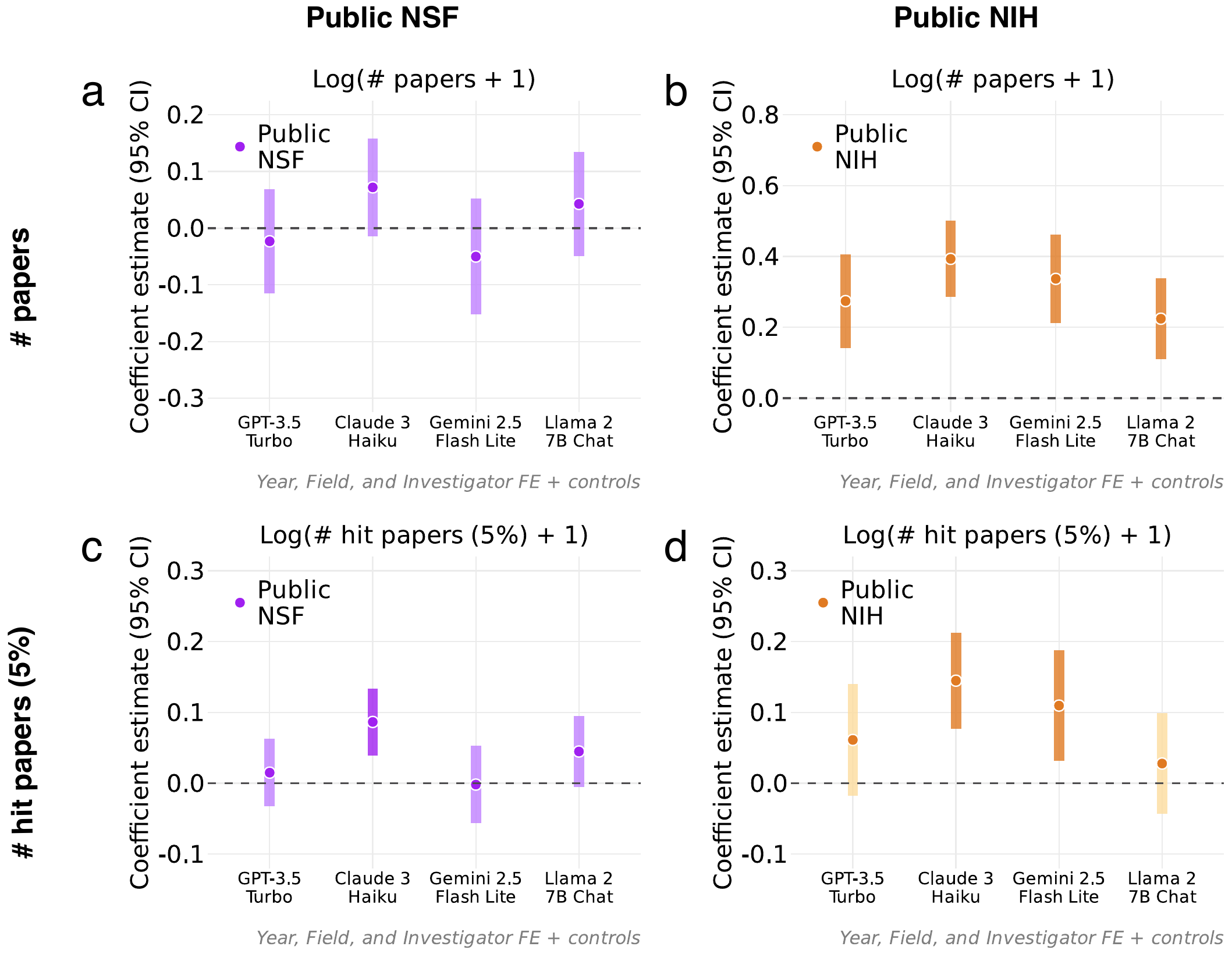}
\caption[LLM use and funding outputs using alternative reference LLMs.]{\textbf{LLM use and funding outputs using alternative reference LLMs.} Regression estimates relating grant-level LLM use ($\alpha$) to the total number of resulting publications for (\textbf{a})~public NSF and (\textbf{b})~public NIH grants, and to the number of hit papers (top 5\%) for (\textbf{c})~public NSF and (\textbf{d})~public NIH grants. For each dataset, estimates are shown separately for each alternative reference LLM (Claude 3 Haiku, Gemini 2.5 Flash Lite, and Llama 2 7B Chat). All regressions include grant start year, field, and investigator fixed effects, as well as controls for funding amount. Points indicate coefficient estimates, and bars denote 95\% confidence intervals.}
\label{fig:Figure_comment2.2_Fig4}
\end{figure}

\newpage
\begin{figure}[htbp!]
\centering
\includegraphics[width=0.7\columnwidth]{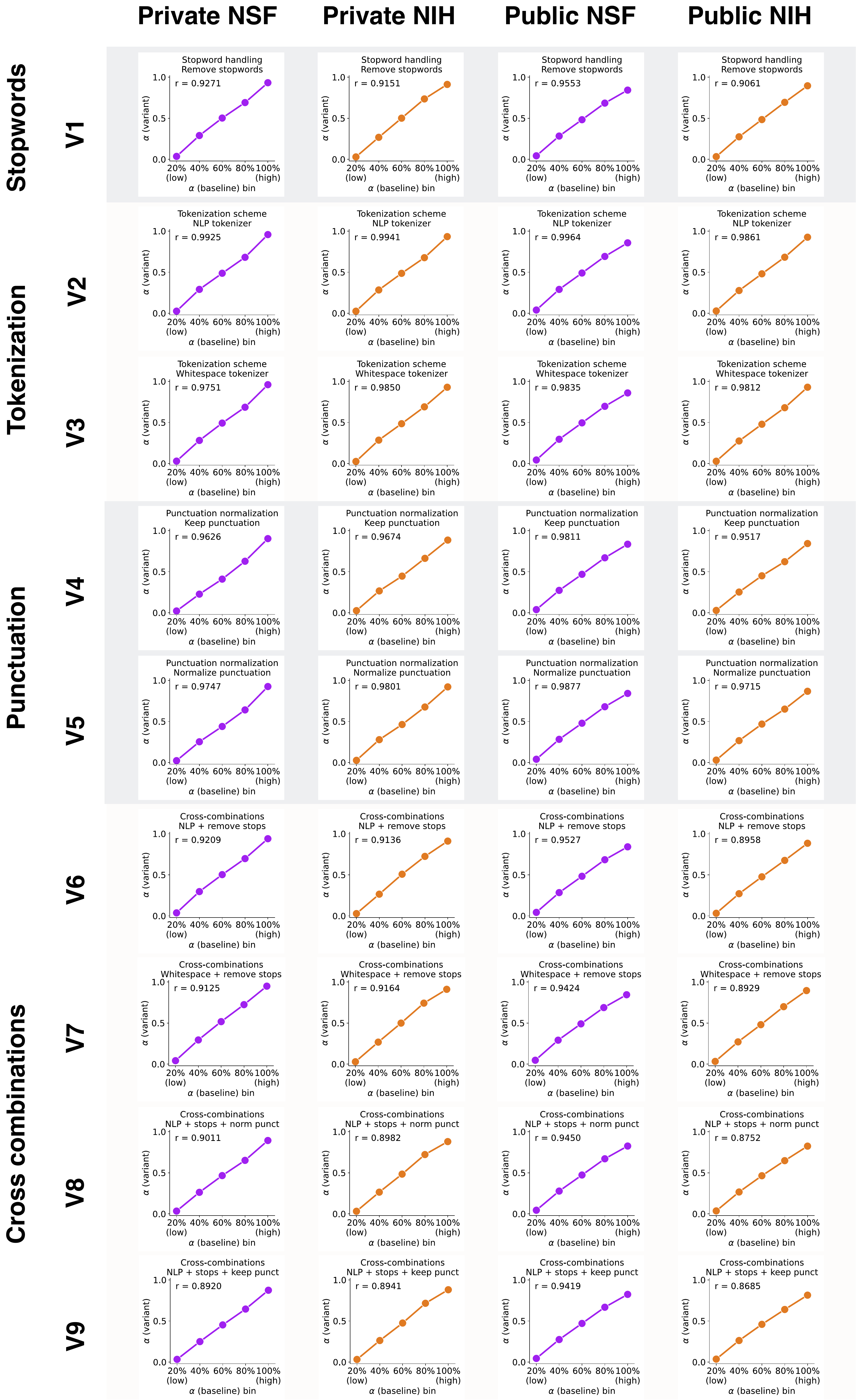}
\caption[Sensitivity of LLM-use estimates to alternative preprocessing pipelines.]{\textbf{Sensitivity of LLM-use estimates to alternative preprocessing pipelines.} Each panel displays binned comparisons of grant-level $\alpha$ estimates between the current preprocessing pipeline ($x$-axis) and an alternative variant ($y$-axis), shown separately for private NSF, private NIH, public NSF, and public NIH datasets (columns). Rows correspond to nine alternative preprocessing pipelines spanning three dimensions: stopword handling (V1), tokenization scheme (V2-V3), punctuation treatment (V4-V5), and cross-dimension combinations (V6-V9).}
\label{fig:Figure_comment2.3_correlation}
\end{figure}

\newpage
\begin{figure}[htbp!]
\centering
\includegraphics[width=0.8\columnwidth]{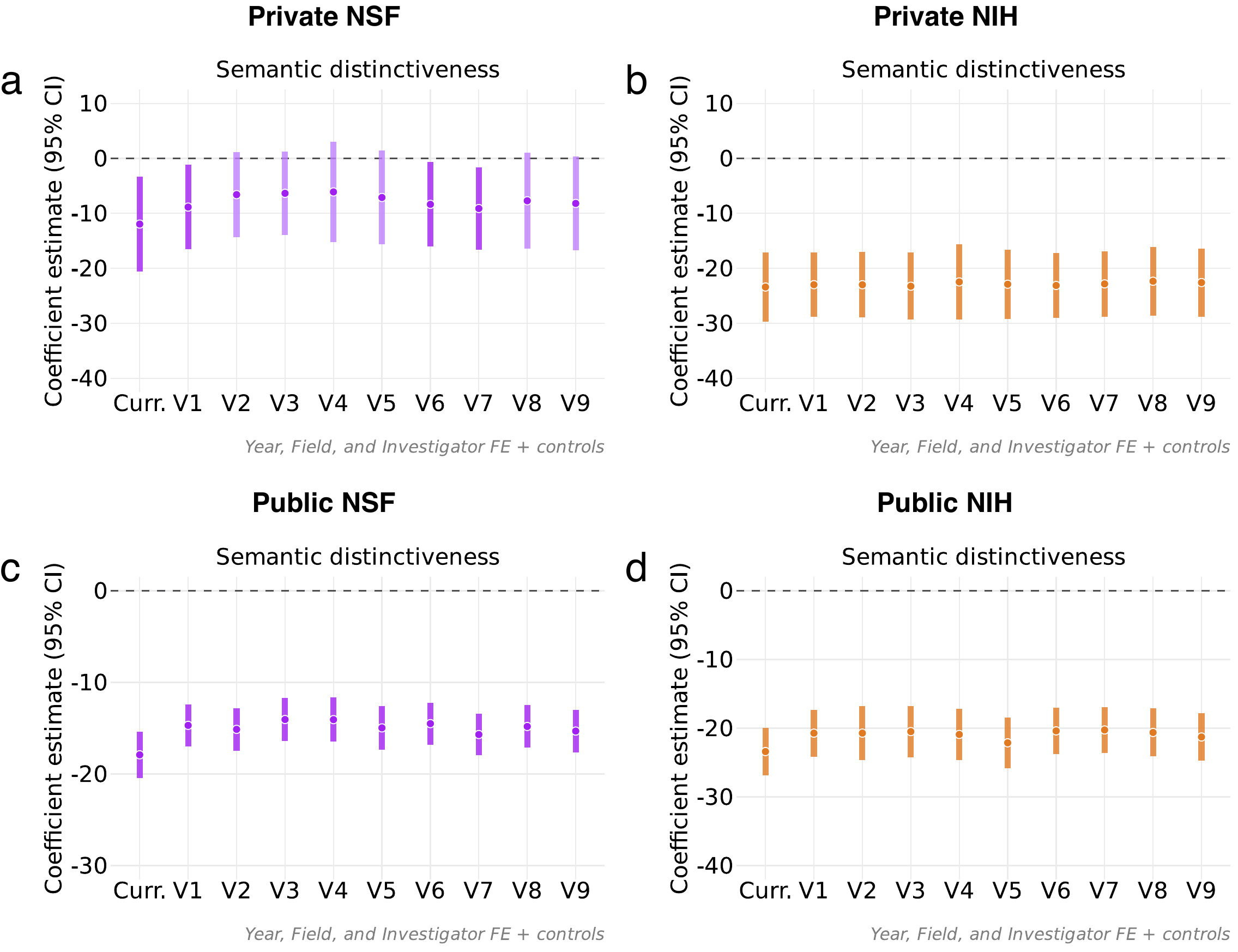}
\caption[LLM use and semantic distinctiveness across alternative preprocessing pipelines.]{\textbf{LLM use and semantic distinctiveness across alternative preprocessing pipelines.} Regression estimates relating grant-level LLM use ($\alpha$) to semantic distinctiveness under the current preprocessing pipeline (Curr.) and nine alternative variants (V1-V9). Results are shown for (\textbf{a})~private NSF proposals, (\textbf{b})~private NIH proposals, (\textbf{c})~public NSF awards, and (\textbf{d})~public NIH awards. All regressions include grant start year, field, and investigator fixed effects, as well as controls for funding amount. Points indicate coefficient estimates, and bars denote 95\% confidence intervals.}
\label{fig:Figure_comment2.3_Fig2}
\end{figure}

\newpage
\begin{figure}[htbp!]
\centering
\includegraphics[width=0.8\columnwidth]{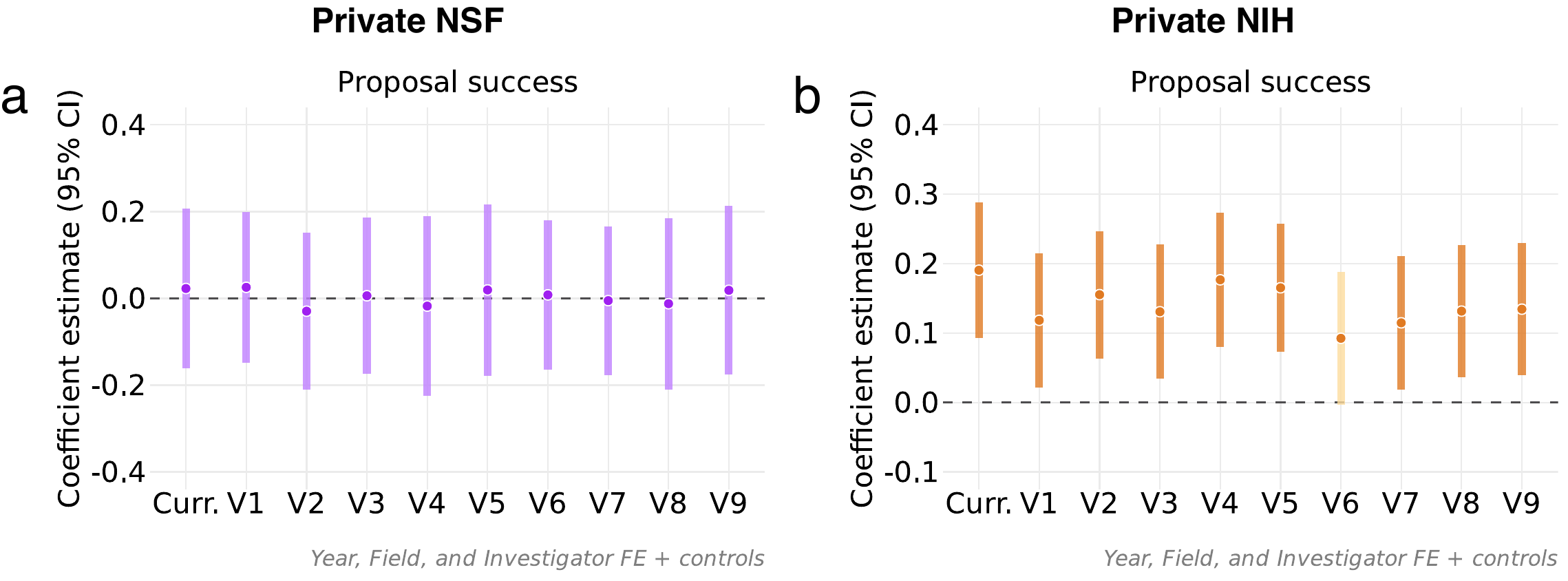}
\caption[LLM use and proposal success across alternative preprocessing pipelines.]{\textbf{LLM use and proposal success across alternative preprocessing pipelines.} Regression estimates relating grant-level LLM use ($\alpha$) to proposal funding success under the current preprocessing pipeline (Curr.) and nine alternative variants (V1-V9). Results are shown for (\textbf{a})~private NSF proposals and (\textbf{b})~private NIH proposals. All regressions include grant start year, field, and investigator fixed effects, as well as controls for funding amount. Points indicate coefficient estimates, and bars denote 95\% confidence intervals.}
\label{fig:Figure_comment2.3_Fig3}
\end{figure}

\newpage
\begin{figure}[htbp!]
\centering
\includegraphics[width=0.8\columnwidth]{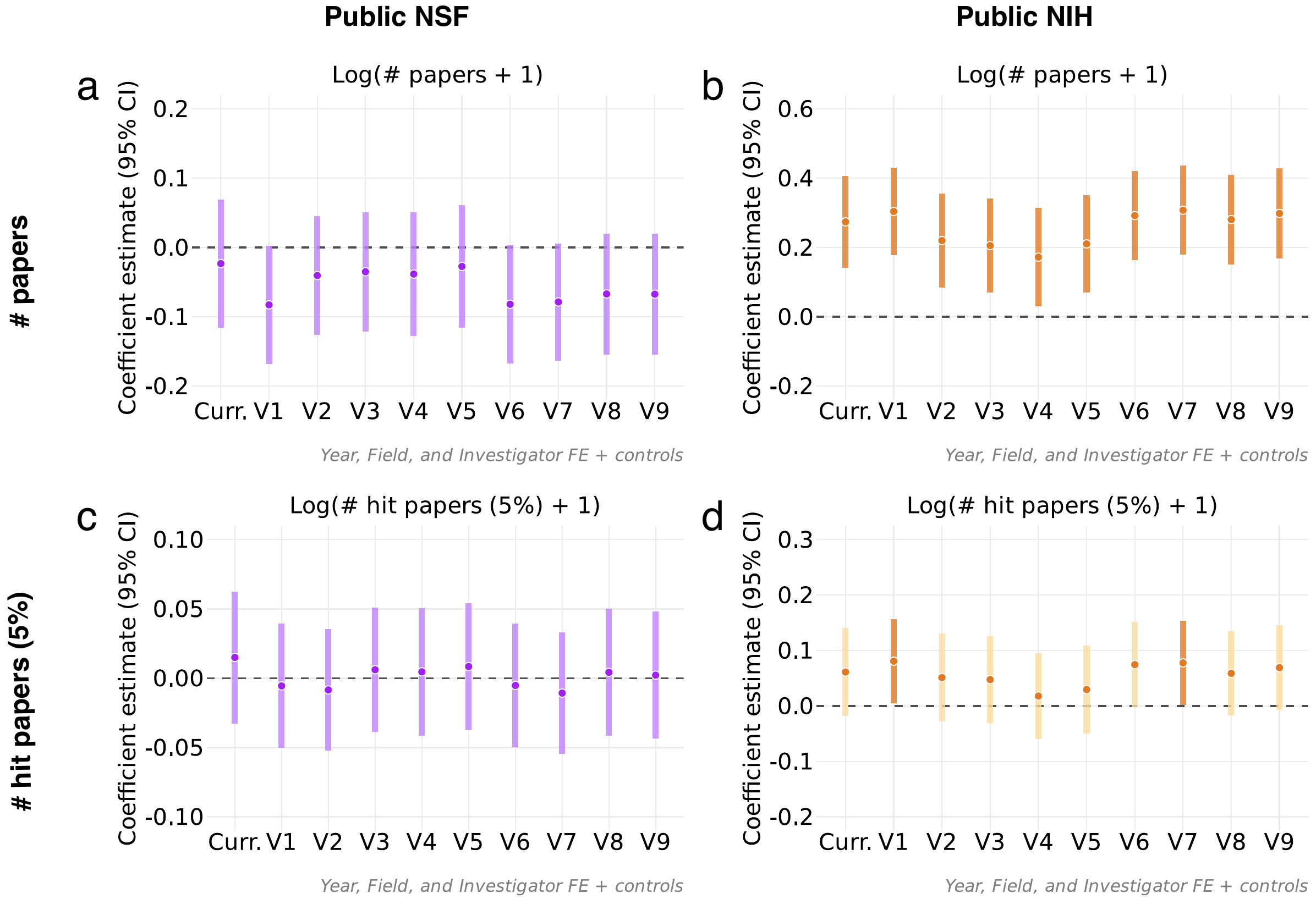}
\caption[LLM use and downstream publication output across alternative preprocessing pipelines.]{\textbf{LLM use and downstream publication output across alternative preprocessing pipelines.} Regression estimates relating grant-level LLM use ($\alpha$) to publication outcomes under the current preprocessing pipeline (Curr.) and nine alternative variants (V1-V9). (\textbf{a}-\textbf{b})~Total publication output for public NSF awards~(\textbf{a}) and public NIH awards~(\textbf{b}). (\textbf{c}-\textbf{d})~High-impact publication output for public NSF awards~(\textbf{c}) and public NIH awards~(\textbf{d}). All regressions include grant start year, field, and investigator fixed effects, as well as controls for funding amount. Points indicate coefficient estimates, and bars denote 95\% confidence intervals.}
\label{fig:Figure_comment2.3_Fig4}
\end{figure}

\newpage
\begin{figure}[htbp!]
\centering
\includegraphics[width=0.8\columnwidth]{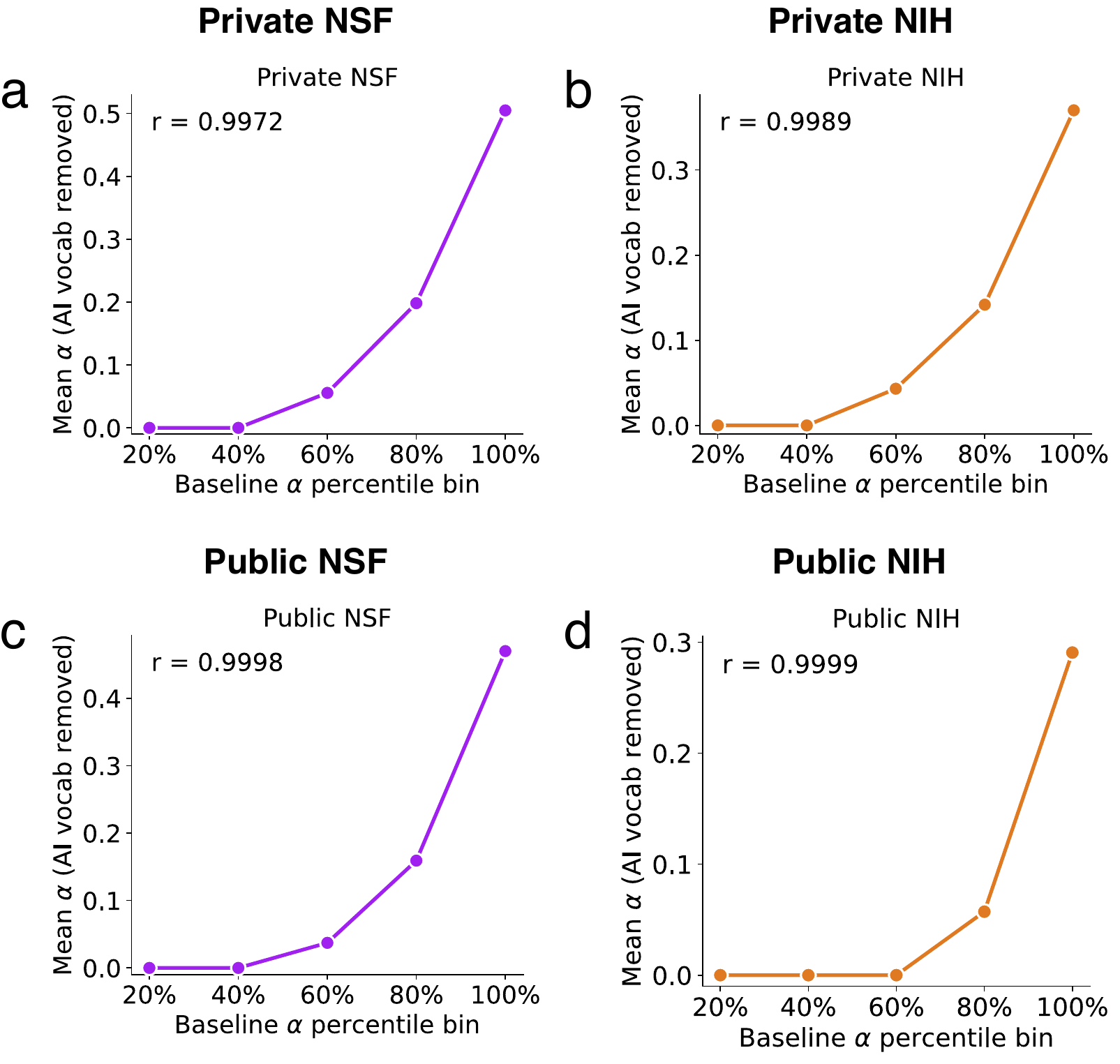}
\caption[LLM-use estimates after removing AI-related scientific vocabulary.]{\textbf{LLM-use estimates after removing AI-related scientific vocabulary.}  This figure assesses whether our measure of individual-level LLM use ($\alpha$) is driven by AI-related scientific terminology. Panels report the relationship between the original $\alpha$ and a recalculated $\alpha$ after removing AI-related terms for private NSF (\textbf{a}), private NIH (\textbf{b}), public NSF (\textbf{c}), and public NIH (\textbf{d}) samples.}
\label{fig:Figure_comment2.6_ngram_correlation}
\end{figure}

\newpage
\begin{figure}[htbp!]
\centering
\includegraphics[width=0.8\columnwidth]{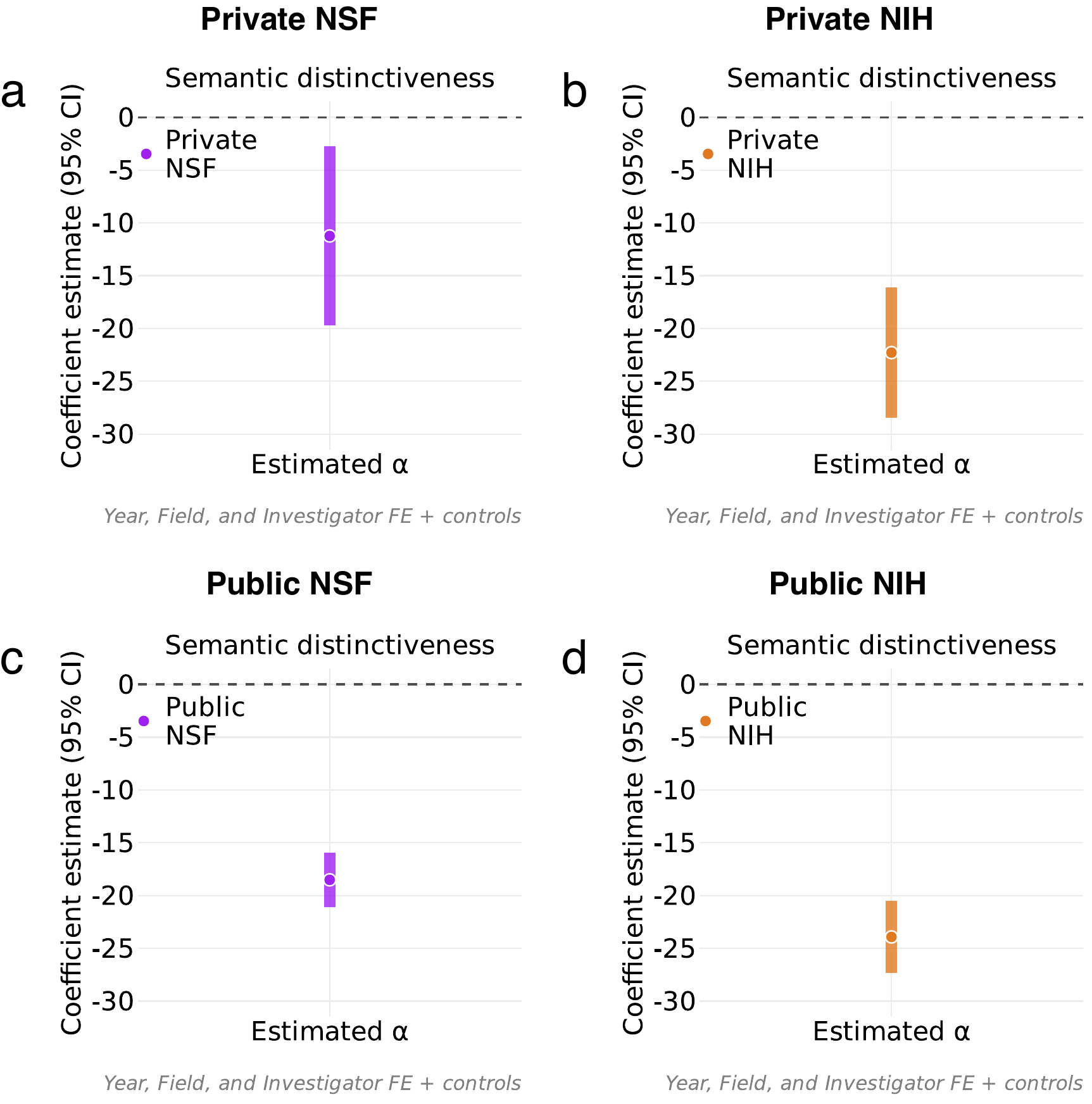}
\caption[LLM use and semantic distinctiveness after removing AI-related scientific vocabulary.]{\textbf{LLM use and semantic distinctiveness after removing AI-related scientific vocabulary.}  (\textbf{a}-\textbf{d}) Regression estimates relating grant-level LLM use ($\alpha$), recalculated after excluding AI-related terms, to semantic distance from abstracts funded in the prior year within the same agency, expressed as within-year percentiles. Panels show results separately for private NSF (\textbf{a}), private NIH (\textbf{b}), public NSF (\textbf{c}), and public NIH (\textbf{d}) grants. All regressions include grant start year, field, and investigator fixed effects, as well as controls for funding amount. Points indicate coefficient estimates, and bars denote 95\% confidence intervals.}
\label{fig:Figure_comment2.6_ngram_Fig2}
\end{figure}

\newpage
\begin{figure}[htbp!]
\centering
\includegraphics[width=0.8\columnwidth]{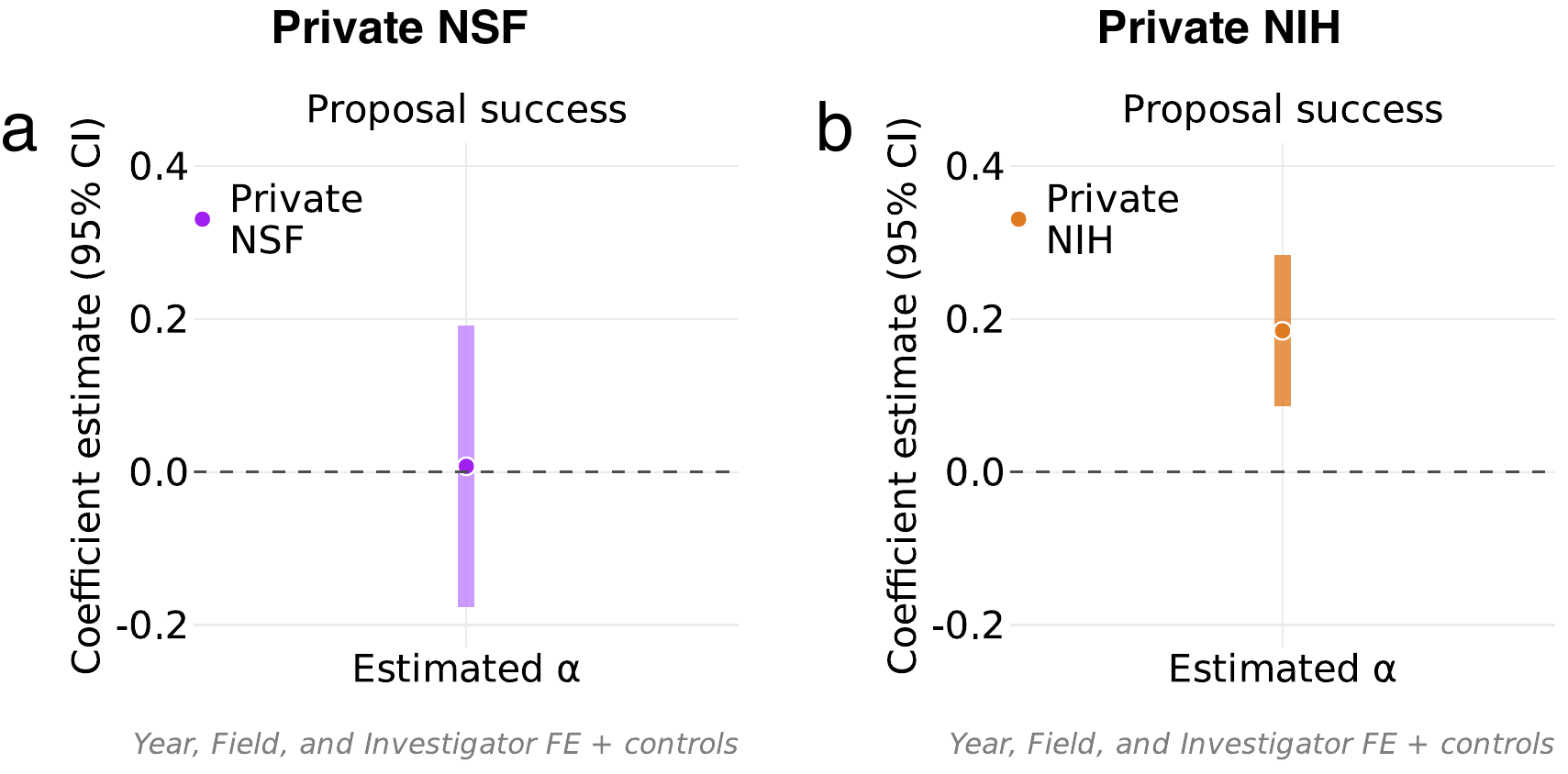}
\caption[LLM use and proposal success after removing AI-related scientific vocabulary.]{\textbf{LLM use and proposal success after removing AI-related scientific vocabulary.}  Based on private NSF and NIH proposal submissions, this figure examines the relationship between LLM use at submission ($\alpha$), recalculated after excluding AI-related terms, and proposal success. (\textbf{a}) Regression estimates for NSF submissions. (\textbf{b}) Corresponding estimates for NIH submissions. All regressions include proposal request start year, field, and investigator fixed effects, as well as controls for requested funding amount. Points indicate coefficient estimates, and bars denote 95\% confidence intervals.}
\label{fig:Figure_comment2.6_ngram_Fig3}
\end{figure}

\newpage
\begin{figure}[htbp!]
\centering
\includegraphics[width=0.8\columnwidth]{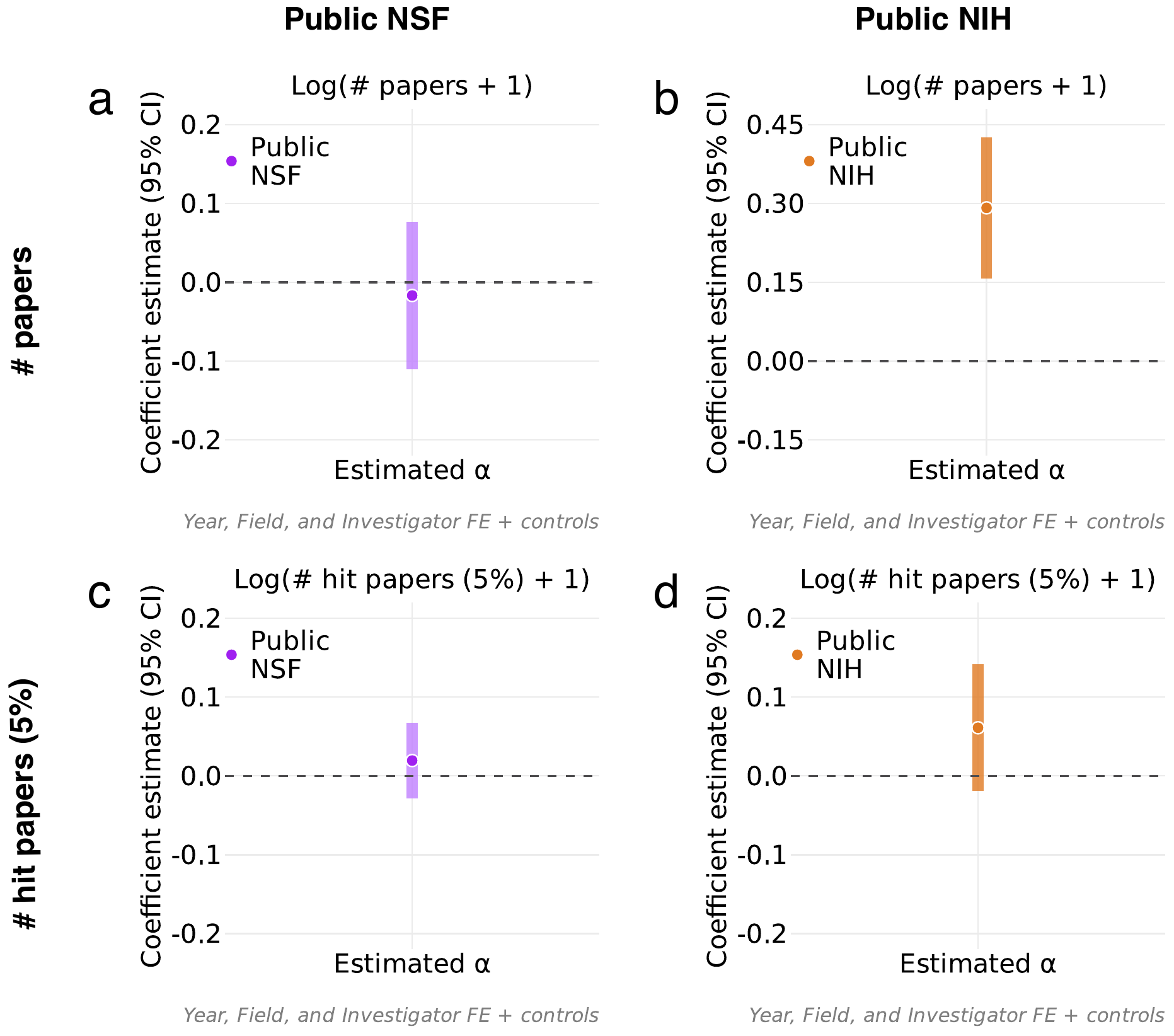}
\caption[LLM use and research funding outputs after removing AI-related scientific vocabulary.]{\textbf{LLM use and research funding outputs after removing AI-related scientific vocabulary.}  (\textbf{a}-\textbf{b}) Regression estimates relating grant-level LLM use ($\alpha$), recalculated after excluding AI-related terms, to the total number of resulting publications for public NSF (\textbf{a}) and public NIH (\textbf{b}) grants. (\textbf{c}-\textbf{d}) Corresponding estimates for high-impact outputs, where a ``hit'' paper is defined as one whose citations fall within the top 5\% of all papers published worldwide in the same year and field. All regressions include grant start year, field, and investigator fixed effects, as well as controls for funding amount. Points indicate coefficient estimates, and bars denote 95\% confidence intervals.}
\label{fig:Figure_comment2.6_ngram_Fig4}
\end{figure}

\newpage
\begin{figure}[htbp!]
\centering
\includegraphics[width=1.0\columnwidth]{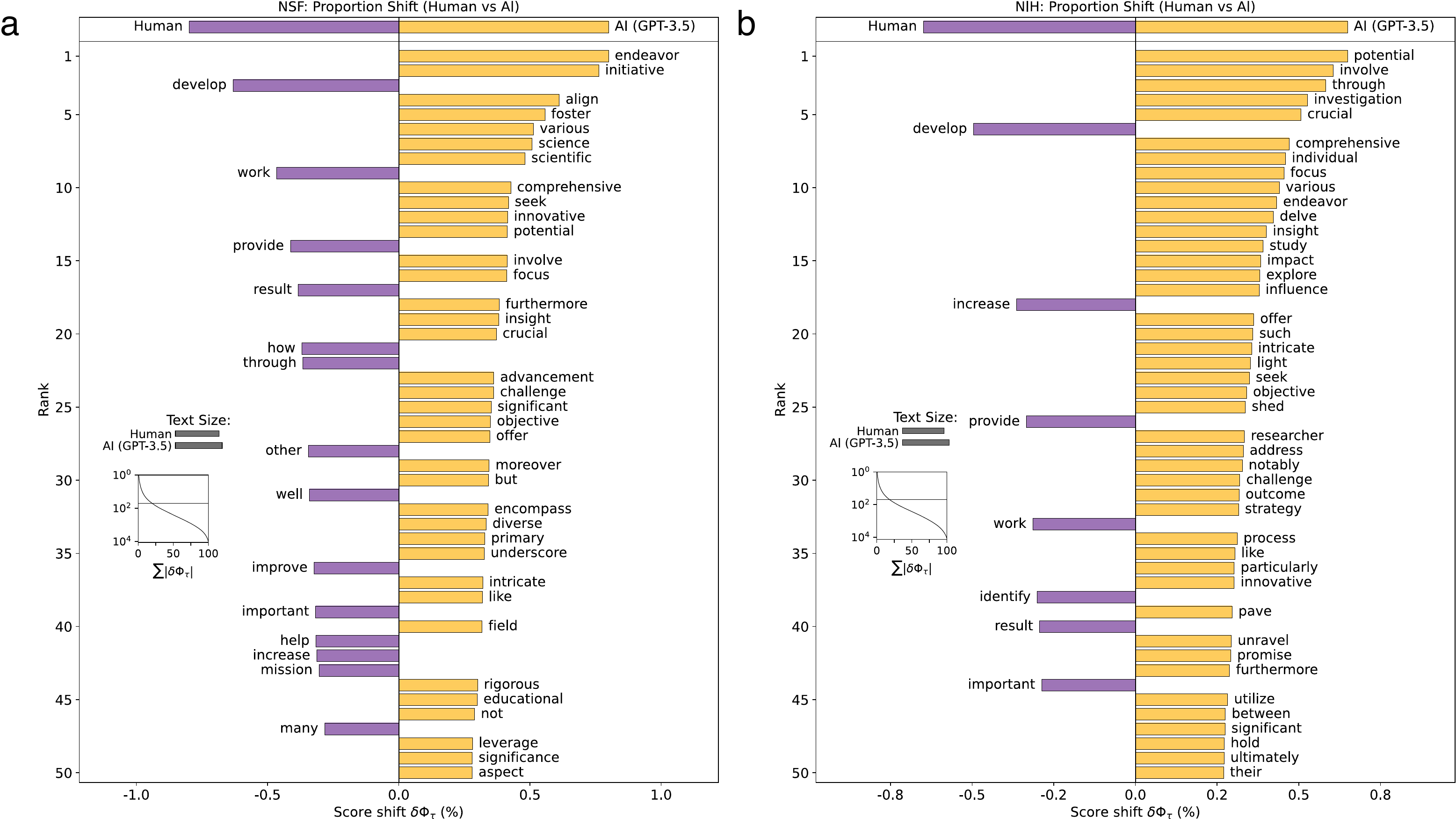}
\caption[Decomposition of the LLM detection signal based on proportion differences.]{\textbf{Decomposition of the LLM detection signal based on proportion differences.} Word-shift graphs rank individual words by their directional contribution to the overall frequency shift between human-written and LLM-modified grant abstracts. Bars extending to the right indicate words that are more frequent in LLM-modified text; bars extending to the left indicate words that are more frequent in human-written text. Results are shown separately for NSF (\textbf{a}) and NIH (\textbf{b}).}
\label{fig:Figure_comment2.9_Proportion}
\end{figure}

\newpage
\begin{figure}[htbp!]
\centering
\includegraphics[width=1.0\columnwidth]{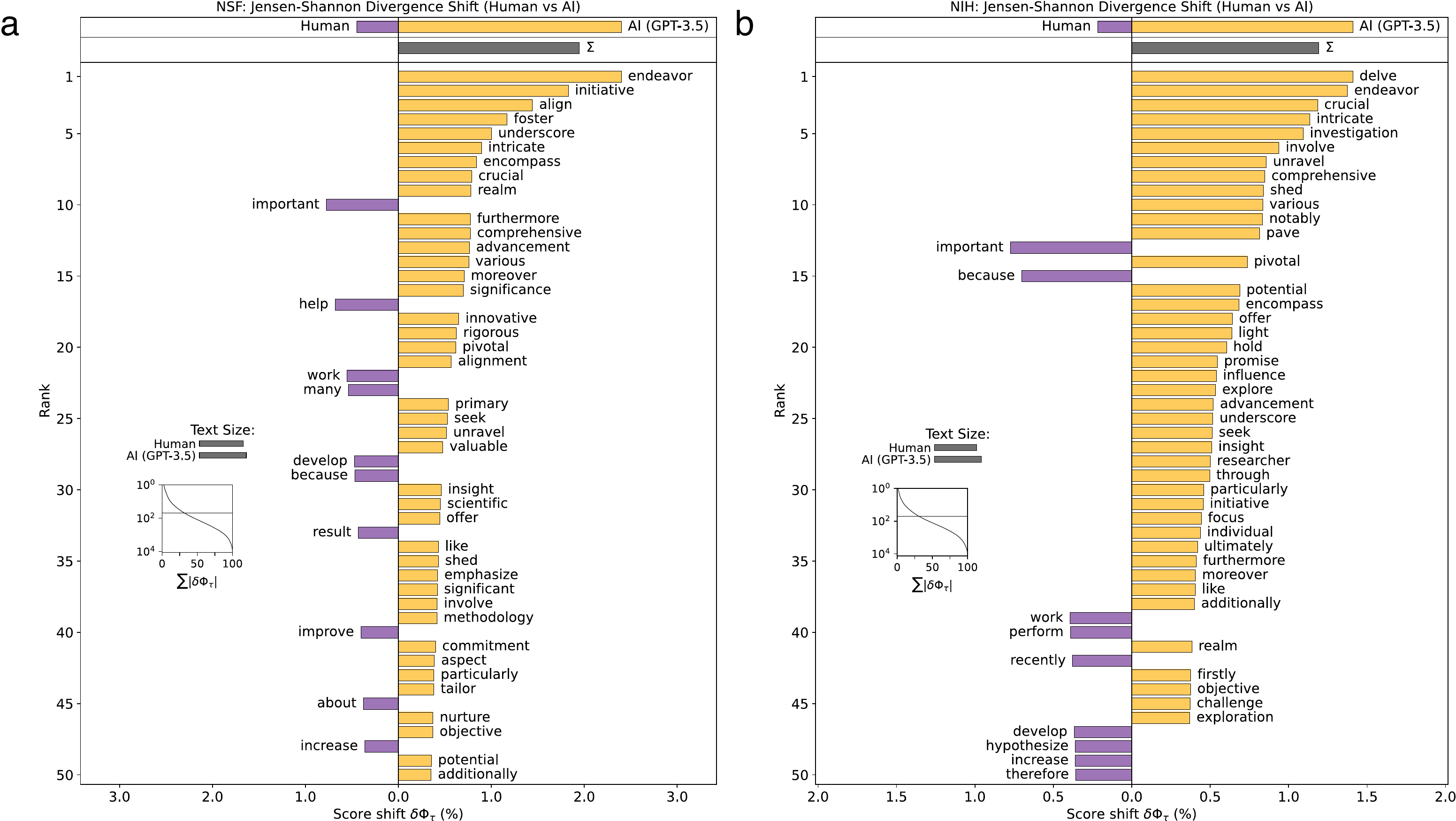}
\caption[Decomposition of the LLM detection signal based on Jensen-Shannon divergence.]{\textbf{Decomposition of the LLM detection signal based on Jensen-Shannon divergence.} Word-shift graphs rank individual words by their contribution to the symmetric Jensen-Shannon divergence between human-written and LLM-modified grant abstracts. Bars extending to the right indicate words whose usage is disproportionately associated with LLM-modified text; bars extending to the left indicate words disproportionately associated with human-written text. Results are shown separately for NSF (\textbf{a}) and NIH (\textbf{b}).}
\label{fig:Figure_comment2.9_JSD}
\end{figure}

\newpage
\begin{figure}[htbp!]
\centering
\includegraphics[width=0.8\columnwidth]{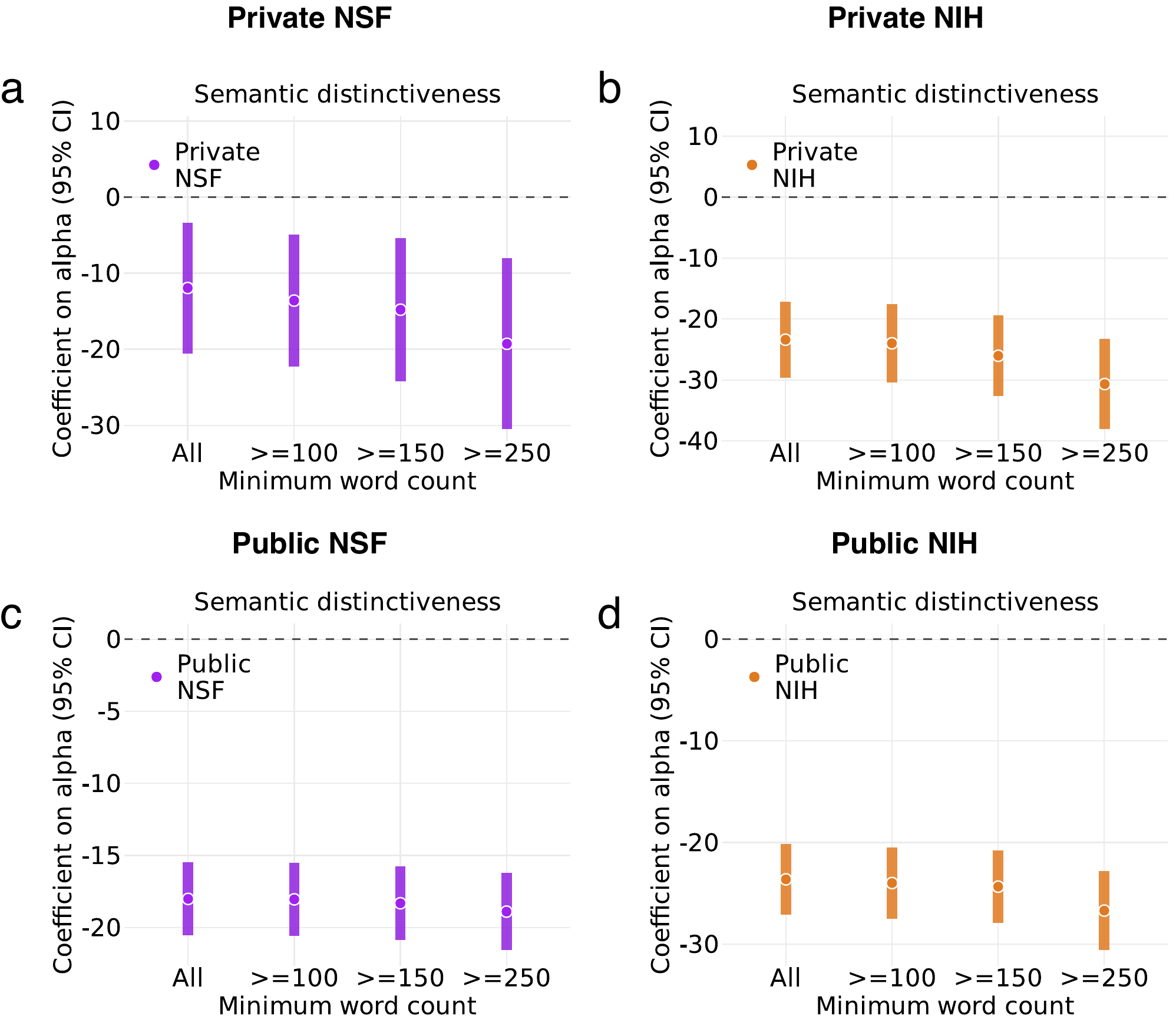}
\caption[LLM use and semantic distinctiveness by abstract length threshold.]{\textbf{LLM use and semantic distinctiveness by abstract length threshold.} Regression estimates relating grant-level LLM use ($\alpha$) to semantic distance
from abstracts funded in the prior year within the same agency, expressed as within-year percentiles,
restricting the sample to abstracts with at least 100, 150, and 250 words.
(\textbf{a}-\textbf{b}) Results for private NSF (\textbf{a}) and private NIH (\textbf{b}) grants.
(\textbf{c}-\textbf{d}) Corresponding estimates for public NSF (\textbf{c}) and public NIH (\textbf{d}) grants.
All regressions include grant start year, field, and investigator fixed effects, as well as controls
for funding amount. Points indicate coefficient estimates, and bars denote 95\% confidence intervals.}
\label{fig:Figure_comment2.4_Fig2}
\end{figure}

\newpage
\begin{figure}[htbp!]
\centering
\includegraphics[width=0.8\columnwidth]{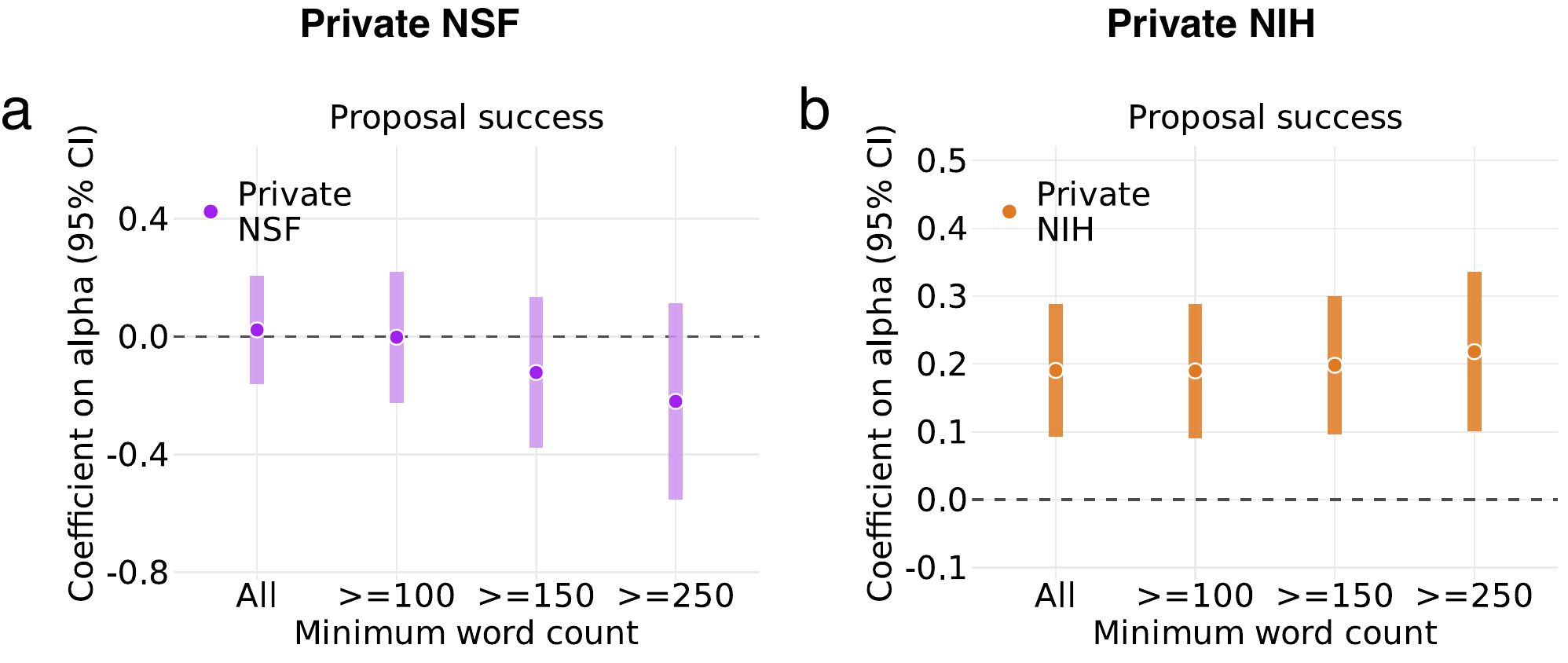}
\caption[LLM use and proposal success by abstract length threshold.]{\textbf{LLM use and proposal success by abstract length threshold.} Regression estimates relating grant-level LLM use ($\alpha$) at submission to proposal
success, restricting the sample to abstracts with at least 100, 150, and 250 words.
(\textbf{a}) Results for NSF submissions.
(\textbf{b}) Corresponding estimates for NIH submissions.
All regressions include proposal request start year, field, and investigator fixed effects, as well
as controls for requested funding amount. Points indicate coefficient estimates, and bars denote
95\% confidence intervals.}
\label{fig:Figure_comment2.4_Fig3}
\end{figure}

\newpage
\begin{figure}[htbp!]
\centering
\includegraphics[width=0.8\columnwidth]{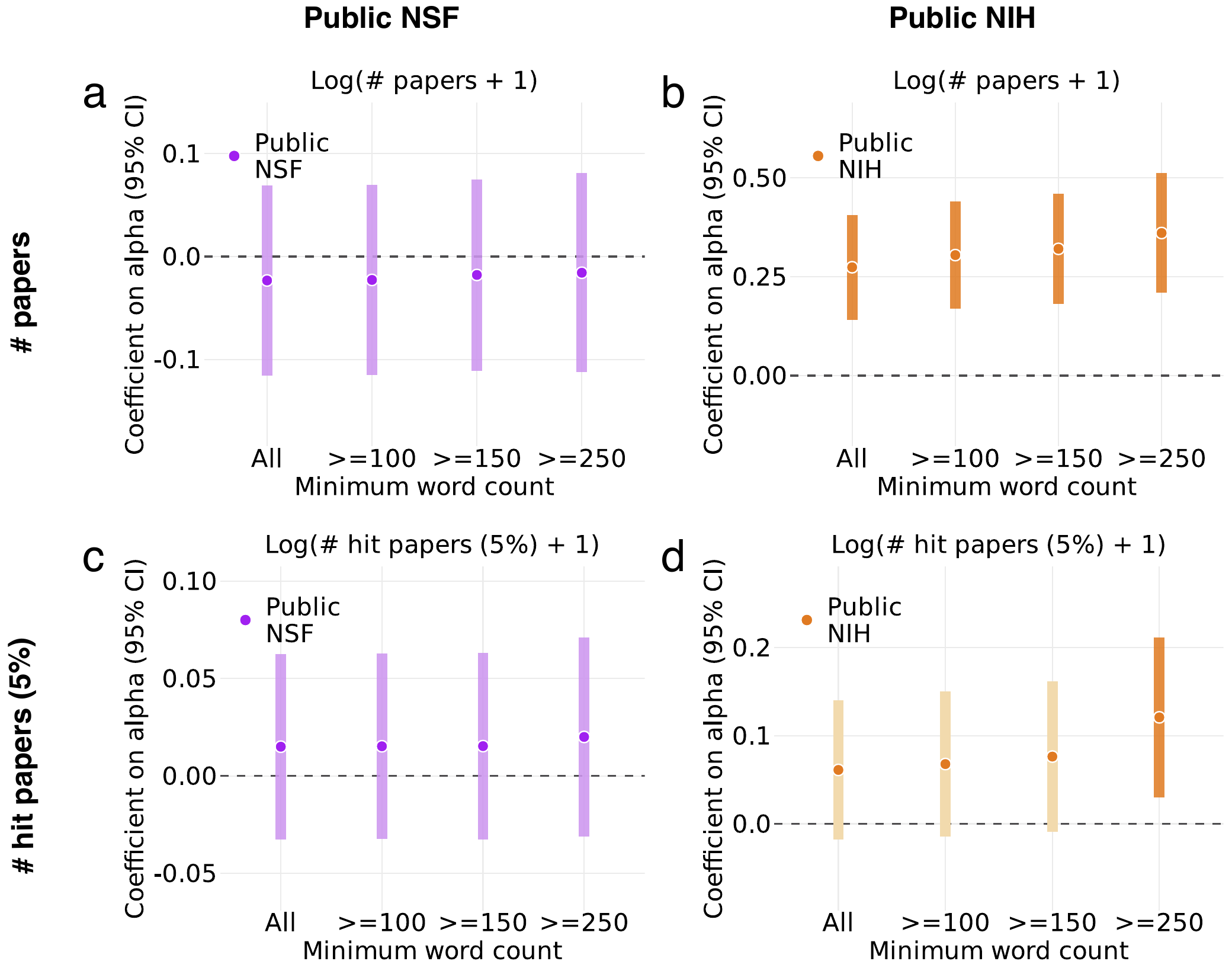}
\caption[LLM use and federal research funding outputs by abstract length threshold.]{\textbf{LLM use and federal research funding outputs by abstract length threshold.} Regression estimates relating grant-level LLM use ($\alpha$) to publication outcomes,
restricting the sample to abstracts with at least 100, 150, and 250 words.
(\textbf{a}-\textbf{b}) Estimates for the total number of resulting publications for NSF (\textbf{a}) and NIH (\textbf{b}) grants.
(\textbf{c}-\textbf{d}) Corresponding estimates for high-impact outputs, where a ``hit'' paper is defined
as one whose citations fall within the top 5\% of all papers published worldwide in the same year
and field, for NSF (\textbf{c}) and NIH (\textbf{d}) grants.
All regressions include grant start year, field, and investigator fixed effects, as well as controls
for funding amount. Points indicate coefficient estimates, and bars denote 95\% confidence intervals.}
\label{fig:Figure_comment2.4_Fig4}
\end{figure}

\newpage
\begin{figure}[htbp!]
\centering
\includegraphics[width=0.8\columnwidth]{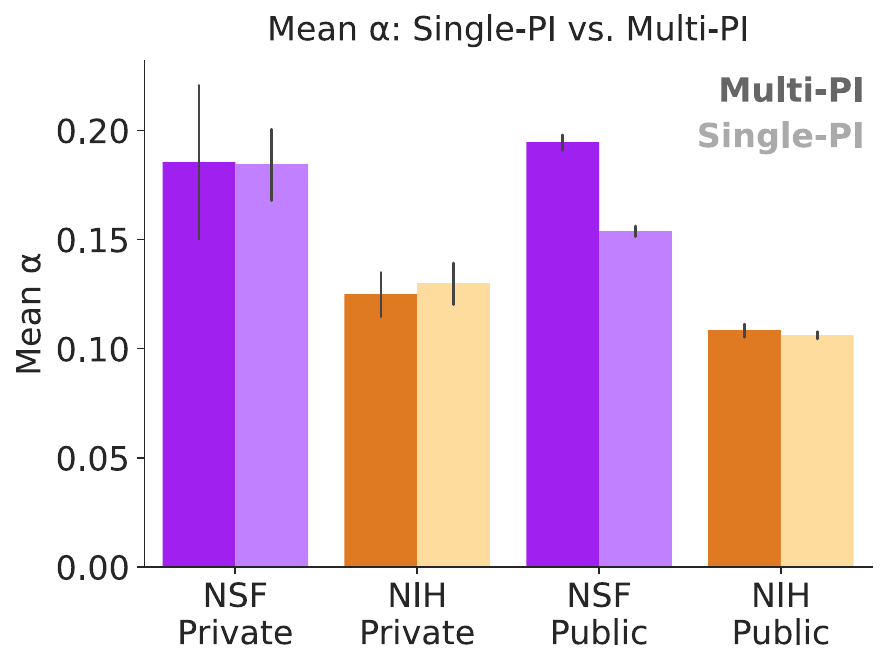}
\caption[Comparison of LLM use between single-investigator and multi-investigator proposals and awards.]{\textbf{Comparison of LLM use between single-investigator and multi-investigator proposals and awards.} Estimated LLM use ($\alpha$) for single-investigator and multi-investigator grants across four datasets.}
\label{fig:Figure_comment2.5_raw_plot}
\end{figure}

\newpage
\begin{figure}[htbp!]
\centering
\includegraphics[width=0.8\columnwidth]{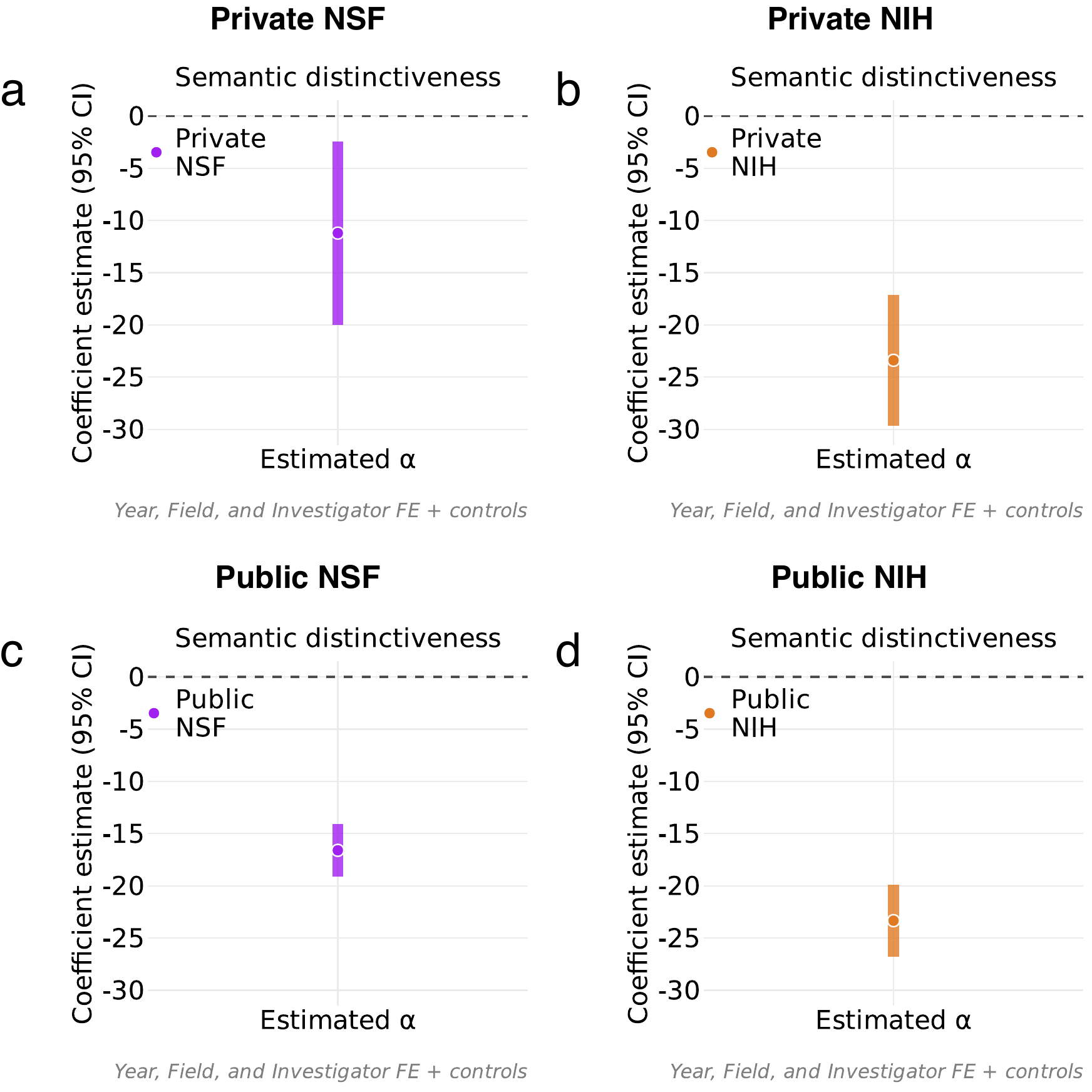}
\caption[LLM use and semantic distinctiveness, controlling for single-investigator versus multi-investigator status.]{\textbf{LLM use and semantic distinctiveness, controlling for single-investigator versus multi-investigator status.} Regression estimates relating grant-level LLM use ($\alpha$) to semantic distance from abstracts funded in the prior year within the same agency, expressed as within-year percentiles. Panels show results separately for private NSF (\textbf{a}), private NIH (\textbf{b}), public NSF (\textbf{c}), and public NIH (\textbf{d}) grants. All regressions include grant start year, field, and investigator fixed effects, as well as controls for funding amount and an indicator for single-investigator versus multi-investigator status. Points indicate coefficient estimates, and bars denote 95\% confidence intervals.}
\label{fig:Figure_comment2.5_Fig2}
\end{figure}

\newpage
\begin{figure}[htbp!]
\centering
\includegraphics[width=0.8\columnwidth]{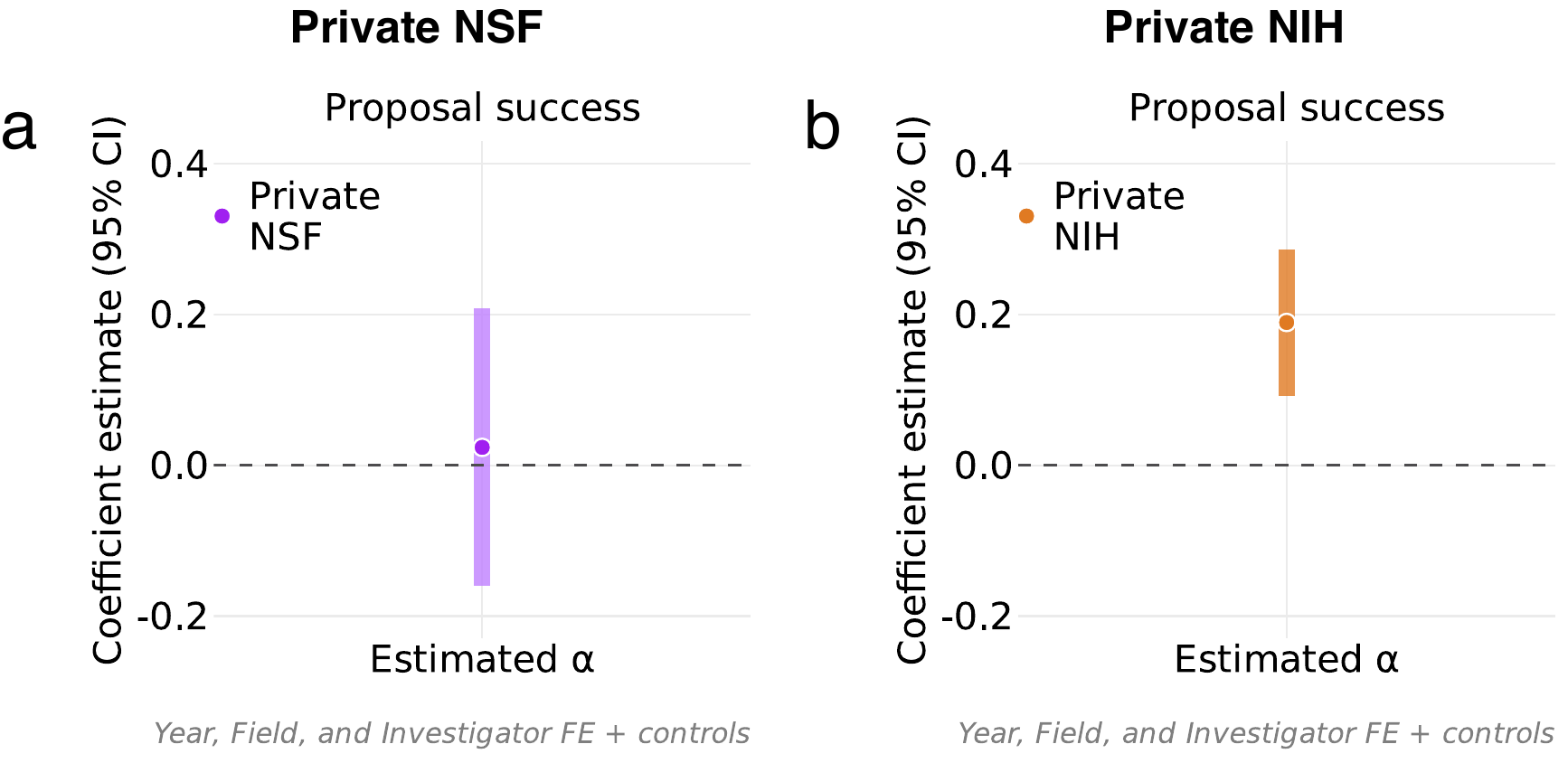}
\caption[LLM use and proposal success, controlling for single-investigator versus multi-investigator status.]{\textbf{LLM use and proposal success, controlling for single-investigator versus multi-investigator status.} Regression estimates relating LLM use at submission ($\alpha$) to proposal success for private NSF (\textbf{a}) and private NIH (\textbf{b}) submissions. All regressions include proposal request start year, field, and investigator fixed effects, as well as controls for requested funding amount and an indicator for single-investigator versus multi-investigator status. Points indicate coefficient estimates, and bars denote 95\% confidence intervals.}
\label{fig:Figure_comment2.5_Fig3}
\end{figure}

\newpage
\begin{figure}[htbp!]
\centering
\includegraphics[width=0.8\columnwidth]{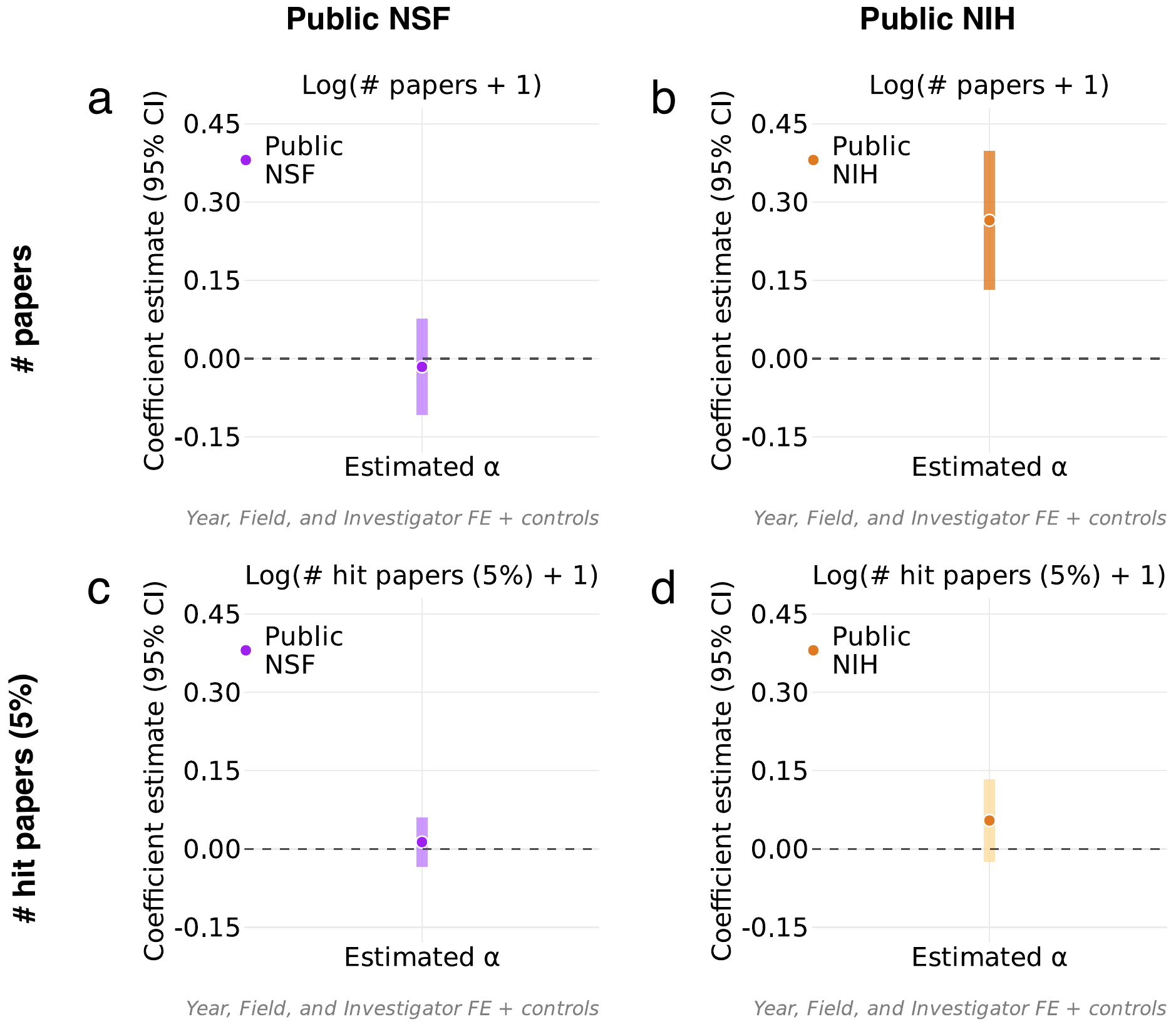}
\caption[LLM use and federal research funding outputs, controlling for single-investigator versus multi-investigator status.]{\textbf{LLM use and federal research funding outputs, controlling for single-investigator versus multi-investigator status.}  (\textbf{a}-\textbf{b}) Regression estimates relating grant-level LLM use ($\alpha$) to the total number of resulting publications for public NSF (\textbf{a}) and public NIH (\textbf{b}) grants. (\textbf{c}-\textbf{d}) Corresponding estimates for high-impact outputs, where a ``hit'' paper is defined as one whose citations fall within the top 5\% of all papers published worldwide in the same year and field, for public NSF (\textbf{c}) and public NIH (\textbf{d}) grants. All regressions include grant start year, field, and investigator fixed effects, as well as controls for funding amount and an indicator for single-investigator versus multi-investigator status. Points indicate coefficient estimates, and bars denote 95\% confidence intervals.}
\label{fig:Figure_comment2.5_Fig4}
\end{figure}

\newpage
\begin{figure}[htbp!]
\centering
\includegraphics[width=0.8\columnwidth]{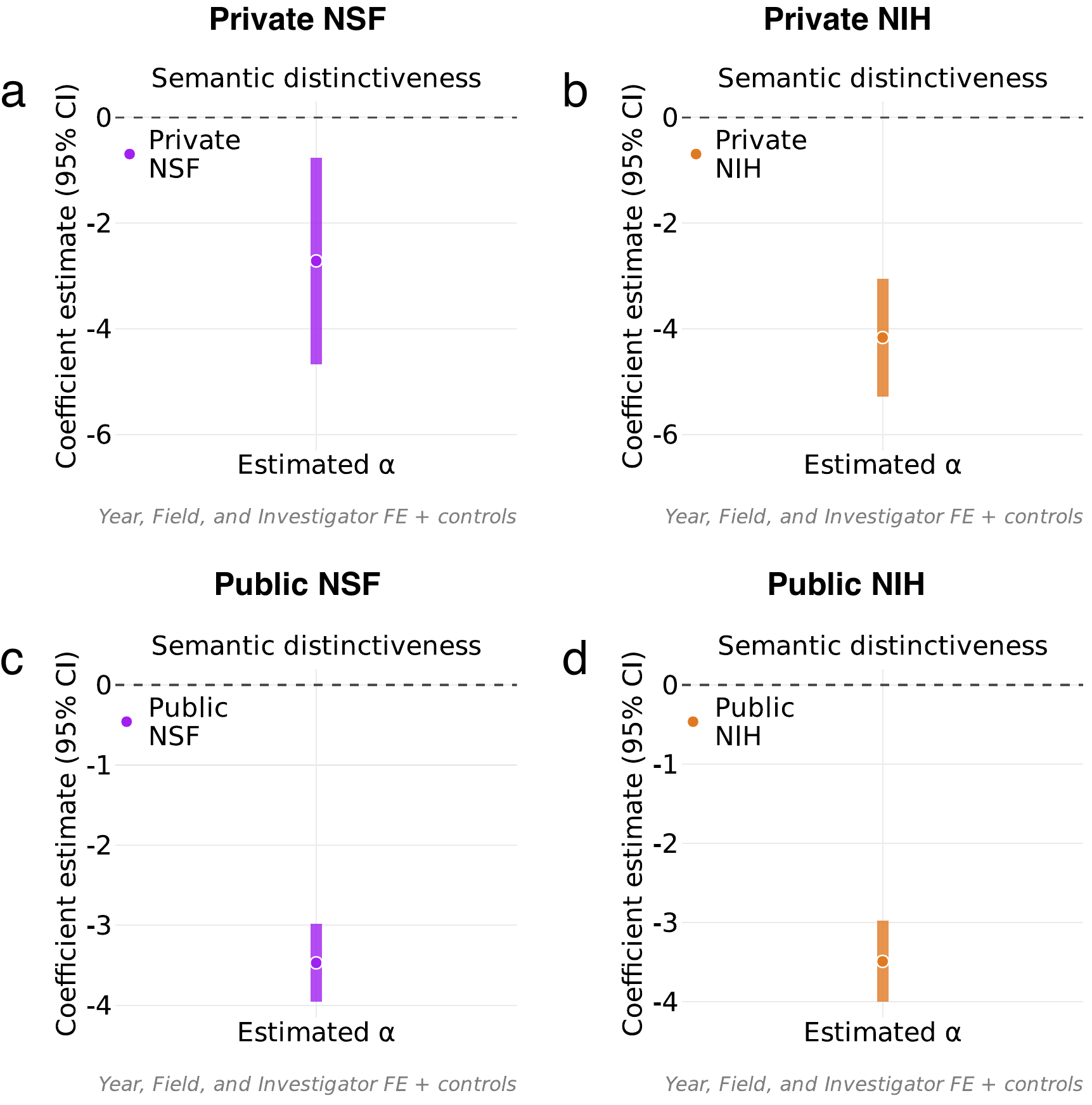}
\caption[LLM use and semantic distinctiveness with standardized predictors.]{\textbf{LLM use and semantic distinctiveness with standardized predictors.}  Regression estimates relating grant-level LLM use ($\alpha$) to semantic distance from abstracts funded in the prior year within the same agency, expressed as within-year percentiles, after standardizing all predictors to $z$-scores.
Panels show results separately for private NSF (\textbf{a}), private NIH (\textbf{b}),
public NSF (\textbf{c}), and public NIH (\textbf{d}) grants. All regressions include grant start year, field, and investigator fixed effects, as well as controls for funding amount. Points indicate coefficient estimates, and bars denote 95\% confidence intervals.}
\label{fig:Figure_comment2.7_Fig2}
\end{figure}

\newpage
\begin{figure}[htbp!]
\centering
\includegraphics[width=0.8\columnwidth]{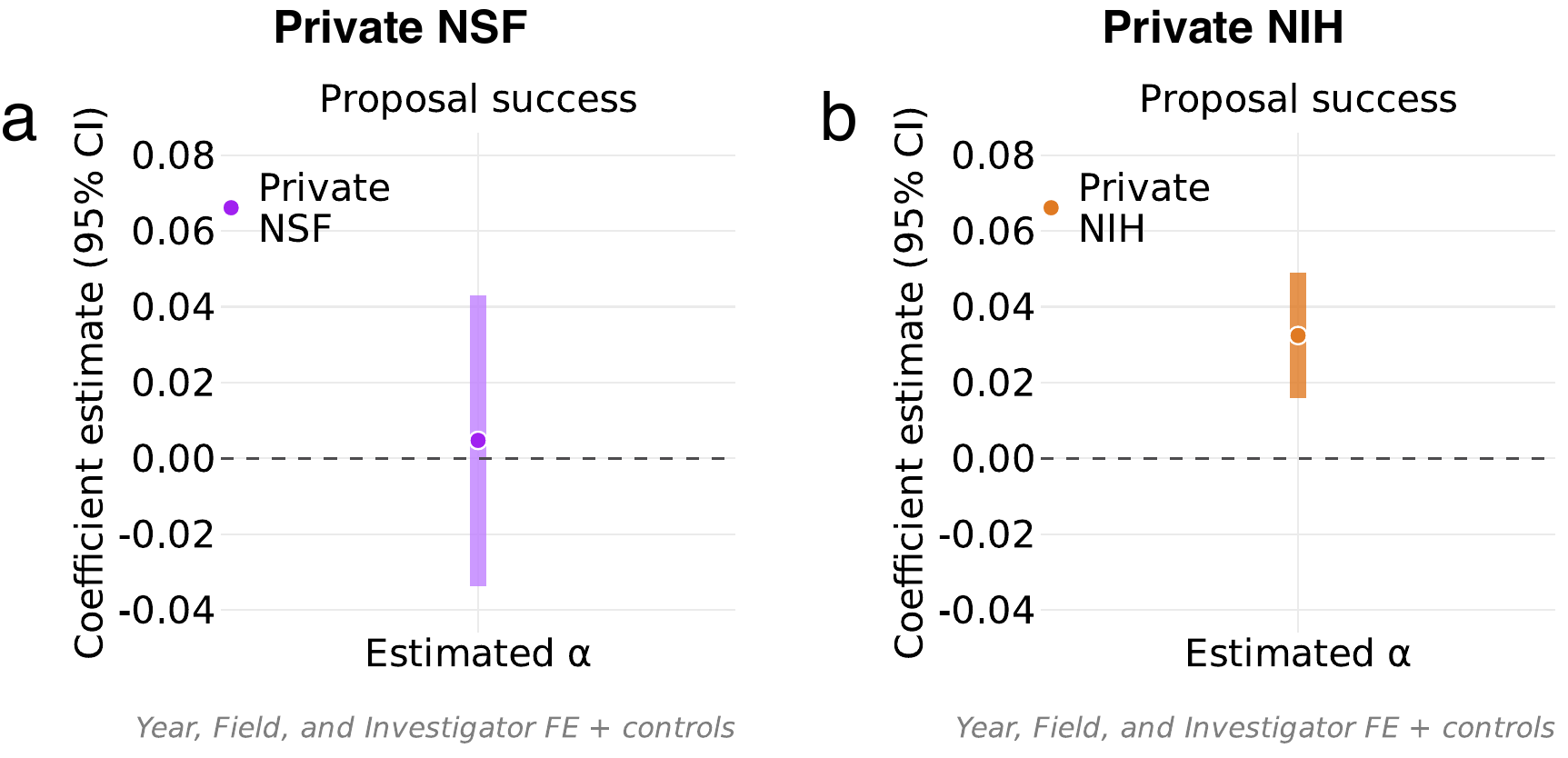}
\caption[LLM use and federal research proposal success with standardized predictors.]{\textbf{LLM use and federal research proposal success with standardized predictors.}  Regression estimates relating proposal-level LLM use ($\alpha$) to proposal success,
after standardizing all predictors to $z$-scores. (\textbf{a})~Estimates for private NSF submissions. (\textbf{b})~Corresponding estimates for private NIH submissions.
All regressions include proposal request start year, field, and investigator fixed effects, as well as controls for requested funding amount. Points indicate coefficient estimates, and bars denote 95\% confidence intervals.}
\label{fig:Figure_comment2.7_Fig3}
\end{figure}

\newpage
\begin{figure}[htbp!]
\centering
\includegraphics[width=0.8\columnwidth]{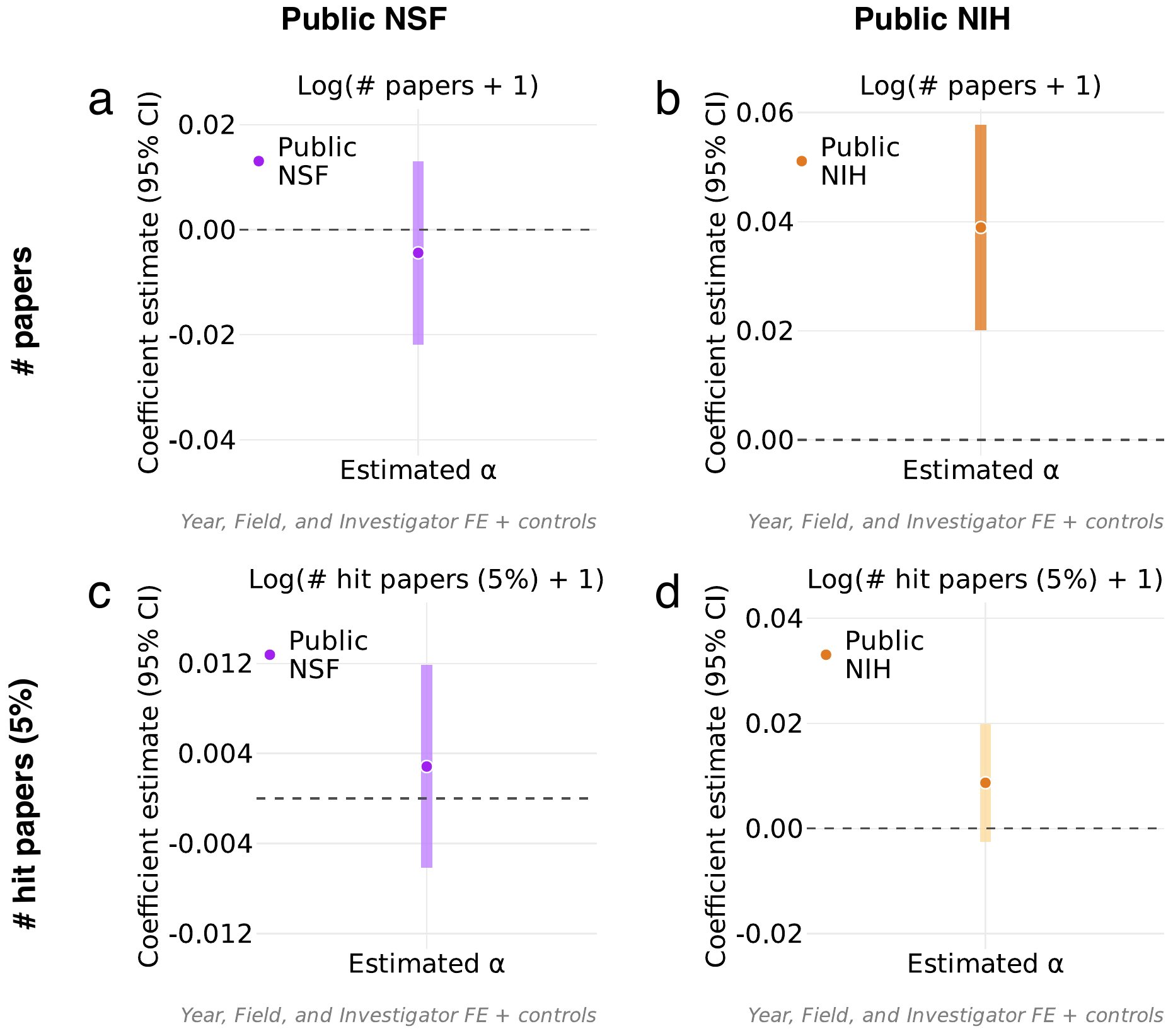}
\caption[LLM use and federal research funding outputs with standardized predictors.]{\textbf{LLM use and federal research funding outputs with standardized predictors.}  Regression estimates relating grant-level LLM use ($\alpha$) to downstream publication
outcomes, after standardizing all predictors to $z$-scores. (\textbf{a}-\textbf{b})~Estimates for total number of resulting publications for public NSF (\textbf{a}) and public NIH (\textbf{b}) grants. (\textbf{c}-\textbf{d})~Corresponding estimates for high-impact outputs, where a ``hit'' paper is defined as one whose citations fall within the top 5\% of all papers published worldwide in the same year and field, for public NSF (\textbf{c}) and public NIH (\textbf{d}). All regressions include grant start year, field, and investigator fixed effects, as well as controls for funding amount. Points indicate coefficient estimates, and bars denote 95\% confidence intervals.}
\label{fig:Figure_comment2.7_Fig4}
\end{figure}

\newpage
\begin{figure}[htbp!]
\centering
\includegraphics[width=0.8\columnwidth]{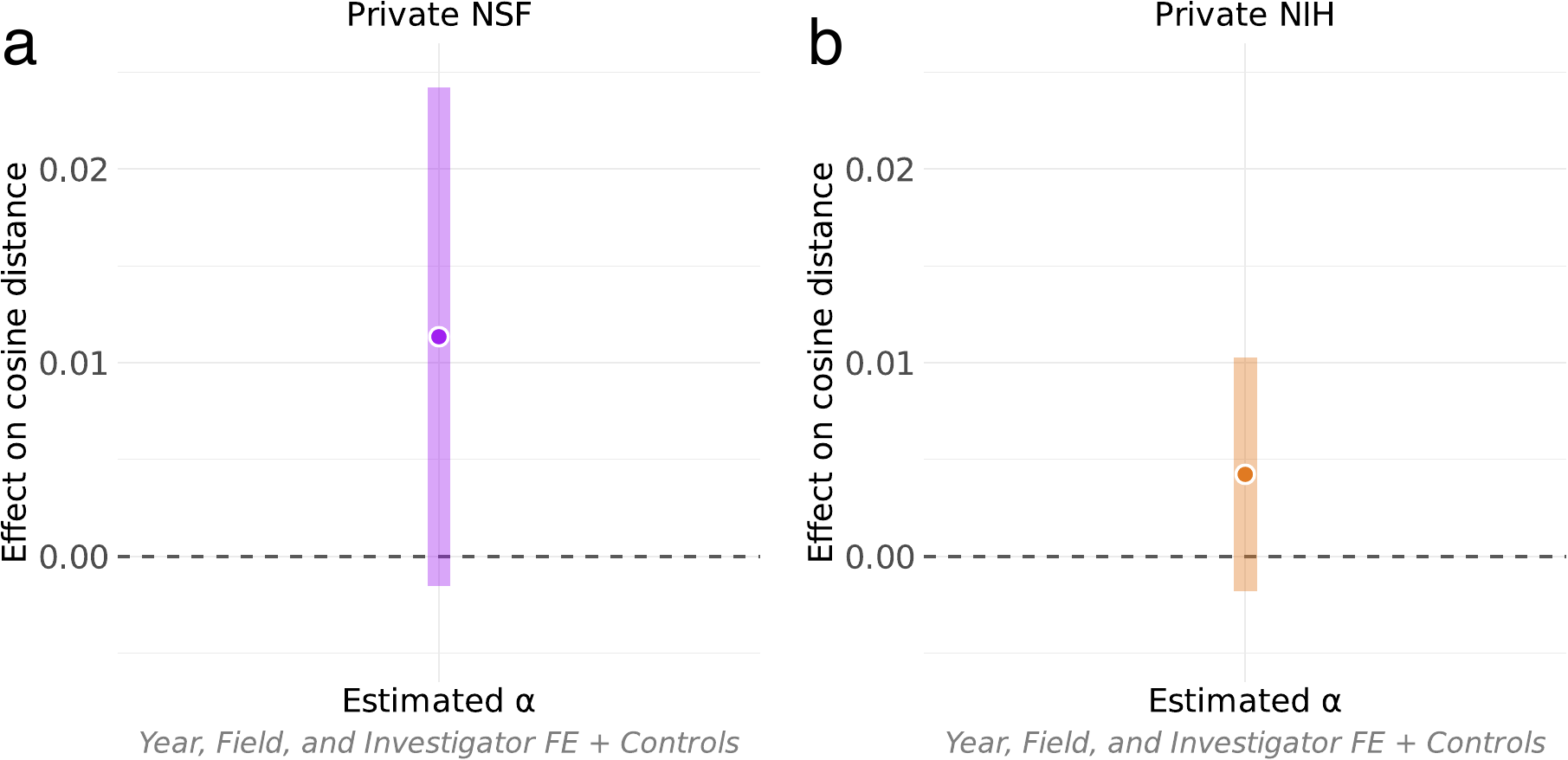}
\caption[Relationship between LLM involvement and within-investigator topical distance.]{\textbf{Relationship between LLM involvement and within-investigator topical distance.}  Coefficient estimates from regressions examining whether proposals with higher estimated LLM involvement ($\alpha$) are more distant from an investigator’s own recent proposals. The dependent variable is the cosine distance between a proposal and the investigator’s proposals submitted within the previous two years within the same agency. Panel (\textbf{a}) reports results for NSF proposals and panel (\textbf{b}) reports results for NIH proposals. Regressions include investigator, year, and field fixed effects and control for proposal funding amount. Error bars denote 95\% confidence intervals clustered at the investigator level.}
\label{fig:Figure_commnet1.3_pivot_within_investigator}
\end{figure}

\newpage
\begin{figure}[htbp!]
\centering
\includegraphics[width=0.8\columnwidth]{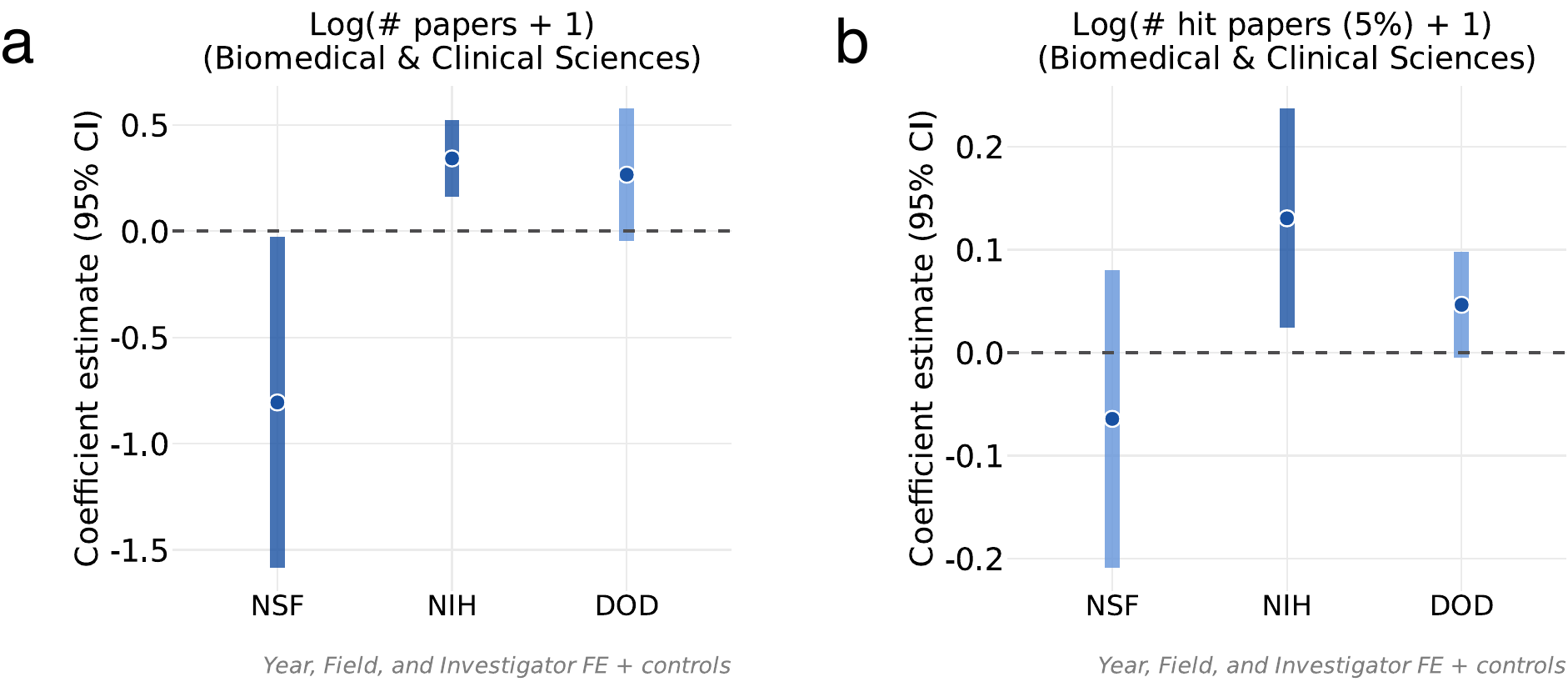}
\caption[LLM use and federal research funding outputs by agency within Biomedical \& Clinical Sciences.]{\textbf{LLM use and federal research funding outputs by agency within Biomedical \& Clinical Sciences.}  (\textbf{a}) Regression estimates relating grant-level LLM use ($\alpha$) to the total number of resulting publications (log(\# papers + 1)) for NSF, NIH, and DOD grants within Biomedical \& Clinical Sciences. (\textbf{b}) Corresponding estimates for high-impact outputs (log(\# hit papers (5\%) + 1)). The analysis is restricted to grants classified under Biomedical \& Clinical Sciences to enable direct cross-agency comparison within the same field. All regressions include grant start year and investigator fixed effects, as well as controls for funding amount. Points indicate coefficient estimates, and bars denote 95\% confidence intervals.}
\label{fig:Figure_comment1.2_Fig4}
\end{figure}

\newpage
\begin{figure}[htbp!]
\centering
\includegraphics[width=0.8\columnwidth]{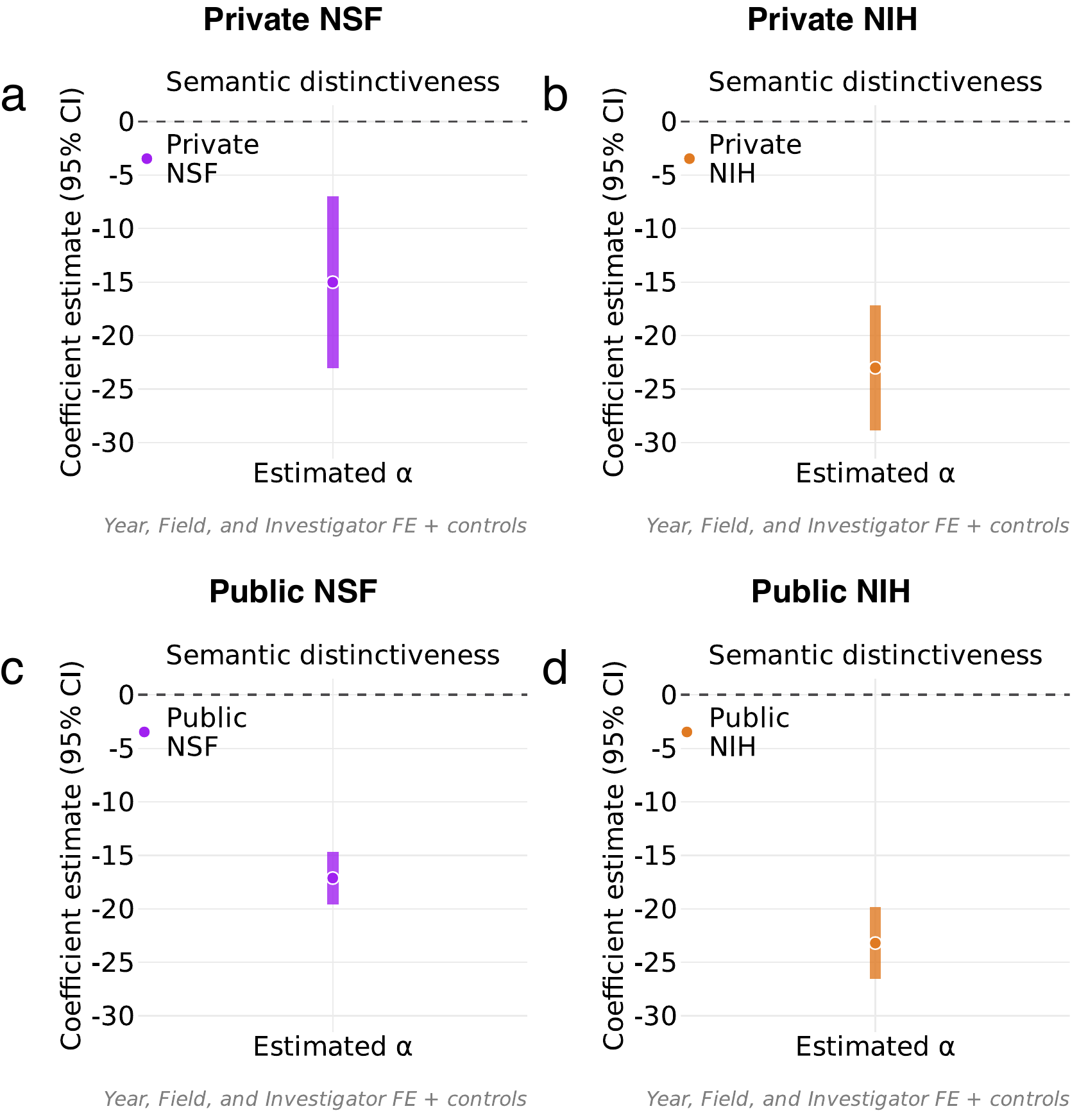}
\caption[LLM use and semantic distinctiveness using single-field unit of analysis.]{\textbf{LLM use and semantic distinctiveness using single-field unit of analysis.} Regression estimates relating grant-level LLM use ($\alpha$) to semantic distance from abstracts funded in the prior year within the same agency, expressed as within-year percentiles, restricting to one field per proposal or award. Panels show results separately for private NSF (\textbf{a}), private NIH (\textbf{b}), public NSF (\textbf{c}), and public NIH (\textbf{d}) grants. All regressions include grant start year, field, and investigator fixed effects, as well as controls for funding amount. Points indicate coefficient estimates, and bars denote 95\% confidence intervals.}
\label{fig:Figure_comment1.7_Fig2}
\end{figure}

\newpage
\begin{figure}[htbp!]
\centering
\includegraphics[width=0.8\columnwidth]{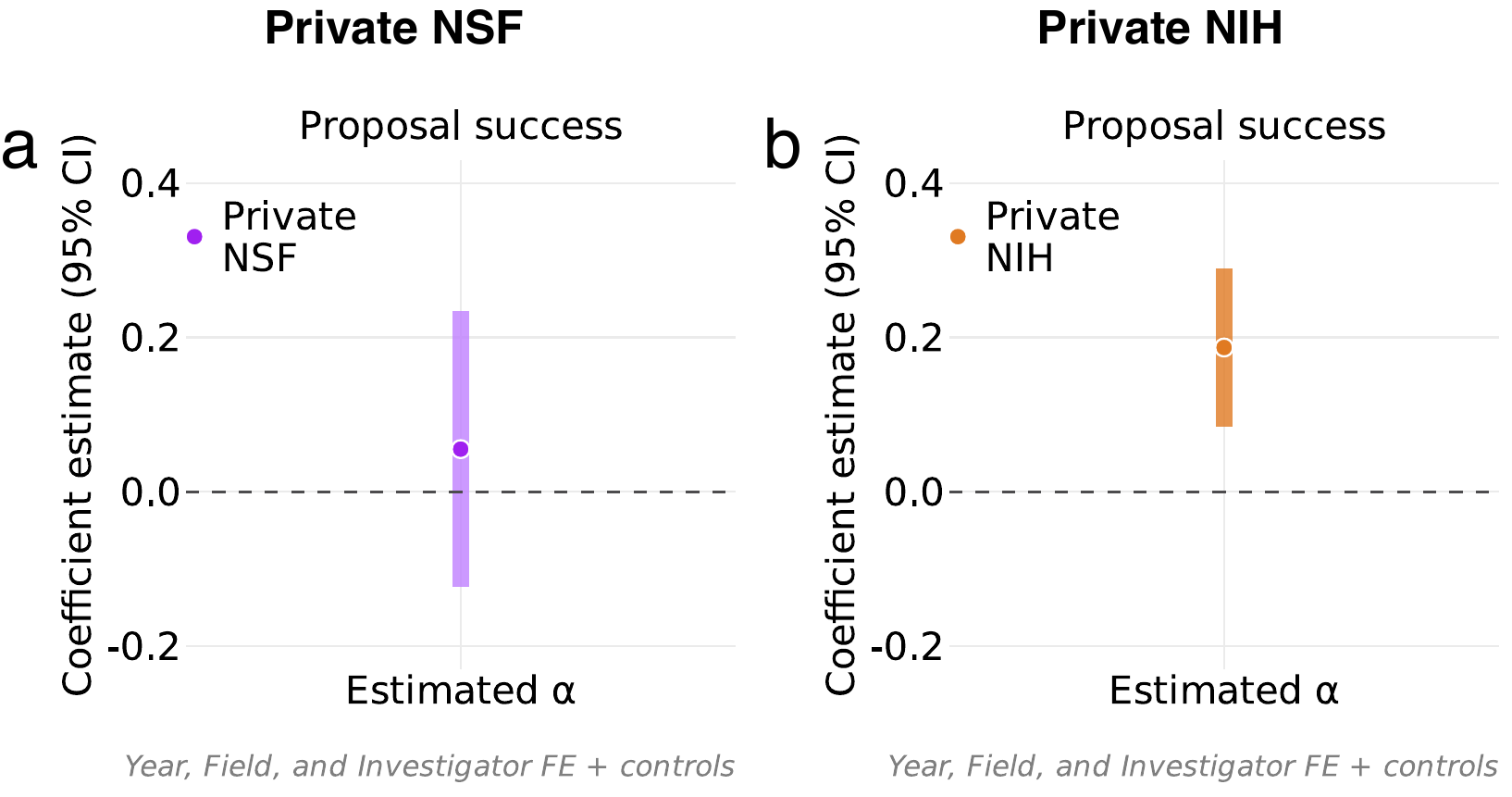}
\caption[LLM use and federal research proposal success using single-field unit of analysis.]{\textbf{LLM use and federal research proposal success using single-field unit of analysis.} Based on private NSF and NIH proposal submissions from two large US R1 universities, this figure examines the relationship between LLM use at submission ($\alpha$) and proposal success, restricting to one field per proposal. (\textbf{a}) Regression estimates for NSF submissions. (\textbf{b}) Corresponding estimates for NIH submissions. All regressions include proposal request start year, field, and investigator fixed effects, as well as controls for requested funding amount. Points indicate coefficient estimates, and bars denote 95\% confidence intervals.}
\label{fig:Figure_comment1.7_Fig3}
\end{figure}

\newpage
\begin{figure}[htbp!]
\centering
\includegraphics[width=0.8\columnwidth]{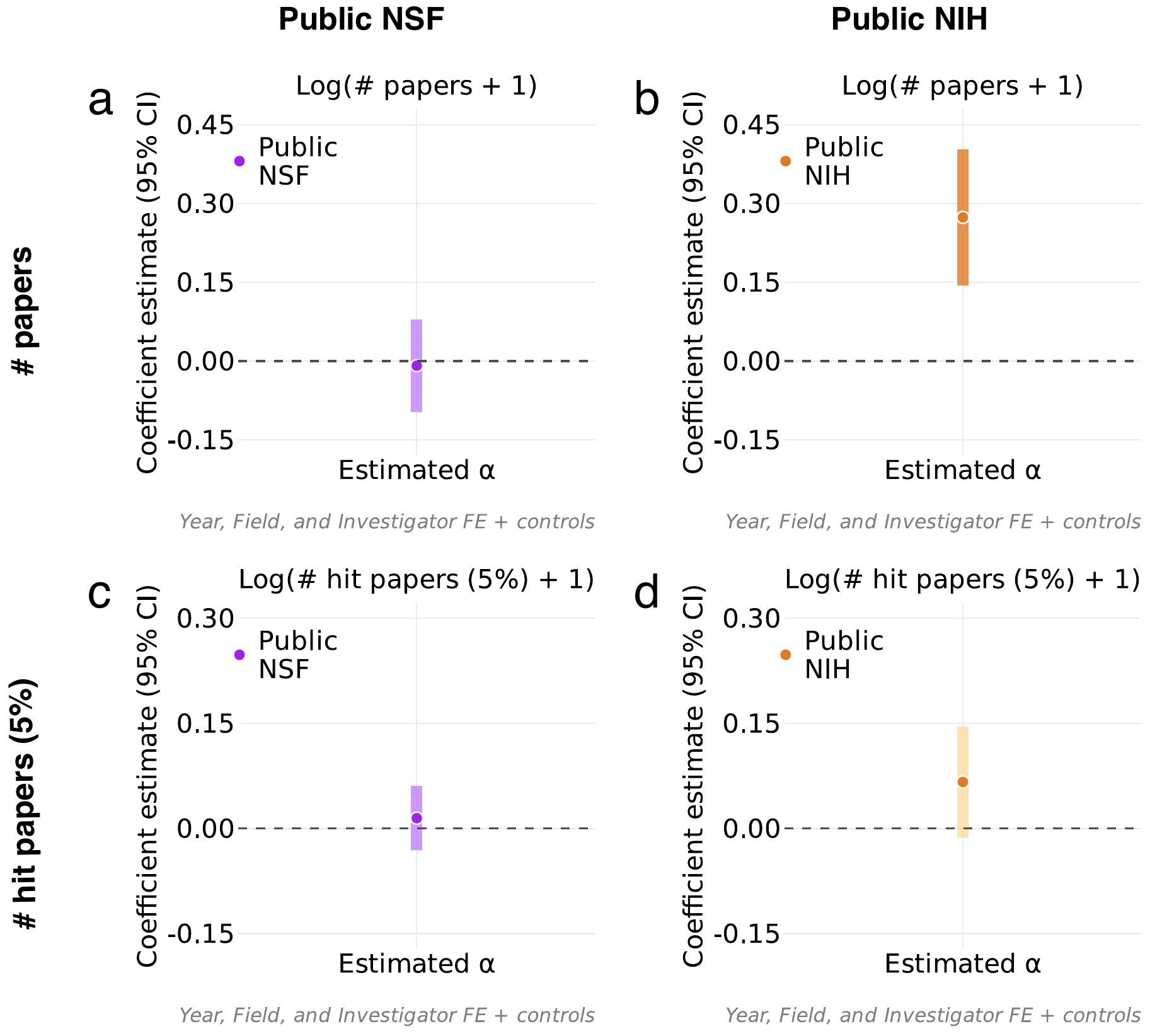}
\caption[LLM use and federal research funding outputs using single-field unit of analysis.]{\textbf{LLM use and federal research funding outputs using single-field unit of analysis.} (\textbf{a}-\textbf{b}) Regression estimates relating grant-level LLM use ($\alpha$) to the total number of resulting publications for NSF (\textbf{a}) and NIH (\textbf{b}) grants, restricting to one field per award. (\textbf{c}-\textbf{d}) Corresponding estimates for high-impact outputs, where a ``hit'' paper is defined as one whose citations fall within the top 5\% of all papers published worldwide in the same year and field. All regressions include grant start year, field, and investigator fixed effects, as well as controls for funding amount. Points indicate coefficient estimates, and bars denote 95\% confidence intervals.}
\label{fig:Figure_comment1.7_Fig4}
\end{figure}

\newpage
\section*{Supplementary Tables}
\addcontentsline{toc}{section}{Supplementary Tables}

\begin{table}[htbp]
\centering
\caption[LLM use and semantic distinctiveness (prior 1 year) in private NSF data.]{\textbf{LLM use and semantic distinctiveness (prior 1 year) in private NSF data.} 
Columns progressively add fixed effects for start year, scientific field, and investigators.}
\label{tab:PrivateNSF-Embedding-1yr}
\begin{tabular}{l *{3}{>{\raggedleft\arraybackslash}p{3.0cm}}}
\toprule
\multicolumn{4}{c}{Private NSF} \\
\cmidrule(lr){1-4}
& \multicolumn{3}{c}{Semantic distinctiveness (prior 1 year)} \\
\cmidrule(lr){2-4}
& Model 1 & Model 2 & Model 3 \\
\midrule
$\alpha$ &  -16.456***  &  -11.2727**  &  -11.9472**  \\
      &  (4.455) &  (3.9927)   &  (4.3752)    \\
log(funding) &  -0.4146  &  -0.3281  &  -1.7646*  \\
          &  (0.9348)  &  (0.9539) &  (0.8016)  \\
start year & No & Yes & Yes \\
field      & No & Yes & Yes \\
investigator         & No & No  & Yes \\
\midrule
$N.$ of Observations &  1001  &  1412  &  1886  \\
$R^2$          &  0.01886  &  0.11690  &  0.72441  \\
\bottomrule
\multicolumn{4}{l}{\footnotesize * p $<$ 0.05, ** p $<$ 0.01, *** p $<$ 0.001.
Standard errors clustered by investigator in parentheses.}
\end{tabular}
\label{table:NSF_private_1y_distance}
\end{table}

\newpage
\begin{table}[htbp]
\centering
\caption[LLM use and semantic distinctiveness (prior 1 year) in private NIH data.]{\textbf{LLM use and semantic distinctiveness (prior 1 year) in private NIH data.} 
Columns progressively add fixed effects for start year, scientific field, and investigators.}
\begin{tabular}{l *{3}{>{\raggedleft\arraybackslash}p{3.0cm}}}
\toprule
\multicolumn{4}{c}{Private NIH} \\
\cmidrule(lr){1-4}
& \multicolumn{3}{c}{Semantic distinctiveness (prior 1 year)} \\
\cmidrule(lr){2-4}
& Model 1 & Model 2 & Model 3 \\
\midrule
$\alpha$ &  -7.974*  &  -10.6152**  &  -23.3958***  \\
      &  (3.7419) &  (3.6906)   &  (3.1939)    \\
log(funding) &  0.0972  &  0.2624  &  -0.4638  \\
          &  (0.4898)  &  (0.4666) &  (0.4191)  \\
start year & No & Yes & Yes \\
field      & No & Yes & Yes \\
investigator         & No & No  & Yes \\
\midrule
$N.$ of Observations &  2532  &  3686  &  8198  \\
$R^2$        &  0.00252  &  0.06034  &  0.53894  \\
\bottomrule
\multicolumn{4}{l}{\footnotesize * p $<$ 0.05, ** p $<$ 0.01, *** p $<$ 0.001.
Standard errors clustered by investigator in parentheses.}
\end{tabular}
\label{table:NIH_private_1y_distance}
\end{table}

\newpage
\begin{table}[htbp]
\centering
\caption[LLM use and semantic distinctiveness (prior 1 year) in public NSF data.]{\textbf{LLM use and semantic distinctiveness (prior 1 year) in public NSF data.} 
Columns progressively add fixed effects for start year, scientific field, and investigators.}
\begin{tabular}{l *{3}{>{\raggedleft\arraybackslash}p{3.0cm}}}
\toprule
\multicolumn{4}{c}{Public NSF} \\
\cmidrule(lr){1-4}
& \multicolumn{3}{c}{Semantic distinctiveness (prior 1 year)} \\
\cmidrule(lr){2-4}
& Model 1 & Model 2 & Model 3 \\
\midrule
$\alpha$ &  -24.539***  &  -18.6069***  &  -18.0163***  \\
      &  (0.871) &  (0.8992)   &  (1.2958)    \\
log(funding) &  1.2935***  &  1.4454***  &  0.7485***  \\
          &  (0.1517)  &  (0.1487) &  (0.184)  \\
start year & No & Yes & Yes \\
field      & No & Yes & Yes \\
investigator         & No & No  & Yes \\
\midrule
$N.$ of Observations &  32278  &  46504  &  60283  \\
$R^2$          &  0.02805  &  0.07757  &  0.78893  \\
\bottomrule
\multicolumn{4}{l}{\footnotesize * p $<$ 0.05, ** p $<$ 0.01, *** p $<$ 0.001.
Standard errors clustered by investigator in parentheses.}
\end{tabular}
\label{table:NSF_public_1y_distance}
\end{table}

\newpage
\begin{table}[htbp]
\centering
\caption[LLM use and semantic distinctiveness (prior 1 year) in public NIH data.]{\textbf{LLM use and semantic distinctiveness (prior 1 year) in public NIH data.} 
Columns progressively add fixed effects for start year, scientific field, and investigators.}
\begin{tabular}{l *{3}{>{\raggedleft\arraybackslash}p{3.0cm}}}
\toprule
\multicolumn{4}{c}{Public NIH} \\
\cmidrule(lr){1-4}
& \multicolumn{3}{c}{Semantic distinctiveness (prior 1 year)} \\
\cmidrule(lr){2-4}
& Model 1 & Model 2 & Model 3 \\
\midrule
$\alpha$ &  -11.093***  &  -17.6547***  &  -23.6247***  \\
      &  (1.1549) &  (1.1611)   &  (1.7726)    \\
log(funding) &  0.9335***  &  0.7318***  &  0.7554**  \\
          &  (0.2288)  &  (0.2048) &  (0.2471)  \\
start year & No & Yes & Yes \\
field      & No & Yes & Yes \\
investigator         & No & No  & Yes \\
\midrule
$N.$ of Observations &  36617  &  54033  &  56413  \\
$R^2$          &  0.00507  &  0.10213  &  0.86886  \\
\bottomrule
\multicolumn{4}{l}{\footnotesize * p $<$ 0.05, ** p $<$ 0.01, *** p $<$ 0.001.
Standard errors clustered by investigator in parentheses.}
\end{tabular}
\label{table:NIH_public_1y_distance}
\end{table}

\newpage
\begin{table}[htbp]
\centering
\caption[LLM use and proposal success in private NSF data.]{\textbf{LLM use and proposal success in private NSF data.} 
Columns progressively add fixed effects for start year, scientific field, and investigators.}
\begin{tabular}{l *{3}{>{\raggedleft\arraybackslash}p{3.0cm}}}
\toprule
\multicolumn{4}{c}{Private NSF} \\
\cmidrule(lr){1-4}
& \multicolumn{3}{c}{Proposal success} \\
\cmidrule(lr){2-4}
& Model 1 & Model 2 & Model 3 \\
\midrule
$\alpha$ &  0.0323  &  -0.0086  &  0.0227  \\
      &  (0.0674) &  (0.0728)   &  (0.0936)    \\
log(funding) &  -0.0728**  &  -0.0846***  &  -0.0292  \\
          &  (0.0225)  &  (0.0214) &  (0.0205)  \\
start year & No & Yes & Yes \\
field      & No & Yes & Yes \\
investigator         & No & No  & Yes \\
\midrule
$N.$ of Observations &  816  &  1145  &  1464  \\
$R^2$          &  0.03969  &  0.06128  &  0.56523  \\
\bottomrule
\multicolumn{4}{l}{\footnotesize * p $<$ 0.05, ** p $<$ 0.01, *** p $<$ 0.001.
Standard errors clustered by investigator in parentheses.}
\end{tabular}
\label{table:NSF_private_success}
\end{table}

\newpage
\begin{table}[htbp]
\centering
\caption[LLM use and proposal success in private NIH data.]{\textbf{LLM use and proposal success in private NIH data.} 
Columns progressively add fixed effects for start year, scientific field, and investigators.}
\begin{tabular}{l *{3}{>{\raggedleft\arraybackslash}p{3.0cm}}}
\toprule
\multicolumn{4}{c}{Private NIH} \\
\cmidrule(lr){1-4}
& \multicolumn{3}{c}{Proposal success} \\
\cmidrule(lr){2-4}
& Model 1 & Model 2 & Model 3 \\
\midrule
$\alpha$ &  0.039  &  0.0599  &  0.1906***  \\
      &  (0.0575) &  (0.0595)   &  (0.0496)    \\
log(funding) &  -0.0522***  &  -0.0502***  &  -0.0443***  \\
          &  (0.0091)  &  (0.0086) &  (0.0096)  \\
start year & No & Yes & Yes \\
field      & No & Yes & Yes \\
investigator         & No & No  & Yes \\
\midrule
$N.$ of Observations &  1783  &  2608  &  5632  \\
$R^2$         &  0.03124  &  0.04564  &  0.37319  \\
\bottomrule
\multicolumn{4}{l}{\footnotesize * p $<$ 0.05, ** p $<$ 0.01, *** p $<$ 0.001.
Standard errors clustered by investigator in parentheses.}
\end{tabular}
\label{table:NIH_private_success}
\end{table}

\newpage
\begin{table}[htbp]
\centering
\caption[LLM use and downstream output (\# papers) in public NSF data.]{\textbf{LLM use and downstream output (\# papers) in public NSF data.} 
Columns progressively add fixed effects for start year, scientific field, and investigators.}
\begin{tabular}{l *{3}{>{\raggedleft\arraybackslash}p{3.0cm}}}
\toprule
\multicolumn{4}{c}{Public NSF} \\
\cmidrule(lr){1-4}
& \multicolumn{3}{c}{Log(\# papers + 1)} \\
\cmidrule(lr){2-4}
& Model 1 & Model 2 & Model 3 \\
\midrule
$\alpha$ &  -0.0981***  &  0.1357***  &  -0.0232  \\
      &  (0.0294) &  (0.0287)   &  (0.0471)    \\
log(funding) &  0.2024***  &  0.1984***  &  0.2636***  \\
          &  (0.0046)  &  (0.0044) &  (0.0075)  \\
start year & No & Yes & Yes \\
field      & No & Yes & Yes \\
investigator         & No & No  & Yes \\
\midrule
$N.$ of Observations &  23741  &  34193  &  43415  \\
$R^2$          &  0.08056  &  0.20724  &  0.84442  \\
\bottomrule
\multicolumn{4}{l}{\footnotesize * p $<$ 0.05, ** p $<$ 0.01, *** p $<$ 0.001.
Standard errors clustered by investigator in parentheses.}
\end{tabular}
\label{table:NSF_public_num_of_papers}
\end{table}

\newpage
\begin{table}[htbp]
\centering
\caption[LLM use and downstream output (\# papers) in public NIH data.]{\textbf{LLM use and downstream output (\# papers) in public NIH data.} 
Columns progressively add fixed effects for start year, scientific field, and investigators.}
\begin{tabular}{l *{3}{>{\raggedleft\arraybackslash}p{3.0cm}}}
\toprule
\multicolumn{4}{c}{Public NIH} \\
\cmidrule(lr){1-4}
& \multicolumn{3}{c}{Log(\# papers + 1)} \\
\cmidrule(lr){2-4}
& Model 1 & Model 2 & Model 3 \\
\midrule
$\alpha$ &  -0.2395***  &  -0.0369  &  0.2739***  \\
      &  (0.0397) &  (0.0406)   &  (0.0676)    \\
log(funding) &  0.1965***  &  0.1749***  &  0.2957***  \\
          &  (0.0047)  &  (0.0046) &  (0.0105)  \\
start year & No & Yes & Yes \\
field      & No & Yes & Yes \\
investigator         & No & No  & Yes \\
\midrule
$N.$ of Observations &  30548  &  45185  &  46577  \\
$R^2$          &  0.07682  &  0.12805  &  0.86251  \\
\bottomrule
\multicolumn{4}{l}{\footnotesize * p $<$ 0.05, ** p $<$ 0.01, *** p $<$ 0.001.
Standard errors clustered by investigator in parentheses.}
\end{tabular}
\label{table:NIH_public_num_of_papers}
\end{table}

\newpage
\begin{table}[htbp]
\centering
\caption[LLM use and downstream output (\# top 5\% hit papers) in public NSF data.]{\textbf{LLM use and downstream output (\# top 5\% hit papers) in public NSF data.} 
Columns progressively add fixed effects for start year, scientific field, and investigators.}
\begin{tabular}{l *{3}{>{\raggedleft\arraybackslash}p{3.0cm}}}
\toprule
\multicolumn{4}{c}{Public NSF} \\
\cmidrule(lr){1-4}
& \multicolumn{3}{c}{Log(\# hit papers(5\%) + 1)} \\
\cmidrule(lr){2-4}
& Model 1 & Model 2 & Model 3 \\
\midrule
$\alpha$ &  -0.0202  &  0.0514***  &  0.015  \\
      &  (0.0124) &  (0.0127)   &  (0.0243)    \\
log(funding) &  0.0537***  &  0.053***  &  0.074***  \\
          &  (0.0025)  &  (0.0025) &  (0.0043)  \\
start year & No & Yes & Yes \\
field      & No & Yes & Yes \\
investigator         & No & No  & Yes \\
\midrule
$N.$ of Observations &  23741  &  34193  &  43415  \\
$R^2$          &  0.03012  &  0.07571  &  0.77834  \\
\bottomrule
\multicolumn{4}{l}{\footnotesize * p $<$ 0.05, ** p $<$ 0.01, *** p $<$ 0.001.
Standard errors clustered by investigator in parentheses.}
\end{tabular}
\label{table:NSF_public_num_of_hit_5pct_papers}
\end{table}

\newpage
\begin{table}[htbp]
\centering
\caption[LLM use and downstream output (\# top 5\% hit papers) in public NIH data.]{\textbf{LLM use and downstream output (\# top 5\% hit papers) in public NIH data.} 
Columns progressively add fixed effects for start year, scientific field, and investigators.}
\begin{tabular}{l *{3}{>{\raggedleft\arraybackslash}p{3.0cm}}}
\toprule
\multicolumn{4}{c}{Public NIH} \\
\cmidrule(lr){1-4}
& \multicolumn{3}{c}{Log(\# hit papers(5\%) + 1)} \\
\cmidrule(lr){2-4}
& Model 1 & Model 2 & Model 3 \\
\midrule
$\alpha$ &  -0.0848***  &  -0.0088  &  0.0612  \\
      &  (0.0188) &  (0.0194)   &  (0.0403)    \\
log(funding) &  0.0833***  &  0.0756***  &  0.1161***  \\
          &  (0.0026)  &  (0.0025) &  (0.0061)  \\
start year & No & Yes & Yes \\
field      & No & Yes & Yes \\
investigator         & No & No  & Yes \\
\midrule
$N.$ of Observations &  30548  &  45185  &  46577  \\
$R^2$          &  0.04699  &  0.07676  &  0.83363  \\
\bottomrule
\multicolumn{4}{l}{\footnotesize * p $<$ 0.05, ** p $<$ 0.01, *** p $<$ 0.001.
Standard errors clustered by investigator in parentheses.}
\end{tabular}
\label{table:NIH_public_num_of_hit_5pct_papers}
\end{table}

\newpage
\begin{table}[htbp]
\centering
\caption[LLM use and downstream output (\# top 1\% hit papers) in public NSF data.]{\textbf{LLM use and downstream output (\# top 1\% hit papers) in public NSF data.} 
Columns progressively add fixed effects for start year, scientific field, and investigators.}
\begin{tabular}{l *{3}{>{\raggedleft\arraybackslash}p{3.0cm}}}
\toprule
\multicolumn{4}{c}{Public NSF} \\
\cmidrule(lr){1-4}
& \multicolumn{3}{c}{Log(\# hit papers(1\%) + 1)} \\
\cmidrule(lr){2-4}
& Model 1 & Model 2 & Model 3 \\
\midrule
$\alpha$ &  -0.0048  &  0.0195**  &  0.0087  \\
      &  (0.0063) &  (0.0064)   &  (0.0129)    \\
log(funding) &  0.02***  &  0.0197***  &  0.0275***  \\
          &  (0.0015)  &  (0.0015) &  (0.0025)  \\
start year & No & Yes & Yes \\
field      & No & Yes & Yes \\
investigator         & No & No  & Yes \\
\midrule
$N.$ of Observations &  23741  &  34193  &  43415  \\
$R^2$          &  0.01641  &  0.03586  &  0.73823  \\
\bottomrule
\multicolumn{4}{l}{\footnotesize * p $<$ 0.05, ** p $<$ 0.01, *** p $<$ 0.001.
Standard errors clustered by investigator in parentheses.}
\end{tabular}
\label{table:NSF_public_num_of_hit_1pct_papers}
\end{table}

\newpage
\begin{table}[htbp]
\centering
\caption[LLM use and downstream output (\# top 1\% hit papers) in public NIH data.]{\textbf{LLM use and downstream output (\# top 1\% hit papers) in public NIH data.} 
Columns progressively add fixed effects for start year, scientific field, and investigators.}
\begin{tabular}{l *{3}{>{\raggedleft\arraybackslash}p{3.0cm}}}
\toprule
\multicolumn{4}{c}{Public NIH} \\
\cmidrule(lr){1-4}
& \multicolumn{3}{c}{Log(\# hit papers(1\%) + 1)} \\
\cmidrule(lr){2-4}
& Model 1 & Model 2 & Model 3 \\
\midrule
$\alpha$ &  -0.0307**  &  -0.0039  &  -0.0221  \\
      &  (0.0107) &  (0.0112)   &  (0.0262)    \\
log(funding) &  0.0342***  &  0.031***  &  0.0435***  \\
          &  (0.0015)  &  (0.0015) &  (0.0036)  \\
start year & No & Yes & Yes \\
field      & No & Yes & Yes \\
investigator         & No & No  & Yes \\
\midrule
$N.$ of Observations &  30548  &  45185  &  46577  \\
$R^2$          &  0.02381  &  0.03764  &  0.81908  \\
\bottomrule
\multicolumn{4}{l}{\footnotesize * p $<$ 0.05, ** p $<$ 0.01, *** p $<$ 0.001.
Standard errors clustered by investigator in parentheses.}
\end{tabular}
\label{table:NIH_public_num_of_hit_1pct_papers}
\end{table}

\newpage
\begin{table}[htbp]
\centering
\caption{\textbf{LLM use and semantic distinctiveness (prior 1 year) with standardized predictors.}
All predictors are converted to $z$-scores before estimation.}
\label{table:Table_comment2.7_Fig2}
\begin{tabular}{lcccc}
\hline
 & \multicolumn{4}{c}{Semantic distinctiveness (prior 1 year, percentile)} \\
 & Private NSF & Private NIH & Public NSF & Public NIH \\
\hline
$\alpha$ & -2.7161**   & -4.1661***  & -3.4698***  & -3.4898***  \\
         & (0.9947)    & (0.5687)    & (0.2496)    & (0.2618)    \\[4pt]
log(funding) & -2.4914* & -0.6565    & 0.9812***   & 0.9423**    \\
             & (1.1317) & (0.5932)   & (0.2412)    & (0.3082)    \\[4pt]
start year   & Yes      & Yes        & Yes         & Yes         \\
field        & Yes      & Yes        & Yes         & Yes         \\
investigator & Yes      & Yes        & Yes         & Yes         \\[4pt]
N            & 1,886    & 8,198      & 60,283      & 56,413      \\
$R^{2}$      & 0.72441  & 0.53894    & 0.78893     & 0.86886     \\
\hline
\multicolumn{5}{l}{\footnotesize * p $<$ 0.05, ** p $<$ 0.01, *** p $<$ 0.001.
Standard errors clustered by investigator in parentheses.}
\end{tabular}
\end{table}

\newpage
\begin{table}[htbp]
\centering
\caption{\textbf{LLM use and proposal success with standardized predictors.}
All predictors are converted to $z$-scores before estimation.}
\label{table:Table_comment2.7_Fig3}
\begin{tabular}{lcc}
\hline
 & \multicolumn{2}{c}{Proposal success} \\
 & Private NSF & Private NIH \\
\hline
$\alpha$ & 0.0047    & 0.0324***  \\
         & (0.0195)  & (0.0084)   \\[4pt]
log(funding) & -0.0436 & -0.0667*** \\
             & (0.0306) & (0.0145)  \\[4pt]
start year   & Yes      & Yes       \\
field        & Yes      & Yes       \\
investigator & Yes      & Yes       \\[4pt]
N            & 1,464    & 5,632     \\
$R^{2}$      & 0.56523  & 0.37319   \\
\hline
\multicolumn{3}{l}{\footnotesize * p $<$ 0.05, ** p $<$ 0.01, *** p $<$ 0.001.
Standard errors clustered by investigator in parentheses.}
\end{tabular}
\end{table}

\newpage
\begin{table}[htbp]
\centering
\caption{\textbf{LLM use and publication outcomes with standardized predictors.}
All predictors are converted to $z$-scores before estimation.}
\label{table:Table_comment2.7_Fig4}
\begin{tabular}{lcccc}
\hline
 & \multicolumn{2}{c}{\# papers} & \multicolumn{2}{c}{\# hit papers (5\%)} \\
 & Public NSF & Public NIH & Public NSF & Public NIH \\
\hline
$\alpha$ & -0.0044    & 0.0389***  & 0.0028     & 0.0087     \\
         & (0.0089)   & (0.0096)   & (0.0046)   & (0.0057)   \\[4pt]
log(funding) & 0.3456*** & 0.369*** & 0.0971***  & 0.1449***  \\
             & (0.0098)  & (0.0131) & (0.0056)   & (0.0076)   \\[4pt]
start year   & Yes       & Yes      & Yes        & Yes        \\
field        & Yes       & Yes      & Yes        & Yes        \\
investigator & Yes       & Yes      & Yes        & Yes        \\[4pt]
N            & 43,415    & 46,577   & 43,415     & 46,577     \\
$R^{2}$      & 0.84442   & 0.86251  & 0.77834    & 0.83363    \\
\hline
\multicolumn{5}{l}{\footnotesize * p $<$ 0.05, ** p $<$ 0.01, *** p $<$ 0.001.
Standard errors clustered by investigator in parentheses.}
\end{tabular}
\end{table}

\newpage
% \bibliographystyle{naturemag}
% \bibliography{references}

\end{document}